\documentclass[twoside]{article}

\usepackage[preprint]{aistats2027}
\usepackage[round]{natbib}

\usepackage{amsmath,amssymb,amsfonts,amsthm,mathtools}
\usepackage{booktabs,multirow,array}
\usepackage{enumitem}
\usepackage{graphicx}
\usepackage{xcolor}
\usepackage{hyperref}
\usepackage{microtype}
\usepackage{algorithm}
\usepackage{algcompatible}
\usepackage{caption}
\usepackage{subcaption}
\usepackage{float}
\usepackage{xspace}
\hypersetup{colorlinks=true,linkcolor=blue,citecolor=blue,urlcolor=blue}

\newtheorem{assumption}{Assumption}
\newtheorem{definition}{Definition}
\newtheorem{proposition}{Proposition}
\newtheorem{lemma}{Lemma}
\newtheorem{theorem}{Theorem}
\newtheorem{corollary}{Corollary}
\newtheorem{remark}{Remark}

\newcommand{\ket}[1]{\left|#1\right\rangle}
\newcommand{\bra}[1]{\left\langle#1\right|}

\newcommand{\icaq}{\emph{ICAQKL}}
\newcommand{\nystrom}{Nystr\"om}

\usepackage{enumitem}

\setlist[itemize]{noitemsep,topsep=2pt}

\usepackage{amsmath,amsfonts,bm}

\def\eqref#1{equation~\ref{#1}}

\def\1{\bm{1}}

\DeclareMathAlphabet{\mathsfit}{\encodingdefault}{\sfdefault}{m}{sl}
\SetMathAlphabet{\mathsfit}{bold}{\encodingdefault}{\sfdefault}{bx}{n}

\begin{document}

\runningtitle{Adaptive Quantum Kernel Acquisition}

\twocolumn[
    \vspace{-1cm}
    
    \aistatstitle{Adaptive Low-Rank Quantum Kernel Acquisition under Finite-Shot Measurements: Perturbation Guarantees for QSVMs}
    
    \vspace{-1cm}
    
    \aistatsauthor{ Duong The Do \And Duong Tung Nguyen }
    
    \aistatsaddress{ School of ECEE \\Arizona State University \\Tempe, AZ 85281, USA \And  School of ECEE \\ Arizona State University \\ Tempe, AZ 85281, USA } 
]

\begin{abstract}
    Quantum kernel methods require estimating entries of a dense Gram matrix through repeated circuit measurements, making kernel construction expensive and statistically noisy. For a dataset of size $n$, dense construction requires $O(n^2)$ kernel queries, while existing approximation methods typically assume inexpensive access to kernel entries or use non-adaptive sampling. We study adaptive low-rank quantum kernel acquisition when the kernel is accessible only through a finite-shot quantum overlap oracle. Our framework, \icaq, uses residual-driven incomplete Cholesky construction to acquire informative kernel columns sequentially and form a positive-semidefinite rank-$r$ surrogate, reducing the training-kernel acquisition cost to $O(nrN_s)$ circuit executions for $r\ll n$, where $N_s$ is the number of shots per queried entry. The central difficulty is that finite-shot errors perturb both the queried kernel values and the adaptive pivot sequence that determines the low-rank factor. We develop a non-asymptotic perturbation framework for this stochastic adaptive process, establishing concentration over the realized queries, stability of noisy pivot selection and factor construction, and propagation of the coupled sampling and approximation errors to the QSVM objective, optimizer, prediction function, and classification margin. The resulting rank--shot conditions characterize when \icaq approaches the full QSVM induced by the same quantum feature map. Experiments on benchmark and biomedical datasets show favorable accuracy--query-cost tradeoffs, competitive performance with the full QSVM, and competitive or improved performance relative to the evaluated Nystr\"om baselines under comparable settings. \icaq is intended to enable measurement-efficient and theoretically controlled use of a selected quantum kernel, rather than to establish quantum advantage over classical kernels.
\end{abstract}

\section{INTRODUCTION}
\label{sec1}
    Kernel methods provide a principled framework for supervised learning, combining nonlinear decision boundaries with convex training and a clear geometric interpretation through reproducing kernel Hilbert spaces (RKHS) \cite{Hofmann2007, Talitckii2024}. 
    In quantum machine learning, this perspective is particularly natural: a quantum feature map embeds a classical input into a quantum state, and pairwise similarities between the resulting states define a kernel that can subsequently be used by a classical kernel learner \cite{Havlicek2019, Schuld2019, Schuld2021}.
    This forms the foundation of quantum kernel methods and quantum support vector machines (QSVMs), a prominent paradigm for near-term quantum-enhanced supervised learning \cite{Havlicek2019, Schuld2019, Schuld2021}.

    A central practical limitation of quantum kernel learning (QKL) is the cost of acquiring the kernel matrix \cite{Thanasilp2024, RodriguezGrasa2025, Coelho2025}. For $n$ training samples, a dense kernel method requires $\Theta(n^2)$ distinct pairwise similarities \cite{Schuld2021}. In the quantum setting, each nontrivial similarity is itself estimated from repeated circuit measurements rather than obtained as an inexpensive deterministic function evaluation. Consequently, with $N_{\mathrm{s}}$ shots per queried entry, dense kernel acquisition incurs $\Theta(n^2N_{\mathrm{s}})$ circuit repetitions in addition to the $O(n^2)$ storage requirement \cite{Havlicek2019}. For normalized fidelity kernels, the diagonal entries are known analytically, but the remaining $n(n-1)/2$ off-diagonal similarities still scale quadratically with the number of training examples. This measurement burden can dominate the cost of the learning pipeline and limits the size of datasets for which a dense QSVM is practical.

    Low-rank kernel approximation provides a natural route to reducing this cost. Classical approaches include randomized matrix approximations, Random Fourier Features, Nystr\"om methods, and Cholesky-type factorizations \cite{Mahoney2011, Rahimi2007, Hayakawa2023,Du2026}. These methods were primarily developed for settings in which kernel entries can be accessed relatively cheaply and accurately. Their use with quantum kernels introduces an additional issue: the approximation algorithm itself determines which costly quantum similarities must be measured, while every such measurement is subject to finite-shot uncertainty. For example, a Nystr\"om approximation queries similarities involving a selected landmark set, whereas pivoted Cholesky selects subsequent columns adaptively according to the residual structure of the approximation. Thus, in QKL, the low-rank approximation  also acts as a \emph{query-allocation mechanism} for a stochastic quantum-kernel oracle.

    Motivated by this observation, we develop \icaq, a finite-shot adaptive low-rank framework for quantum kernel acquisition based on incomplete Cholesky factorization. We emphasize that pivoted incomplete Cholesky itself is a classical matrix approximation technique; our contribution is its formulation and analysis under adaptive finite-shot quantum-kernel access. Rather than constructing the dense Gram matrix, \icaq~queries only the kernel entries required by the adaptively selected pivot columns and forms the rank-$r$ positive-semidefinite surrogate 
    \begin{align} 
        \widetilde{\mathbf K}_r^{Q,(N_{\mathrm{s}})} = \widehat{\mathbf G}_r^{(N_{\mathrm{s}})} \big( \widehat{\mathbf G}_r^{(N_{\mathrm{s}})} \big)^\top . 
    \end{align} 
    Under symmetry, caching, and the analytically known fidelity diagonal, a rank-$r$ construction requires at most $O(nr)$ fresh quantum-kernel queries, compared with $\Theta(n^2)$ for dense acquisition. When $r\ll n$, this reduces the corresponding measurement count from $\Theta(n^2N_{\mathrm{s}})$ to $O(nrN_{\mathrm{s}})$.

    The main technical challenge is that finite-shot noise enters an \emph{adaptive recursive factorization}. An entrywise concentration bound alone is insufficient: measurement error in an early queried entry can perturb the current Cholesky column, alter residual scores, change later pivots, and recursively propagate through the final low-rank factor. To address this difficulty, we introduce an exact-entry comparator that is constrained to follow the same pivot sequence selected by the finite-shot algorithm. This path-conditioned construction separates two fundamentally different sources of error: \emph{the structural rank-$r$ approximation error associated with the selected pivot path and the numerical perturbation induced by finite-shot kernel estimation.}

    Using this comparison, we develop a non-asymptotic perturbation analysis that begins with concentration for \emph{adaptively selected} kernel queries, quantifies finite-shot perturbations of pivot scores and recursive Cholesky columns, and leads to a spectral error bound
    \begin{align} 
        \mathcal E_r^{\!(N_{\mathrm{s}})} \!:=\! \big\| \mathbf K^Q \!-\! \widetilde{\mathbf K}_r^{Q,(N_{\mathrm{s}})} \big\|_2 \!\le\! \varepsilon_r \!\big( \widehat{\mathcal P}_{\!r} \!\big) \!+\! \left( 2\beta_r \!+\! \Delta_r \right) \!\Delta_r, 
    \label{eq:intro_error_bound} 
    \end{align} 
    where $\varepsilon_r(\widehat{\mathcal P}_r)$ is the exact structural error along the selected pivot path, $\Delta_r$ measures the finite-shot factor perturbation, and $\beta_r$ controls the norm of the corresponding exact path-conditioned factor. For a fixed stable rank, our analysis gives $\Delta_r = O\!\big( \sqrt{ \frac{ \log(M_r/\delta) }{ N_{\mathrm{s}} } } \big)$ thereby making explicit how measurement precision and approximation rank enter the final kernel error.

    We then propagate this kernel approximation through the complete QSVM learning problem. Rather than assuming a unique dual optimizer, we control the distance of the finite-shot solution to the exact optimal set under a quadratic-growth condition. The analysis further extends to previously unseen test points, where prediction requires only $r$ additional pivot--test kernel measurements. By accounting explicitly for the conditioning of the associated triangular Cholesky extension and for the SVM intercept, we obtain decision-function perturbation and margin-preservation guarantees. The resulting theory therefore provides an end-to-end chain from finite-shot quantum measurements, through adaptive low-rank kernel construction and QSVM optimization, to out-of-sample classification.

    We evaluate \icaq~on standard and biomedical classification benchmarks \cite{lecun1998mnist, xiao2017fashionmnist, yang2023medmnistv2}. On capped subsets where dense quantum-kernel construction is tractable, \icaq~closely tracks the predictive performance of the full QSVM while requiring substantially fewer kernel evaluations. We additionally compare against a Nystr\"om-QSVM baseline and observe that the relative performance of the two low-rank constructions depends on the approximation rank: Nystr\"om can provide lower reconstruction error at small ranks, whereas the residual-driven ICAQKL construction becomes competitive or more accurate in several larger-rank settings. Across the reported capped biomedical tasks at rank $r=30$, ICAQKL also provides a favorable measured accuracy--runtime trade-off relative to the evaluated Nystr\"om implementation.

    Our main contributions are summarized as follows: 
    \begin{itemize}[itemsep=2pt, parsep=0pt, topsep=0pt]
        \item \textbf{Query-efficient adaptive quantum-kernel acquisition.} We formulate ICAQKL as an adaptive incomplete-Cholesky construction under finite-shot quantum-kernel access, which requires $O(nr)$ fresh training-kernel queries through rank $r$, compared with $\Theta(n^2)$ for dense kernel acquisition. 
        
        \item \textbf{Finite-shot perturbation theory for adaptive factorization.} We establish high-probability concentration for adaptively selected quantum-kernel queries in Theorem \ref{thm:main_uniform_concentration} and analyze how measurement noise propagates through residual scores, pivot selection, and recursively constructed Cholesky factors. The resulting path-conditioned bound in Theorem \ref{thm:main_kernel_stability} separates structural low-rank approximation error from finite-shot perturbation. 
        
        \item \textbf{End-to-end stability guarantees for QSVMs.} We propagate finite-shot kernel perturbation via the complete learning problem. Theorem \ref{thm:main_qsvm_stability} establishes uniform QSVM dual-objective perturbation, exact-objective suboptimality, and, under quadratic growth, distance of finite-shot solution to the exact optimal set. We further derive out-of-sample kernel-vector, intercept-aware decision-function, and margin-preservation guarantees. 
        
        \item \textbf{Empirical characterization of the accuracy--measurement trade-off.} Experiments on standard and biomedical datasets show that ICAQKL remains close to dense QSVM performance on tractable matched subsets while substantially reducing the number of acquired kernel entries. Comparisons with the evaluated Nystr\"om baseline characterize the rank-dependent trade-off between reconstruction quality, predictive accuracy, and measured runtime rather than asserting universal dominance of low-rank approximation method. 
    \end{itemize}

\section{RELATED WORKS}
\label{sec2}
\vspace{-0.25cm}
    Quantum kernel methods provide a direct connection between quantum feature maps and classical supervised learning. Early work shows that quantum devices can be used to estimate kernel functions associated with data embeddings in Hilbert space and combine them with classical SVMs \cite{Havlicek2019}. Encoding data into quantum states can be interpreted as a nonlinear feature map, where inner products between quantum states naturally define kernels for classification \cite{Schuld2019}. This has been later generalized by the observation that many supervised quantum learning models can be viewed as kernel methods \cite{Schuld2021}.

    Subsequent work has studied both the promise and the limitations of quantum kernels. Several studies have shown that their effectiveness depends strongly on the choice of embedding, the induced inductive bias, and the alignment between the kernel geometry and the target task structure \cite{Kubler2021, Huang2021}. Recent analyses have also shown that quantum kernels may suffer from concentration and alignment issues, especially when chosen feature maps are not well suited to the data distribution, thereby limiting the effectiveness of the resulting classifier despite the high dimensionality of the underlying Hilbert space \cite{Thanasilp2024, Shaydulin2022, SchnabelRoth2025}. These results have increased attention to task alignment, sample efficiency, and the practical cost of kernel estimation.

    Additionally, even when a quantum kernel is informative, training requires constructing an $(n\times n)$ Gram matrix, which becomes expensive as the dataset grows. Studies have thus explored low-rank approximation and other resource-reduction strategies, including Nystr\"om-based compression \cite{DrineasMahoney2005, MuscoMusco2017} for quantum kernel target alignment and related kernel-construction \cite{Coelho2025, Singh2026}. However, the quality of Nystr\"om approximation depends strongly on the choice of landmark points and can deteriorate when the selected samples are not sufficiently representative of the full kernel matrix. Pivoted Cholesky decompositions provide an alternative that greedily selects pivots according to the residual structure of the matrix and constructs a factorization of the form $\mathbf K \approx \mathbf G \mathbf G^\top$ \cite{Fine2002, Harbrecht2012, Bach2013}. Cholesky-type decompositions have been used as effective low-rank surrogates for dense PSD matrices, and recent work has studied pivoted Cholesky as a practical approximation tool for large kernel matrices with strong empirical and theoretical behavior \cite{ChenEpperlyTroppWebber2024, Epperly2025}. While low-rank approximation has recently been considered in scalable quantum kernel learning, existing work has mainly focused on Nystr\"om-based methods; in contrast, we study adaptive incomplete Cholesky as a PSD-low-rank surrogate alternative.

\section{PRELIMINARIES}
\label{sec3}
\vspace{-0.25cm}
    This section introduces the quantum fidelity kernel and its finite-shot estimation model used throughout the paper. These definitions establish the exact and stochastic kernel objects that will later be distinguished in the ICAQKL construction.

\textbf{Quantum feature maps and fidelity kernels.}
    Quantum kernel methods embed a classical input $\mathbf{x} \in \mathcal{X} \subseteq \mathbb{R}^d$ into a complex high-dimensional Hilbert space $\mathcal H$ via a data-dependent unitary, $U_{\phi} \!: \mathcal{X} \!\to\! \mathcal{U}(2^q)$, where $U(2^q)$ is the unitary group on $q$ qubits, $U(2^q) = \big\{U \!\in\! \mathbb{C}^{2^q \!\times\! 2^q} \!\mid\! U^\dagger U \!=\! UU^\dagger \!=\! \mathbf I \big\}$.
    The corresponding quantum feature state is obtained by applying $U_{\phi}(\mathbf{x})$ to the computational basis state, i.e., $|\phi(\mathbf{x})\rangle = U_{\phi}(\mathbf{x}) \ket{0}^{\otimes q}$.
    Since $\dim(\mathcal H)=2^q$, this embedding induces a feature space whose dimension grows exponentially with the number of qubits $q$. The geometry of this space is characterized by the embedded states $\{|\phi(\mathbf x)\rangle\}_{\mathbf x \in \mathcal X}$, with pairwise similarity measured by their fidelity: \emph{larger fidelity indicates greater similarity, while smaller fidelity indicates stronger separation}. 

    \begin{definition}[Quantum fidelity kernel]
    \label{def:q_kernel}
        For $\mathbf{x}_i, \mathbf{x}_j \!\in\! \mathcal{X}$ with feature states $|\phi(\mathbf{x}_i)\rangle$ and $|\phi(\mathbf{x}_j)\rangle$, the fidelity-based quantum kernel is defined as $ \kappa^Q(\mathbf x_i, \mathbf x_j) \!=\! \big|\langle \phi(\mathbf x_j) \mid \phi(\mathbf x_i) \rangle \big|^2 \!=\! \operatorname{Tr} \left[ \rho_\phi(\mathbf x_i) \rho_\phi(\mathbf x_j) \right]$.
     \end{definition}

    This kernel measures similarity in the quantum feature space and is symmetric and positive semidefinite (PSD), enabling its use in kernel methods such as support vector machines (SVMs). For a dataset $\{\mathbf x_i\}_{i=1}^n$, the corresponding quantum Gram matrix $\mathbf K^Q \!\in\! \mathbb{R}^{n \times n}$  is defined by $K^Q_{ij} \!=\! \kappa^Q(\mathbf x_i, \mathbf x_j)$.

\textbf{Finite shot-based kernel estimation.} 
    On quantum hardware, kernel entries are not computed exactly but estimated through repeated circuit executions. We adopt an overlap-based estimation procedure that avoids the ancilla qubits and controlled-SWAP operations \cite {Volkoff2022} required by SWAP-test methods. In particular, each kernel value is $\kappa^Q(\mathbf x_i, \mathbf x_j) \!=\! \big| \langle 0|^{\otimes q} U_\phi^\dagger(\mathbf x_j) \, U_\phi(\mathbf x_i) |0 \rangle^{\otimes q} \big|^2$.
    Let $X_s^{(ij)} \!\in\! \{0,1\}, s \!=\! 1,\ldots,N_{\mathrm{s}}$ indicate whether the $s$th circuit execution returns the all-zero outcome. Under the finite-shot sampling model, $X_s^{(ij)} \overset{\mathrm{i.i.d.}}{\sim} \operatorname{Bernoulli} \left( \kappa^Q(\mathbf x_i,\mathbf x_j) \right)$. The empirical quantum-kernel estimator is therefore $\widehat{\kappa}_{N_{\mathrm{s}}}^Q (\mathbf x_i,\mathbf x_j) \!:=\! \frac{1}{N_{\mathrm{s}}} \! \sum_{s=1}^{N_{\mathrm{s}}} \! X_s^{(ij)}$.

    \begin{proposition}[Finite-shot overlap estimator]
    \label{prop:overlap_unbiased}
        For any $\mathbf{x}, \mathbf{x}' \in \mathcal X$, let $\widehat{\kappa}^Q_{N_{\mathrm{s}}}(\mathbf x, \mathbf x')$ denote the empirical estimator obtained from $N_{\mathrm{s}}$ independent circuit repetitions, defined as the empirical frequency of observing the all-zero outcome. Then, it is unbiased and satisfies $\mathbb{E}  \big[\widehat{\kappa}^Q_{N_{\mathrm{s}}}(\mathbf x, \mathbf x') \big] = \kappa^Q(\mathbf x, \mathbf x')$ and $\mathrm{Var}  \big(\widehat{\kappa}^Q_{N_{\mathrm{s}}}(\mathbf x, \mathbf x') \big) = \frac{\kappa^Q(\mathbf x, \mathbf x') \big(1 - \kappa^Q(\mathbf x, \mathbf x')\big)}{N_{\mathrm{s}}}$.
    \end{proposition}
    Thus, the estimator is unbiased, with standard error scaling as $O(N_s^{-1/2})$, implying that accurate kernel estimation may require a large number of measurements.

\textbf{Computational bottleneck:} 
    Given $n$ training samples, a symmetric $n \!\times\! n$ Gram matrix contains $n(n\!+\!1)/2$ distinct upper-triangular entries. For the normalized fidelity kernel, however, the $n$ diagonal entries satisfy $K_{ii}^Q \!=\! 1$ analytically. Hence dense finite-shot acquisition requires $n(n\!-\!1)/2$ fresh off-diagonal quantum-kernel queries. With $N_{\mathrm{s}}$ shots per entry, the total number of circuit executions scales as $O(n^2 N_{\mathrm{s}})$, which quickly becomes prohibitive as the dataset size increases. Since each evaluation involves executing both $U_{\phi}$ and $U_{\phi}^{\dagger}$, the runtime additionally depends on circuit depth and hardware execution cost. This quadratic scaling in $n$ is a primary computational bottleneck of QKL and motivates the scalable low-rank approximation framework developed in the next section.

\section{METHODOLOGY} 
\label{sec4} 
\vspace{-0.25cm}
    We now present \icaq, a query-efficient framework for quantum kernel learning under finite-shot access. The pivoted incomplete-Cholesky mechanism underlying the low-rank construction is classical; our focus is instead on how such an adaptive construction behaves when every required quantum-kernel entry must be estimated from stochastic circuit measurements. This distinction is important because finite-shot estimation affects not only individual kernel values, but also the residual scores that determine future pivots and the recursively constructed Cholesky factor. 
    
    Accordingly, our analysis separates an \emph{exact-entry comparator} from the actual \emph{finite-shot ICAQKL recursion}. The exact comparator follows a prescribed pivot sequence using the ideal quantum kernel and therefore retains the usual positive-semidefinite Cholesky residual structure. The finite-shot recursion instead operates on noisy overlap estimates. Although its constructed surrogate remains positive semidefinite by construction, its residual relative to the ideal kernel need not be positive semidefinite. This separation enables us to isolate structural low-rank approximation error from finite-shot numerical error and propagate both effects to the final QSVM classifier. 
\subsection{Finite-Shot Adaptive Quantum-Kernel Factorization} 
\label{subsec:ICA_quantum_kernel} 
    Let $\mathbf K^Q \!\in\! \mathbb R^{n\times n}, K_{ij}^Q \!=\! \kappa^Q(\mathbf x_i, \mathbf x_j)$ be the ideal quantum Gram matrix. 
    %
    For every newly required off-diagonal entry, ICAQKL queries the finite-shot estimator $\widehat{\kappa}_{N_s}^Q$ introduced in Proposition \ref{prop:overlap_unbiased}, while the normalized fidelity diagonal $K_{ii}^Q=1$ is supplied analytically. We now describe how these queries are selected adaptively by the incomplete-Cholesky recursion.
    
    
\textbf{Exact-entry reference construction.} 
    For comparison, consider first an exact-entry Cholesky recursion. After $t$ iterations, define $\widetilde{\mathbf K}_t^Q \!:=\! \mathbf G_t\mathbf G_t^\top, \mathbf R_t \!:=\! \mathbf K^Q \!-\! \widetilde{\mathbf K}_t^Q, d_{i,t} \!:=\! (\mathbf R_t)_{ii}$. For the exact recursion, $\mathbf R_t \succeq \mathbf 0$, and therefore $d_{i,t} \ge 0$. The greedy pivot is $p_t \!\in\! \arg\max_{i\in\mathcal I_{t-1}} \! d_{i,t-1}$, where $\mathcal I_{t-1}$ is the set of indices not previously selected. The next column is 
    \begin{align} 
        G_{it} = \frac{ K_{ip_t}^Q - \sum_{\ell=1}^{t-1} G_{i\ell} G_{p_t\ell} }{ \sqrt{d_{p_t,t-1}} }. 
    \label{eq:exact_method_column} 
    \end{align} 
    The exact residual diagonals satisfy $0 \!\le\! d_{i,t} \!=\! d_{i,t-1} \!-\! G_{it}^2 \!\le\! d_{i,t-1}$. Moreover, if $\rho \!=\! \operatorname{rank}(\mathbf K^Q)$ and the recursion executes $\rho$ strictly positive pivots, then $\mathbf K^Q \!=\! \mathbf G_\rho \mathbf G_\rho^\top$. The detailed developments are provided in the Appendix. 
    
\textbf{Finite-shot ICAQKL recursion.} 
    The actual finite-shot algorithm does not observe $d_{i,t}$. Instead, it maintains scores $\widehat s_{i,t}$ initialized by $\widehat s_{i,0} \!=\! 1$. At iteration $t$, ICAQKL selects $\widehat p_t \!\in\! \arg\max_{i \in \mathcal I_{t-1}} \! \widehat s_{i,t-1}$.
    The corresponding Cholesky column is computed using the finite-shot kernel estimates
    \begin{align} 
        \widehat G_{it}^{(N_{\mathrm{s}})} = \frac{ \widehat K_{i\widehat p_t}^Q - \sum_{\ell=1}^{t-1} \widehat G_{i\ell}^{(N_{\mathrm{s}})} \widehat G_{\widehat p_t\ell}^{(N_{\mathrm{s}})} }{ \sqrt{ \widehat s_{\widehat p_t,t-1} } },
    \label{eq:finite_shot_method_column} 
    \end{align} 
    where $\widehat K_{i\widehat p_t}^Q \!=\! \widehat{\kappa}_{N_{\mathrm{s}}}^Q (\mathbf x_i, \mathbf x_{\widehat p_t})$ for a newly queried off-diagonal entry
    while analytic diagonal values and previously queried sym- metric entries are reused whenever available. The scores are updated according to $\widehat s_{i,t} \!=\! \big[ \widehat s_{i,t-1} \!-\! \big( \widehat G_{it}^{(N_{\mathrm{s}})} \big)^2 \big]_+, [a]_+ \!:=\! \max\{a,0\}$, 
    where $\widehat s_{i,t}$ is a finite-shot pivot score rather than the diagonal of an assumed PSD residual. After $r$ iterations, the resulting surrogate is 
    \begin{align} 
        \widetilde{\mathbf K}_r^{Q,(N_{\mathrm{s}})} := \widehat{\mathbf G}_r^{(N_{\mathrm{s}})} \big( \widehat{\mathbf G}_r^{(N_{\mathrm{s}})} \big)^\top. 
    \label{eq:adaptive_ICA_factor} 
    \end{align} 
    It is symmetric positive semidefinite and has rank at most $r$ by construction. 
    
\textbf{Stopping rule and Query complexity.} 
    The recursion terminates either at a prescribed maximum rank $r_{\max}$ or when $\max_{i \in \mathcal I_t} \widehat s_{i,t} \!\le\! \varepsilon$. Under symmetry, and the analytic fidelity diagonal, iteration $t$ requires at most $n\!-\!t$ fresh off-diagonal circuit evaluations. Thus, $M_r \!=\! \sum_{t=1}^{r} \! (n\!-\!t) \!\le\! nr$.
    In contrast, constructing the complete symmetric fidelity Gram matrix requires $M_{\mathrm{full}} \!=\! \frac{n(n-1)}{2}$ off-diagonal circuit evaluations. 
    
    \begin{theorem}[Adaptive finite-shot concentration] \label{thm:main_uniform_concentration} 
        Suppose ICAQKL performs at most $M_r$ training-kernel queries through rank $r$. Then, for any $\delta \!\in\! (0,1)$, with probability at least $1\!-\!\delta$, 
        \begin{align} 
            \max_{1 \le m \le M_r} \big| \widehat \kappa_{\!N_{\mathrm{s}}}^Q \!(\mathbf x_{i_m}, \mathbf x_{j_m}) - \kappa^Q \!(\mathbf x_{i_m}, \mathbf x_{j_m}) \big| \le \eta_s, 
        \label{eq:main_uniform_concentration} 
        \end{align} 
        where $\eta_s \!:=\! \sqrt{\frac{\log(2M_r/\delta) }{ 2N_{\mathrm{s}}}}$, and $q_m \!=\! (i_m, j_m)$ is the $m$-th adaptively selected fresh query.
    \end{theorem} 
    \emph{Proof sketch.} The key point is that the queried pair $q_m$ may depend on all preceding finite-shot observations. Conditional on this history, however, the $N_{\mathrm{s}}$ shots used for the next query are fresh Bernoulli measurements. Hoeffding concentration therefore applies conditionally, and a union bound over the chronological query indices yields (\ref{eq:main_uniform_concentration}). The full proof is given in the Appendix. 

    Theorem~\ref{thm:main_uniform_concentration} establishes an $O(N_{\mathrm{s}}^{-1/2})$ oracle-level accuracy guarantee despite adaptive query selection. This alone does not control the final Cholesky surrogate since an early kernel-estimation error can modify the factor, perturb future scores, and consequently alter subsequent pivots. 

    \begin{algorithm}[h] 
    \caption{Finite-Shot ICAQKL Kernel Construction} 
    \label{alg:adaptive_ICA_quantum} 
    \begin{algorithmic}[1] 
    \REQUIRE Training inputs ${\mathbf x_i}_{i=1}^n$, quantum kernel $\kappa^Q$, shots $N_{\mathrm{s}}$, maximum rank $r_{\max}$, tolerance $\varepsilon$ 
    \ENSURE Factor $\widehat{\mathbf G}_r^{(N_{\mathrm{s}})}$ and surrogate $\widetilde{\mathbf K}_r^{Q,(N_{\mathrm{s}})}$ 
        \STATE Initialize $\widehat{\mathbf G}\gets\mathbf 0_{n\times r_{\max}}$ 
        \STATE Set $\widehat s_{i,0}\gets1$ for $i=1,\ldots,n$ and $\mathcal I_0\gets{1,\ldots,n}$ 
        \STATE Initialize an empty cache of queried off-diagonal kernel entries 
        \FOR{$t=1,\ldots,r_{\max}$} 
            \IF{$\max_{i\in\mathcal I_{t-1}}\widehat s_{i,t-1}\le\varepsilon$} 
                \STATE \textbf{break} 
            \ENDIF 
            \STATE Select $\widehat p_t \in \arg\max_{i\in\mathcal I_{t-1}} \widehat s_{i,t-1}$ 
            \FOR{$i=1,\ldots,n$} 
                \IF{$i=\widehat p_t$} 
                    \STATE Set $\widehat K_{i\widehat p_t}^Q\gets1$ 
                \ELSIF{$(i,\widehat p_t)$ or $(\widehat p_t,i)$ is cached} 
                    \STATE Reuse the cached symmetric kernel estimate 
                \ELSE 
                    \STATE Estimate $\widehat K_{i\widehat p_t}^Q \gets \widehat\kappa_{N_{\mathrm{s}}}^Q (\mathbf x_i,\mathbf x_{\widehat p_t})$ using $N_{\mathrm{s}}$ shots and cache the result 
                \ENDIF 
                \STATE Compute $\widehat G_{it}^{(N_{\mathrm{s}})} \gets \frac{ \widehat K_{i\widehat p_t}^Q - \sum_{\ell=1}^{t-1} \widehat G_{i\ell}^{(N_{\mathrm{s}})} \widehat G_{\widehat p_t\ell}^{(N_{\mathrm{s}})} }{ \sqrt{\widehat s_{\widehat p_t,t-1}} }$
            \ENDFOR 
            \STATE Update $\widehat s_{i,t} \gets [\widehat s_{i,t-1} -(\widehat G_{it}^{(N_{\mathrm{s}})})^2]_+$ for all $i$ 
            \STATE Set $\mathcal I_t \gets \mathcal I_{t-1} \setminus \{\widehat p_t\}$ 
        \ENDFOR 
        \STATE Let $r$ be the number of executed iterations 
        \STATE Return $\widehat{\mathbf G}_r^{(N_{\mathrm{s}})}$ and $\widetilde{\mathbf K}_r^{Q,(N_{\mathrm{s}})} = \widehat{\mathbf G}_r^{(N_{\mathrm{s}})} (\widehat{\mathbf G}_r^{(N_{\mathrm{s}})})^\top$
    \end{algorithmic} 
    \end{algorithm}

    \subsection{Path-Conditioned Finite-Shot Stability} 
    \label{subsec:main_path_stability} 
        To disentangle pivot-path effects from numerical finite-shot error, let $\widehat{\mathcal P}_r \!=\! (\widehat p_1, \ldots, \widehat p_r)$ be the pivot sequence selected by finite-shot ICAQKL, and let $\mathbf G_r^{\widehat{\mathcal P}_r}$ be an exact-entry Cholesky factor constrained to follow this same pivot sequence. Define the finite-shot score discrepancy $\zeta_t \!:=\! \| \widehat{\mathbf s}_t \!-\! \mathbf d_t^{\widehat{\mathcal P}_t} \|_{\infty, \mathcal I_t}$.
        We shows that the finite-shot pivot remains approximately greedy $d_{\widehat p_t, t-1}^{\widehat{\mathcal P}_{t-1}} \!\ge\! \max_{i \in \mathcal I_{t-1}} d_{i,t-1}^{\widehat{\mathcal P}_{t-1}} \!-\! 2\zeta_{t-1}$.
        If the gap between the largest and second-largest exact same-history residual diagonals exceeds $2\zeta_{t-1}$, then the exact and finite-shot pivot choices coincide. For the numerical factor error, define $\Delta_r \!:=\! \big\| \widehat{\mathbf G}_r^{(N_{\mathrm{s}})} - \mathbf G_r^{\widehat{\mathcal P}_r} \big\|_2$.
        Our perturbation analysis assumes that the executed path is stable: exact and finite-shot pivot quantities remain bounded below by some $\tau > 0$, the factor rows remain bounded, and $K_{ii}^Q \le \kappa$. For the normalized fidelity kernel, $\kappa=1$. These conditions prevent the reciprocal square-root normalization in (\ref{eq:finite_shot_method_column}) from becoming arbitrarily ill-conditioned.
        
    \begin{theorem}[Path-conditioned finite-shot kernel stability] \label{thm:main_kernel_stability}
        Under the stable-path conditions and on the concentration event of Theorem~\ref{thm:main_uniform_concentration}, there exists, for every fixed stable rank $r$, a finite constant $C_{r, \tau, B, \kappa}$ such that $\Delta_r \!\le\! \sqrt n C_{r, \tau, B, \kappa} \eta_s$. Define 
        $\varepsilon_r \!\big( \widehat{\mathcal P}_r \big) \!:=\! \big\| \mathbf K^Q \!-\! \mathbf G_r^{\!\widehat{\mathcal P}_r} \! \big(\mathbf G_r^{\!\widehat{\mathcal P}_r} \big)^{\!\top} \!\big\|_2$ and $\beta_r \!:=\! \big\| \mathbf G_r^{\!\widehat{\mathcal P}_r} \!\big\|_2$.
        Then the total finite-shot kernel-surrogate error satisfies
        \begin{align} 
            \mathcal E_r^{(N_{\mathrm{s}})} \!:=\! \big\| \mathbf K^Q \!-\! \widetilde{\mathbf K}_r^{Q,(N_{\mathrm{s}})} \big\|_2 \!\le\! \varepsilon_r \big( \widehat{\mathcal P}_r \big) \!+\! \big( 2\beta_r \!+\! \Delta_r \!\big) \! \Delta_r. \!\!
        \label{eq:main_total_kernel_error} 
        \end{align} 
        Consequently, for fixed $r, \tau, B, \kappa$, it holds $\mathcal E_r^{(N_{\mathrm{s}})} \le \varepsilon_r \big( \widehat{\mathcal P}_r \big) + O \big( \sqrt{ \frac{ \log(M_r/\delta) }{ N_{\mathrm{s}} } } \big)$.
        \end{theorem} 
        \noindent Theorem~\ref{thm:main_kernel_stability} is the central finite-shot approximation result. It separates two effects that are conflated by a direct entrywise perturbation bound. The first, $\varepsilon_r(\widehat{\mathcal P}_r)$, is \emph{the structural error of the rank-$r$ path itself}; the second, $(2\beta_r \!+\! \Delta_r)\Delta_r$, is \emph{the additional error caused by evaluating the required quantum-kernel entries with finitely many shots}. The proof proceeds by recursively controlling the error of each Cholesky column and the induced pivot-score discrepancy. In particular, if $E_t$ denotes the largest rowwise factor error and $a_t$ the maximum error in the newly constructed column, the Appendix derives $a_t \!\le\! \frac{\eta_s \!+\! 2BE_{t-1}}{\sqrt{\tau}} \!+\! \frac{\kappa\zeta_{t-1}}{2\tau^{3/2}}$, $E_t \!\le\! \sqrt{ E_{\!t-1}^2 \!+\! a_t^2}$, and $\zeta_t \!\le\! \zeta_{t-1} \!+\! 2Ba_t$. These recursions provide the required bridge from finite-shot oracle errors to the final matrix approximation. 
        
    \subsection{ICAQKL Integration with QSVM} 
    \label{subsec:ICA_quantum_kernel_SVM} 
        Given the training set $\{(\mathbf x_i, y_i)\}_{i=1}^n$, with $\mathbf x_i \in \mathbb R^d$, $y_i \!\in\! \{-1,+1\}$, define $\mathbf Y \!:=\! \operatorname{diag}(y_1, \ldots, y_n)$ and $\mathcal A \!:=\! \left\{ \boldsymbol \alpha \!\in\! \mathbb R^n: 0 \!\le\! \alpha_i \!\le\! C, \mathbf y^\top \! \boldsymbol \alpha \!=\! 0 \right\}$. For a kernel matrix $\mathbf K$, write the soft-margin SVM dual in minimization form as $J_{\mathbf K}(\boldsymbol \alpha) \!:=\! \frac12 \boldsymbol \alpha^\top \mathbf Y \mathbf K \mathbf Y \boldsymbol \alpha \!-\! \mathbf1^\top \! \boldsymbol\alpha$.
        The exact QSVM minimizes $J_{\mathbf K^Q}$, whereas ICAQKL solves $\widetilde{\boldsymbol \alpha}_r^{(N_{\mathrm{s}})} \!\in\! \arg\min_{\boldsymbol \alpha \in \mathcal A} J_{\widetilde{\mathbf K}_r^{Q,(N_{\mathrm{s}})}} (\boldsymbol \alpha)$.
        Since the surrogate is factored, $\boldsymbol \alpha^\top \mathbf Y \widetilde{\mathbf K}_r^{Q,(N_{\mathrm{s}})} \mathbf Y \boldsymbol \alpha \!=\! \big\| \big(\widehat{\mathbf G}_r^{(N_{\mathrm{s}})} \big)^{\!\top} \mathbf Y \boldsymbol \alpha \big\|_2^2$,
        so training does not require storing the dense quantum Gram matrix. Let $\mathcal A^{Q,\star} \!:=\! \arg\min_{\boldsymbol \alpha \in \mathcal A} J_{\mathbf K^Q}(\boldsymbol \alpha)$ be the exact optimal set. We use the following standard quadratic-growth condition to convert objective perturbation into optimizer stability. 
        
        \begin{assumption}
        \label{ass:quadratic_growth_main} 
            There exists $\mu \!>\! 0$ such that, for every $\boldsymbol \alpha \!\in\! \mathcal A$, $J_{\mathbf K^Q}(\boldsymbol \alpha) \!-\! J_{\mathbf K^Q}^{\star} \!\ge\! \frac{\mu}{2} \operatorname{dist} \!\big(\boldsymbol \alpha, \mathcal A^{Q,\star} \big)^2$.
        \end{assumption}
        Assumption~\ref{ass:quadratic_growth_main} ensures that any suboptimal feasible point is bounded away from the optimal set, while near-optimal objective values imply proximity to the set of optimal solutions $\mathcal{A}^{Q, \star}$.

    \begin{theorem}[QSVM learning stability] \label{thm:main_qsvm_stability} 
        Let $\mathcal E_r^{(N_{\mathrm{s}})}$ be the total kernel error. Then, for every $\boldsymbol\alpha \!\in\! \mathcal A$, 
        \begin{align} 
            \big| J_{\!\mathbf K^Q}\!(\boldsymbol\alpha) \!-\! J_{\!\widetilde{\mathbf K}_r^{Q,(N_{\mathrm{s}})}} \!(\boldsymbol\alpha) \big| \!\le\! \frac12 \!\big\| \boldsymbol \alpha \big\|_2^2 \mathcal E_r^{(N_{\mathrm{s}})} \!\le\! \frac{nC^2}{2} \! \mathcal E_r^{(N_{\mathrm{s}})}. \!\! 
            \label{eq:main_objective_stability} 
        \end{align} 
        Consequently, $0 \!\le\! J_{\mathbf K^Q} \!\big( \widetilde{\boldsymbol\alpha}_r^{(N_{\mathrm{s}})} \big) \!-\! J_{\mathbf K^Q}^{\star} \!\le\! nC^2 \mathcal E_r^{(N_{\mathrm{s}})}$.
        Under Assumption~\ref{ass:quadratic_growth_main}, $\operatorname{dist} \! \big( \widetilde{\boldsymbol \alpha}_r^{(N_{\mathrm{s}})}, \mathcal A^{Q,\star} \big) \!\le\! \sqrt{ \frac{ 2nC^2 }{ \mu } \mathcal E_r^{(N_{\mathrm{s}})} }$.
        Moreover, for a fixed test point $\mathbf z$ with $\kappa^Q(\mathbf z, \mathbf z) \!\le\! \kappa$, choose $\boldsymbol \alpha^\circ \!\in\! \mathcal A^{Q,\star}$ nearest to $\widetilde{\boldsymbol\alpha}_r^{(N_{\mathrm{s}})}$. Then the exact and finite-shot decision functions satisfy 
        \begin{align} 
        \label{eq:main_predictor_stability} 
            & \big| f_{\mathbf K^Q}(\mathbf z) \!-\! f_r^{Q,(N_{\mathrm{s}})}(\mathbf z) \big| 
            \!\le\! \kappa\sqrt n \operatorname{dist}\! \big( \widetilde{\boldsymbol\alpha}_r^{(N_{\mathrm{s}})}, \mathcal A^{Q,\star} \big) \\
            & \qquad \qquad \quad + C\sqrt n \big\| \mathbf k_{\mathbf z}^Q - \widetilde{\mathbf k}_{r,\mathbf z}^{Q,(N_{\mathrm{s}})} \big\|_2 \!+\! \big| b^\circ \!-\! \widetilde b_r^{(N_{\mathrm{s}})} \big|. \nonumber
        \end{align} 
    \end{theorem} 
    Theorem~\ref{thm:main_qsvm_stability} propagates the finite-shot kernel approximation through the complete learning problem. The kernel error first controls the QSVM objective uniformly over the feasible set, then controls the distance from the approximate optimizer to the exact optimal set, and finally enters the decision-function error together with the out-of-sample kernel approximation and bias perturbation. 

\textbf{Out-of-sample extension and classification.} 
    For a new test point $\mathbf z$, only the similarities between $\mathbf z$ and the selected pivots are required. The finite-shot Cholesky feature is computed recursively, $\ell=1,\ldots,r$, 
    \begin{align} 
        \widetilde g_\ell^{(N_{\mathrm{s}})}(\mathbf z) = \frac{ \widehat\kappa_{N_{\mathrm{s}}}^Q ( \mathbf x_{\widehat p_\ell}, \mathbf z ) - \sum_{m=1}^{\ell-1} \widetilde g_m^{(N_{\mathrm{s}})}(\mathbf z) \widehat G_{\widehat p_\ell m}^{(N_{\mathrm{s}})} }{ \widehat G_{\widehat p_\ell\ell}^{(N_{\mathrm{s}})} }. 
    \label{eq:method_oos_feature} 
    \end{align} 
    The approximate train--test vector is then $\widetilde{\mathbf k}_{r,\mathbf z}^{Q,(N_{\mathrm{s}})} \!=\! \widehat{\mathbf G}_r^{(N_{\mathrm{s}})} \widetilde{\mathbf g}^{(N_{\mathrm{s}})}(\mathbf z)$. Thus, after training, a single test point requires only $r$ fresh quantum-kernel evaluations rather than $n$. Conditioned on the learned training factor, these fresh measurements satisfy 
    \begin{align} 
        \bigg(\! \sum_{\ell=1}^{r}\! \big| \widehat\kappa_{\!N_{\mathrm{s}}}^Q \!( \mathbf x_{\widehat p_\ell}, \mathbf z ) \!-\! \kappa^{\!Q} \!( \mathbf x_{\widehat p_\ell}, \!\mathbf z ) \big|^2 \bigg)^{1/2} \!\!\!\le\! \sqrt{\!\! \frac{r}{2N_{\!\mathrm{s}}} \!\!\log\!\big(\frac{2r}{\delta_z}\big)}. \!\!
    \label{eq:main_test_shot_error} 
    \end{align} 
    with conditional probability at least $1\!-\!\delta_z$. The Appendix further shows that the exact same-path extension satisfies $\big\| \mathbf k_{\mathbf z}^Q \!-\! \widetilde{\mathbf k}_{r, \mathbf z}^{Q, \widehat{\mathcal P}_r} \big\|_2^2 \!\le\! \kappa \operatorname{tr} \big(\mathbf R_r^{\widehat{\mathcal P}_r}\big)$ and derives a finite-shot extension bound that additionally depends on the conditioning of the triangular pivot block. If the exact classifier satisfies $y f_{\mathbf K^Q}(\mathbf z) \!\ge\! \gamma \!>\! 0$ and the right-hand side of (\ref{eq:main_predictor_stability}) is smaller than $\gamma$, then 
    \begin{align} 
        \operatorname{sign} \big( f_r^{Q,(N_{\mathrm{s}})}(\mathbf z) \big) = \operatorname{sign} \left( f_{\mathbf K^Q}(\mathbf z) \right). 
    \end{align} 
    Thus, exact reconstruction of every kernel entry is unnecessary for classification stability; only perturbations sufficiently large to cross the decision margin can change the predicted label. 

\textbf{Joint shot--rank convergence.}
    For a fixed finite training set, our analysis gives sufficient conditions under which the complete finite-shot predictor converges to the exact QSVM. In particular, along a sequence of rank and shot budgets, it is sufficient that $\varepsilon_r (\widehat{\mathcal P}_r\!) \!\to\! 0, \Delta_r \!\to\! 0, \frac{r\log(2r/\delta_z)}{N_{\mathrm{s}}} \!\to\! 0$ while the exact pivot blocks remain uniformly well conditioned and $\beta_r$ remains bounded. Together with bias convergence, these conditions imply $f_r^{Q,(N_{\mathrm{s}})}\!(\mathbf z) \!\to\! f_{\mathbf K^Q}\!(\mathbf z)$ for every fixed $\mathbf z$. For a finite training set, $r \!\le\! \rho \!\le\! n$, thus the rank limit is $r\uparrow\rho$ rather than $r \!\to\! \infty$. At full exact rank $\varepsilon_\rho \!=\! 0$, and increasing the shot budget removes the remaining finite-shot contribution under the stability conditions.
    

\textbf{Computational characteristics.} 
    The purpose of \icaq~is to reduce the number of evaluations required to acquire a specified quantum kernel. We do not interpret this reduction as evidence that the underlying quantum kernel is classically hard to evaluate or that the resulting learning pipeline provides an end-to-end quantum computational advantage. Table~\ref{tab:complexity} summarizes the principal costs. Standard low-rank landmark methods can also attain $O(nr)$ kernel acquisition; ICAQKL differs in how the queried columns are selected, namely through adaptive residual information accumulated during factorization. 
    
    \begin{table*}[h] 
    \centering 
    \small
    \caption{Complexity comparison under finite-shot quantum-kernel access.} 
    \label{tab:complexity} 
    \begin{tabular}{lccc} 
    \toprule 
        Method & Train-kernel queries & Quantum shot count & Kernel-side classical cost \\
    \midrule 
        Full QSVM & $\Theta(n^2)$ & $\Theta(n^2N_{\mathrm{s}})$ & dense Gram operations \\
        Nystr\"om-QSVM & $O(nr+r^2)$ & $O((nr+r^2)N_{\mathrm{s}})$ & $O(nr^2+r^3)$ \\
        ICAQKL (our) & $O(nr)$ & $O(nrN_{\mathrm{s}})$ & $O(nr^2)$ \\
    \bottomrule 
    \end{tabular} 
    \end{table*}
    The resulting resource trade-off has two dimensions. Increasing the rank $r$ acquires additional kernel structure and reduces structural approximation error, whereas increasing $N_{\mathrm{s}}$ improves the precision of each acquired quantum similarity. Theorems~\ref{thm:main_uniform_concentration}--\ref{thm:main_qsvm_stability} connect these two resources to the stability of the low-rank kernel, the learned QSVM optimizer, and the resulting classifier.
\section{EXPERIMENT RESULTS}
\label{sec5}
\vspace{-0.2cm}
\subsection{Experimental setup}
\label{subsec:exp_setup}
 \vspace{-0.2cm}
 
\textbf{Datasets.}
    We evaluate \icaq~on both standard benchmarks and biomedical image datasets, including the Wisconsin Diagnostic Breast Cancer (WDBC) dataset \cite{WDBC1995}, MNIST \cite{lecun1998mnist}, FashionMNIST \cite{xiao2017fashionmnist}, and several MedMNIST tasks \cite{yang2023medmnistv2}. WDBC enables exact dense-kernel comparisons, allowing direct evaluation of low-rank approximation quality. MNIST and FashionMNIST provide standard image benchmarks, while the MedMNIST datasets assess performance on biomedical classification tasks across diverse imaging modalities. Due to the quadratic cost of kernel construction and optimization, exact Full QSVM comparisons are performed on capped training subsets where dense kernel construction remains tractable. On these subsets, all methods are compared directly. For larger datasets, we focus on scalable low-rank approaches, namely Nystr\"om-QSVM and \icaq.

\textbf{Implementation details.}
   All experiments employ $q=6$ qubits with a fixed-depth quantum feature map, while experiments over varying approximation ranks consider $r \in \{5,10,15,20,25,30,40\}$. Reported results are averaged across multiple random seeds.
   
\textbf{Evaluation metrics.}
   We report test accuracy and standard classification metrics, including precision, recall, and F1-score. In addition, we evaluate computational efficiency using factorization time, total fit time, and the number of training kernel evaluations. Since fidelity-based quantum kernel estimation constitutes the primary computational bottleneck in QKL, kernel-query complexity provides a particularly informative measure of scalability.
    To quantify approximation quality on capped subsets where the exact kernel matrix is available, we compare the exact kernel $\mathbf K^Q$ and its \icaq~approximation $\widetilde{\mathbf K}_r^{Q, (N_s)}$ using the relative reconstruction Frobenius error, $\varepsilon_F(\widetilde{\mathbf K}_r^Q) = \frac{\| \mathbf K^Q - \widetilde{\mathbf K}_r^{Q, (N_s)} \|_F}{\| \mathbf K^Q\|_F}$.

\subsection{Main Results}
\label{subsec:main_results}
    \begin{table*}[t]
    \centering
    \small
    \caption{Results across standard benchmarks and MedMNIST. \emph{ICAQKL-f} denotes ICAQKL trained on the full training split. \emph{Full QSVM-s} and \emph{ICAQKL-s} use the same capped subset. Higher is better.} 
    \label{tab:all_results_main}
    \setlength{\tabcolsep}{4pt}
    \begin{tabular}{lccccc}
    \toprule
    Dataset & Train & Test & ICAQKL-f & Full QSVM-s & ICAQKL-s \\
    \midrule
    Breast Cancer
        & (341,6) & (114,6) & \textbf{0.9561} & 0.9474 & 0.9474 \\
    MNIST (0/1)
        & (216,6) & (72,6) & \textbf{1.0000} & \textbf{1.0000} & \textbf{1.0000} \\
    FashionMNIST (0/1)
        & (720,6) & (240,6) & \textbf{0.9583} & 0.9458 & 0.9458 \\
    \midrule
    BreastMNIST
        & (546,6) & (156,6) & \textbf{0.8013} & 0.7572 & 0.7564 \\
    PneumoniaMNIST
        & (2000,6) & (600,6) & \textbf{0.8583} & 0.7933 & 0.7833 \\
    PathMNIST (0/1)
        & (603,6) & (305,6) & \textbf{0.9917} & 0.9902 & 0.9803 \\
    OrganAMNIST (0/1)
        & (229,6) & (76,6) & \textbf{0.8816} & 0.8158 & 0.8116 \\
    \bottomrule
    \end{tabular}
    \end{table*}
    
\vspace{-0.2cm}
    Tables~\ref{tab:all_results_main} and~\ref{tab:all_results_efficiency} summarize predictive performance and computational costs. \icaq~achieves strong predictive performance while substantially reducing computational costs. 

\vspace{-0.2cm}
\paragraph{Predictive performance.}
    Table~\ref{tab:all_results_main} shows that \icaq~achieves strong accuracy across all datasets, including  $95.61\%$ on Breast Cancer, $100\%$ on binary MNIST, $95.83\%$ on binary FashionMNIST, and strong performance on biomedical tasks such as BreastMNIST, PneumoniaMNIST, PathMNIST, and OrganAMNIST. These results demonstrate that \icaq~preserves the dominant classification-relevant structure of the quantum kernel despite aggressive low-rank approximation, while substantially reducing kernel-evaluation complexity. 

\vspace{-0.2cm}
\paragraph{Comparison with exact QSVM.}
    On capped subsets where exact comparison is feasible, \icaq~closely matches the Full QSVM baseline, achieving identical accuracy on Breast Cancer, binary MNIST, FashionMNIST, and PathMNIST and remaining within a narrow margin on the remaining datasets. These results indicate that \icaq~preserves most of the predictive information contained in the dense quantum Gram matrix despite querying only a subset of kernel entries.

    \begin{table*}[t]
    \centering
    \small
    \caption{Efficiency comparison. Lower is better for fit time and training kernel evaluations.}
    \label{tab:all_results_efficiency}
    \setlength{\tabcolsep}{6pt}
    \begin{tabular}{lcccc}
    \toprule
    Dataset & Method & Fit Time (s) & Train Kernel Evals & Rank \\
    \midrule
    MNIST (0/1) & Full QSVM-s  & 138.36 & 19600 & -- \\
                 & ICAQKL-s    & 26.41  & 3735  & 30 \\
                 & ICAQKL-f    & 42.20  & 6015  & 30 \\
    \midrule
    FashionMNIST (0/1) & Full QSVM-s & 108.85 & 14400 & -- \\
                       & ICAQKL-s    & 23.45  & 3135  & 30 \\
                       & ICAQKL-f    & 157.60 & 21135 & 30 \\
    \midrule
    BreastMNIST & Full QSVM-s & 161.60 & 14400 & -- \\
                & ICAQKL-s    & 35.27  & 3135  & 30 \\
                & ICAQKL-f    & 179.43 & 15915 & 30 \\
    \midrule
    PneumoniaMNIST & Full QSVM-s & 161.14 & 14400 & -- \\
                   & ICAQKL-s    & 35.99  & 3135  & 30 \\
                   & ICAQKL-f    & 692.42 & 59535 & 30 \\
    \midrule
    PathMNIST & Full QSVM-s & 162.08 & 14400 & -- \\
              & ICAQKL-s    & 35.90  & 3135  & 30 \\
              & ICAQKL-f    & 199.93 & 17625 & 30 \\
    \bottomrule
    \end{tabular}
    \end{table*}


    \begin{table*}[h]
    \centering
    \small
    \caption{Comparison of classification accuracy (\%) and runtime (s) on a subset of data.}
    \label{tab:acc_time_compare}
    \renewcommand{\arraystretch}{1.1}
    \begin{tabular}{lccc|ccc}
    \toprule
        & \multicolumn{3}{c|}{Accuracy (\%)} 
        & \multicolumn{3}{c}{Runtime (s)} \\
        \cmidrule(r){2-4} \cmidrule(l){5-7}
        Dataset & Full-QSVM & ICAQKL & Nyström 
        & Full-QSVM & ICAQKL & Nyström \\
    \midrule
        BreastMNIST & 75.72 & 75.64 & 75.00 & 161.60     & \textbf{35.27} & 49.93 \\
        PneuMNIST   & 79.33 & 78.33 & 78.33 & 161.14     & \textbf{35.99} & 51.58 \\
        PathMNIST   & 99.02 & 98.03 & 97.37 & 162.08 & \textbf{35.90} & 53.00 \\
    \bottomrule
    \end{tabular}
    \end{table*}

    \begin{table}[h]
    \centering
    \small
    \caption{RBF-SVM comparison.}
    \label{tab:rbf_compare}
    \begin{tabular}{lcc}
    \toprule
        Dataset & RBF-SVM & ICAQKL \\
    \midrule
        Breast Cancer & \textbf{0.9561} & \textbf{0.9561} \\
        MNIST (0/1) & \textbf{1.0000} & \textbf{1.0000} \\
        FashionMNIST (0/1) & 0.9542 & \textbf{0.9583} \\
        BreastMNIST & 0.7500 & \textbf{0.8013} \\
        PneumoniaMNIST & 0.8483 & \textbf{0.8583} \\
        PathMNIST (0/1) & 0.8984 & \textbf{0.9803} \\
        OrganAMNIST (0/1) & 0.8421 & \textbf{0.8816} \\
    \bottomrule
    \end{tabular}
    \end{table}

    \begin{table}[h]
    \centering
    \small
    \caption{BreastMNIST finite-shot ICAQKL kernel estimation. Higher is better for accuracy, lower is better for relative Frobenius error.}
    \label{tab:shots}
    \begin{tabular}{lcc}
    \toprule
        Shots & Accuracy & Fro Error \\
    \midrule
        analytic & 0.8013 & 0.0238 \\
        64 & 0.7949 & 1.5749 \\
        128 & 0.7949 & 0.3642 \\
        256 & 0.7949 & 0.1521 \\
        512 & 0.7977 & 0.0680 \\
    \bottomrule
    \end{tabular}
    \end{table}
    
\vspace{-0.3cm}
\paragraph{Computational efficiency.}
    Table~\ref{tab:all_results_efficiency} shows that \icaq~reduces fit time by approximately $4\times$--$5\times$ relative to capped Full QSVM, while substantially reducing the number of kernel evaluations. On MedMNIST datasets, Full QSVM requires $14{,}400$ kernel evaluations, whereas \icaq~uses only $3{,}135$. These results are consistent with the theoretical  $O(nr)$ query complexity and demonstrate that \icaq~effectively alleviates the kernel-construction bottleneck.


\vspace{-0.3cm}
\paragraph{Comparison with classical kernels.}
    Table~\ref{tab:rbf_compare} compares \icaq~with a classical RBF-SVM on binary tasks. \icaq~matches the classical baseline on Breast Cancer and binary MNIST, while outperforming it on several biomedical image datasets, including BreastMNIST, PneumoniaMNIST, PathMNIST, and OrganAMNIST. These results indicate that the low-rank quantum kernel remains competitive with strong classical kernels on medical image classification tasks.

\vspace{-0.3cm}
\paragraph{Finite-shot kernel estimation.}
    Table~\ref{tab:shots} shows that 
    increasing the shot budget from $64$ to $512$ substantially reduces the relative Frobenius error, from $1.5749$ to $0.0680$, while the classification accuracy remains comparatively stable. For an individual quantum-kernel entry, the finite-shot estimator has variance $O(N_{\mathrm{s}}^{-1})$ and hence standard deviation $O(N_{\mathrm{s}}^{-1/2})$, as established in Proposition~\ref{prop:overlap_unbiased}. The observed reduction in reconstruction error is therefore qualitatively consistent with decreasing finite-shot sampling uncertainty as the shot budget increases. While test accuracy changes only slightly across shot levels, the kernel reconstruction error decreases consistently with increasing shot budget, indicating improved kernel fidelity and convergence toward the analytic kernel. These results suggest that reconstruction error provides a more reliable measure of convergence than accuracy.

    \begin{figure}[t]
    \centering
    \begin{subfigure}[t]{0.3\linewidth}
        \centering
        \includegraphics[width=\linewidth]{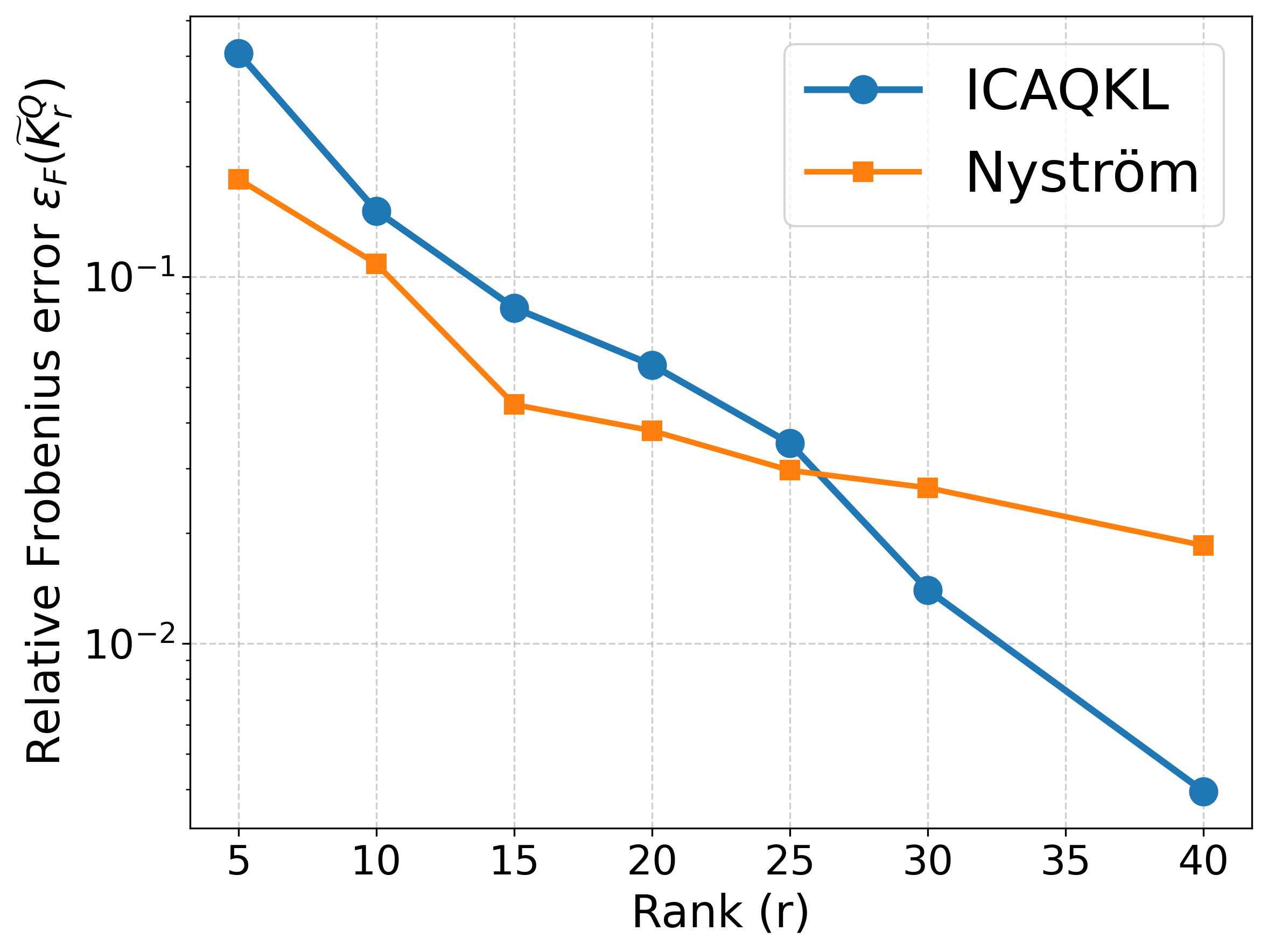}
        \caption{Error vs rank}
        \label{fig:rank_fro}
    \end{subfigure}
    \hfill
    \begin{subfigure}[t]{0.3\linewidth}
        \centering
        \includegraphics[width=\linewidth]{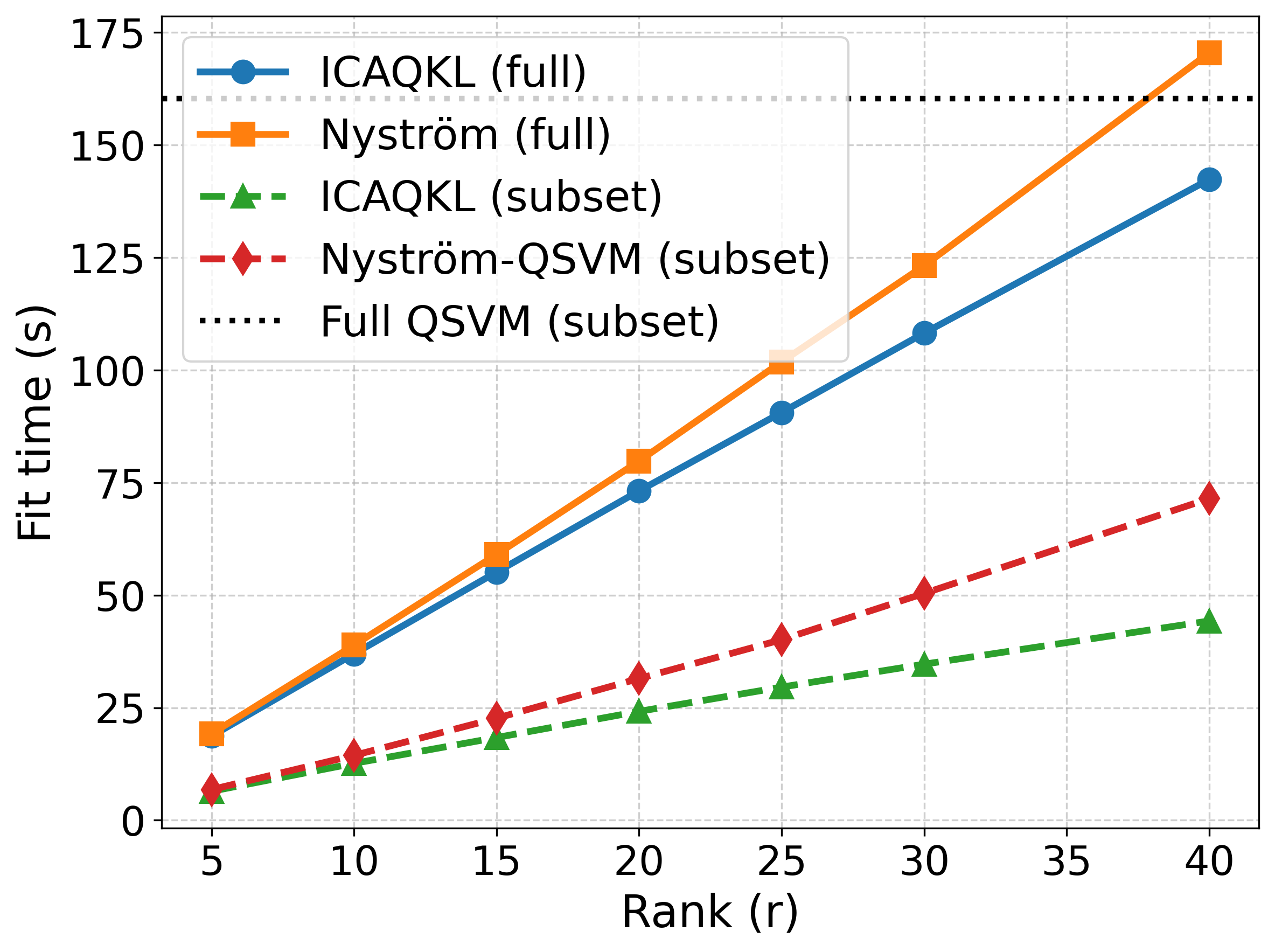}
        \caption{Time vs rank}
        \label{fig:rank_runtime}
    \end{subfigure}
    \hfill
    \begin{subfigure}[t]{0.3\linewidth}
        \centering
        \includegraphics[width=\linewidth]{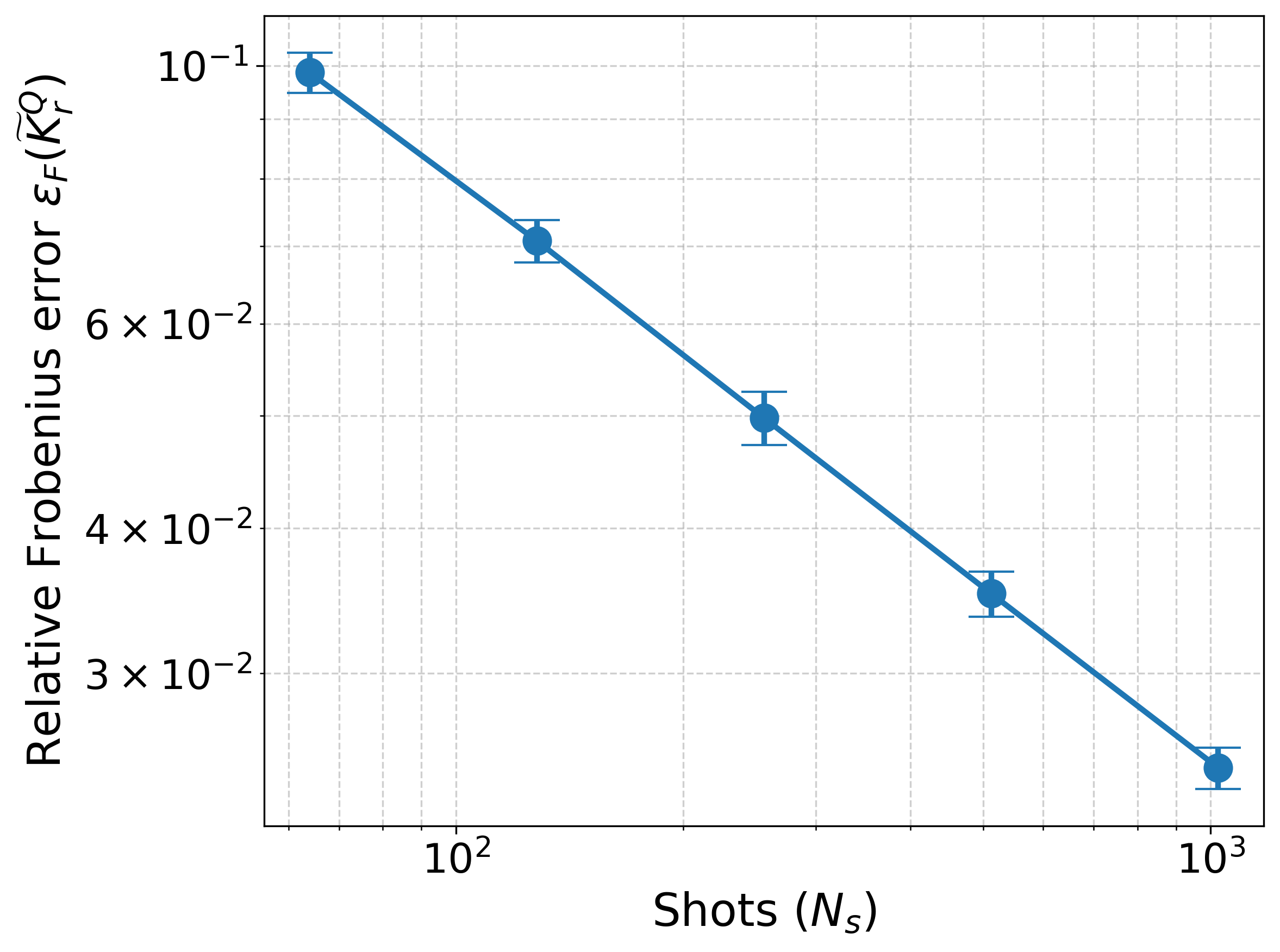}
        \caption{Error vs shots}
        \label{fig:fro_vs_shots}
    \end{subfigure} 
    \caption{Rank vs. shot results. (a) Approximation error decreases with rank. (b) Runtime grows approximately linearly with rank. (c) Kernel estimation error decreases with increasing shot budget.}
    \label{fig:ablation_all}
    \end{figure}

\vspace{-0.3cm}
\paragraph{Rank and shot ablation.}
    Fig.~\ref{fig:ablation_all} summarizes the effect of approximation rank and shot budget on kernel quality and computational cost. Fig.~\ref{fig:rank_fro} shows the relative Frobenius error as a function of rank. Both \icaq~and \nystrom~exhibit monotonic error decay as the rank increases, consistent with the expected convergence of low-rank approximations. \nystrom~ performs slightly better at small ranks due to its global landmark-based approximation, while \icaq~achieves lower reconstruction error at larger ranks due to adaptive residual-driven pivot selection. 

    Fig.~\ref{fig:rank_runtime} reports runtime as a function of rank. Both \icaq~and \nystrom~exhibit approximately linear growth with rank, consistent with the expected $O(nr)$ scaling. \icaq~consistently achieves lower runtime than \nystrom, particularly at larger ranks where the additional $O(r^2)$ landmark-kernel overhead of \nystrom~becomes increasingly significant. The Full QSVM baseline appears as a constant horizontal line since its cost is independent of rank. This highlights the key advantage of low-rank methods: they provide a tunable tradeoff between computational cost and approximation accuracy, while avoiding the quadratic complexity of the dense kernel approach.

    Fig.~\ref{fig:fro_vs_shots} illustrates the effect of shot noise on kernel estimation. The relative Frobenius error decreases steadily as the number of shots increases, following the expected $O(N_s^{-1/2})$ scaling of the estimator variance. As the shot budget increases, the variance of the kernel estimates decreases, leading to progressively more accurate approximations of the analytic kernel. The error bars remain small across all shot levels, indicating stable estimation. These results confirm that finite-shot quantum kernel estimation converges reliably toward the analytic kernel as measurement noise decreases.

    Table~\ref{tab:acc_time_compare} compares the classification accuracy and runtime of Full-QSVM, \icaq, and \nystrom~methods on subsets of the BreastMNIST, PneuMNIST, and PathMNIST datasets. \icaq~achieves predictive performance close to Full-QSVM while significantly reducing runtime. For example, on PathMNIST, \icaq~achieves $98.03\%$ accuracy versus $99.02\%$ for Full-QSVM, while decreasing runtime from $162.08\,\mathrm{s}$ to $35.90\,\mathrm{s}$. Similar trends are observed across the remaining datasets. Compared with the \nystrom~QSVM, \icaq~consistently achieves slightly higher accuracy together with lower runtime. These results suggest that adaptive residual-driven pivot selection preserves the dominant structure of the dense quantum kernel matrix more effectively than fixed landmark-based approximation while substantially reducing kernel-evaluation complexity.

\section{CONCLUSION}
\label{sec6}
We presented \icaq, a query-efficient framework for scalable quantum kernel learning based on adaptive incomplete Cholesky approximation. By selectively querying kernel entries, 
\icaq~reduces quantum evaluation complexity from $O(n^2)$ to $O(nr)$ while preserving predictive performance. 
We established end-to-end theoretical guarantees linking finite-shot estimation and low-rank approximation to  QSVM stability and prediction. We empirically showed that \icaq~closely matches exact QSVM while outperforming \nystrom-based methods at comparable ranks and achieving strong accuracy-cost tradeoffs. 
Empirically, \icaq~closely matches exact QSVM on tractable subsets and consistently outperforms \nystrom-based methods at comparable rank, achieving strong accuracy-cost tradeoffs. 
These results highlight the importance of adaptive, query-efficient kernel construction for scalable quantum machine learning. Future work includes incorporating hardware noise models and developing adaptive rank selection strategies.

\bibliographystyle{plainnat}
\bibliography{References}

@inproceedings{Bach2013,
  author    = {Francis Bach},
  title     = {Sharp Analysis of Low-Rank Kernel Matrix Approximations},
  booktitle = {Proceedings of the International Conference on Learning Theory (COLT)},
  volume    = {30},
  pages     = {1--25},
  year      = {2013},
  publisher = {JMLR}
}

@article{ChenEpperlyTroppWebber2024,
  author    = {Yifan Chen and Ethan N. Epperly and Joel A. Tropp and Robert J. Webber},
  title     = {Randomly Pivoted Cholesky: Practical Approximation of a Kernel Matrix with Few Entry Evaluations},
  journal   = {Communications on Pure and Applied Mathematics},
  volume    = {78},
  number    = {},
  pages     = {995--1041},
  year      = {2025},
  doi       = {10.1002/cpa.22234}
}

@misc{Coelho2025,
  author        = {Rodrigo Coelho and Georg Kruse and Andreas Rosskopf},
  title         = {Quantum-Efficient Kernel Target Alignment},
  year          = {2025},
  month         = {Feb.},
  eprint        = {2502.08225},
  archivePrefix = {arXiv},
  primaryClass  = {quant-ph},
  doi           = {10.48550/arXiv.2502.08225},
  url           = {https://arxiv.org/abs/2502.08225}
}

@article{DrineasMahoney2005,
  author    = {Petros Drineas and Michael W. Mahoney},
  title     = {On the Nystr{\"o}m Method for Approximating a Gram Matrix for Improved Kernel-Based Learning},
  journal   = {Journal of Machine Learning Research},
  volume    = {6},
  number    = {72},
  pages     = {2153--2175},
  year      = {2005},
  month     = {Dec.},
  url       = {https://jmlr.org/papers/v6/drineas05a.html}
}

@inproceedings{Du2026,
  author    = {Jinda Du and Jian Hou and Huaqiang Yuan},
  title     = {Improving Nystr{\"o}m Spectral Clustering with Unsupervised Vector Quantization and Incomplete Cholesky Decomposition},
  booktitle = {PRICAI 2025: Trends in Artificial Intelligence},
  series    = {Lecture Notes in Computer Science},
  volume    = {16451},
  pages     = {155--170},
  year      = {2026},
  publisher = {Springer, Singapore},
  doi       = {10.1007/978-981-95-7075-1_10},
  url       = {https://doi.org/10.1007/978-981-95-7075-1_10}
}

@article{Epperly2025,
  author    = {Ethan N. Epperly and Joel A. Tropp and Robert J. Webber},
  title     = {Embrace Rejection: Kernel Matrix Approximation by Accelerated Randomly Pivoted Cholesky},
  journal   = {SIAM Journal on Matrix Analysis and Applications},
  volume    = {46},
  number    = {4},
  pages     = {2527--2557},
  year      = {2025},
  doi       = {10.1137/24M1699048},
  url       = {https://doi.org/10.1137/24M1699048}
}

@article{Fine2002,
  author    = {Shai Fine and Katya Scheinberg},
  title     = {Efficient SVM Training Using Low-Rank Kernel Representations},
  journal   = {Journal of Machine Learning Research},
  volume    = {2},
  number    = {Mar},
  pages     = {243--264},
  year      = {2002},
  month     = {Mar.},
  url       = {https://jmlr.org/papers/v2/fine01a.html}
}

@article{Harbrecht2012,
  author    = {Helmut Harbrecht and Michael Peters and Reinhold Schneider},
  title     = {On the Low-Rank Approximation by the Pivoted Cholesky Decomposition},
  journal   = {Applied Numerical Mathematics},
  volume    = {62},
  number    = {4},
  pages     = {428--440},
  year      = {2012},
  month     = {Apr.},
  doi       = {10.1016/j.apnum.2011.10.001},
  url       = {https://doi.org/10.1016/j.apnum.2011.10.001}
}

@article{Havlicek2019,
  author    = {Vojt{\v{e}}ch Havl{\'i}{\v{c}}ek and Antonio D. C{\'o}rcoles and Kristan Temme and Aram W. Harrow and Abhinav Kandala and Jerry M. Chow and Jay M. Gambetta},
  title     = {Supervised Learning with Quantum-Enhanced Feature Spaces},
  journal   = {Nature},
  volume    = {567},
  number    = {7747},
  pages     = {209--212},
  year      = {2019},
  month     = {Mar.},
  doi       = {10.1038/s41586-019-0980-2},
  url       = {https://doi.org/10.1038/s41586-019-0980-2}
}

@misc{Hayakawa2023,
  author        = {Satoshi Hayakawa and Harald Oberhauser and Terry Lyons},
  title         = {Sampling-Based Nystr{\"o}m Approximation and Kernel Quadrature},
  year          = {2023},
  eprint        = {2301.09517},
  archivePrefix = {arXiv},
  primaryClass  = {math.NA},
  doi           = {10.48550/arXiv.2301.09517},
  url           = {https://arxiv.org/abs/2301.09517}
}

@article{Hofmann2007,
  author    = {Thomas Hofmann and Bernhard Sch{\"o}lkopf and Alex Smola},
  title     = {Kernel Methods in Machine Learning},
  journal   = {Annals of Statistics},
  volume    = {36},
  number    = {3},
  pages     = {1171--1220},
  year      = {2008},
  month     = {Jun.},
  doi       = {10.1214/009053607000000677},
  url       = {https://doi.org/10.1214/009053607000000677}
}

@inproceedings{Kubler2021,
  author    = {Jonas M. K{\"u}bler and Simon Buchholz and Bernhard Sch{\"o}lkopf},
  title     = {The Inductive Bias of Quantum Kernels},
  booktitle = {Proceedings of the 35th International Conference on Neural Information Processing Systems},
  pages     = {12661--12673},
  year      = {2021},
  publisher = {Curran Associates, Inc.},
}

@article{Mahoney2011,
  author    = {Michael W. Mahoney},
  title     = {Randomized Algorithms for Matrices and Data},
  journal   = {Foundations and Trends in Machine Learning},
  volume    = {3},
  number    = {2},
  pages     = {123--224},
  year      = {2011},
  month     = {Feb.},
  doi       = {10.1561/2200000035},
  url       = {https://doi.org/10.1561/2200000035}
}

@inproceedings{MuscoMusco2017,
  author    = {Cameron Musco and Christopher Musco},
  title     = {Recursive Sampling for the Nystrom Method},
  booktitle = {Advances in Neural Information Processing Systems},
  volume    = {30},
  year      = {2017},
  publisher = {Curran Associates, Inc.},
}

@inproceedings{Rahimi2007,
  author    = {Ali Rahimi and Benjamin Recht},
  title     = {Random Features for Large-Scale Kernel Machines},
  booktitle = {Advances in Neural Information Processing Systems},
  volume    = {20},
  pages = {1177–1184},
  numpages = {8},
  year      = {2007},
  publisher = {Curran Associates, Inc.},
}

@article{RodriguezGrasa2025,
  author    = {Pablo Rodriguez-Grasa and Yue Ban and Mikel Sanz},
  title     = {Neural Quantum Kernels: Training Quantum Kernels with Quantum Neural Networks},
  journal   = {Physical Review Research},
  volume    = {7},
  number    = {2},
  pages     = {023269},
  numpages  = {16},
  year      = {2025},
  month     = {Jun.},
  doi       = {10.1103/xphb-x2g4},
  url       = {https://link.aps.org/doi/10.1103/xphb-x2g4}
}

@article{SchnabelRoth2025,
  author    = {Jan Schnabel and Marco Roth},
  title     = {Quantum Kernel Methods under Scrutiny: A Benchmarking Study},
  journal   = {Quantum Machine Intelligence},
  volume    = {7},
  number    = {58},
  year      = {2025},
  month     = {Apr.},
  doi       = {10.1007/s42484-025-00273-5},
  url       = {https://doi.org/10.1007/s42484-025-00273-5}
}

@misc{Schuld2021,
  author        = {Maria Schuld},
  title         = {Supervised Quantum Machine Learning Models Are Kernel Methods},
  year          = {2021},
  month         = {Jan.},
  eprint        = {2101.11020},
  archivePrefix = {arXiv},
  primaryClass  = {quant-ph},
  doi           = {10.48550/arXiv.2101.11020},
  url           = {https://arxiv.org/abs/2101.11020}
}

@article{Schuld2019,
  author    = {Maria Schuld and Nathan Killoran},
  title     = {Quantum Machine Learning in Feature Hilbert Spaces},
  journal   = {Physical Review Letters},
  volume    = {122},
  number    = {4},
  pages     = {040504},
numpages = {6},
  year      = {2019},
  month     = {Feb.},
  doi       = {10.1103/PhysRevLett.122.040504},
  url       = {https://link.aps.org/doi/10.1103/PhysRevLett.122.040504}
}

@article{Huang2021,
  author    = {Hsin-Yuan Huang and Michael Broughton and Masoud Mohseni and Ryan Babbush and Sergio Boixo and Hartmut Neven and Jarrod R. McClean},
  title     = {Power of Data in Quantum Machine Learning},
  journal   = {Nature Communications},
  volume    = {12},
  number    = {2631},
  year      = {2021},
  month     = {May},
  doi       = {10.1038/s41467-021-22539-9},
  url       = {https://doi.org/10.1038/s41467-021-22539-9}
}

@misc{Singh2026,
  author        = {Utkarsh Singh and Jean-Fr{\'e}d{\'e}ric Laprade and Aaron Z. Goldberg and Khabat Heshami},
  title         = {A Resource Efficient Quantum Kernel},
  year          = {2026},
  eprint        = {2507.03689},
  archivePrefix = {arXiv},
  primaryClass  = {quant-ph},
  doi           = {10.48550/arXiv.2507.03689},
  url           = {https://arxiv.org/abs/2507.03689}
}

@article{Talitckii2024,
  author    = {Aleksandr Talitckii and Brendon Colbert and Matthew M. Peet},
  title     = {Efficient Convex Algorithms for Universal Kernel Learning},
  journal   = {Journal of Machine Learning Research},
  volume    = {25},
  number    = {203},
  pages     = {1--40},
  year      = {2024},
  url       = {https://www.jmlr.org/papers/v25/23-0528.html}
}

@article{Thanasilp2024,
  author    = {Supanut Thanasilp and Samson Wang and M. Cerezo and Zo{\"e} Holmes},
  title     = {Exponential Concentration in Quantum Kernel Methods},
  journal   = {Nature Communications},
  volume    = {15},
  number    = {5200},
  year      = {2024},
  month     = {Jun.},
  doi       = {10.1038/s41467-024-49287-w},
  url       = {https://doi.org/10.1038/s41467-024-49287-w}
}

@article{Volkoff2022,
  author    = {T. J. Volkoff and Yi{\u{g}}it Suba{\c{s}}{\i}},
  title     = {Ancilla-Free Continuous-Variable {SWAP} Test},
  journal   = {Quantum},
  volume    = {6},
  pages     = {800},
  year      = {2022},
  month     = {Sep.},
  doi       = {10.22331/q-2022-09-08-800},
  url       = {https://doi.org/10.22331/q-2022-09-08-800}
}

@article{yang2023medmnistv2,
  author    = {Jiancheng Yang and Rui Shi and Donglai Wei and others},
  title     = {MedMNIST v2: A Large-Scale Lightweight Benchmark for 2D and 3D Biomedical Image Classification},
  journal   = {Scientific Data},
  volume    = {10},
  number    = {1},
  pages     = {41},
  year      = {2023},
  month     = {Jan.},
  doi       = {10.1038/s41597-022-01721-8},
  url       = {https://doi.org/10.1038/s41597-022-01721-8}
}

@article{lecun1998mnist,
  title   = {Gradient-Based Learning Applied to Document Recognition},
  author  = {LeCun, Yann and Bottou, L{\'e}on and Bengio, Yoshua and Haffner, Patrick},
  journal = {Proceedings of the IEEE},
  volume  = {86},
  number  = {11},
  pages   = {2278--2324},
  year    = {1998},
  doi={10.1109/5.726791}}

@misc{xiao2017fashionmnist,
  author        = {Han Xiao and Kashif Rasul and Roland Vollgraf},
  title         = {Fashion-MNIST: A Novel Image Dataset for Benchmarking Machine Learning Algorithms},
  year          = {2017},
  eprint        = {1708.07747},
  archivePrefix = {arXiv},
  primaryClass  = {cs.LG},
  doi           = {10.48550/arXiv.1708.07747},
  url           = {https://arxiv.org/abs/1708.07747}
}

@misc{WDBC1995,
  author       = {Wolberg, William H. and Street, W. Nick and Mangasarian, Olvi L.},
  title        = {Wisconsin Diagnostic Breast Cancer (WDBC)},
  year         = {1993},
  howpublished = {UCI Machine Learning Repository},
  doi          = {10.24432/C5DW2B},
  url          = {https://doi.org/10.24432/C5DW2B}
}

@article{Shaydulin2022,
  author    = {Ruslan Shaydulin and Stefan Wild},
  title     = {Importance of Kernel Bandwidth in Quantum Machine Learning},
  journal   = {Phys. Rev. A},
  volume    = {106},
  issue     = {4},
  pages     = {042407},
  numpages  = {11},
  year      = {2022},
  month     = {Oct.},
  doi       = {10.1103/PhysRevA.106.042407},
  url       = {https://link.aps.org/doi/10.1103/PhysRevA.106.042407}
}

\clearpage
\appendix
\thispagestyle{empty}

\onecolumn

\section{MORE COMPREHENSIVE PRELIMINARIES}

\subsection{Classical Support Vector Machines}
\label{subsec:CSVMs}
    We consider a supervised classification dataset $\mathcal D \!=\! \{(\mathbf{x}_i, y_i)\}_{i=1}^n$, where $\mathbf{x}_i \!\in\! \mathcal X \!\subseteq\! \mathbb R^d$ is a $d$-dimension input vector and $y_i \!\in\! \left\{-1, +1\right\}$ denotes the corresponding class label. The goal of SVM is to seek an affine decision function, $f(\mathbf{x}) \!=\! \mathbf{w}^\top \mathbf{x} \!+\! b$, whose induced separating hyperplane achieves a large margin between the two classes. Here, 
    $\mathbf{w}\in\mathbb{R}^d$ defines the normal direction of the separating hyperplane, and $b\in\mathbb{R}$ is a bias term. To accommodate non-separable or noisy observations, slack variables $\boldsymbol \xi \!=\! (\xi_1, \ldots, \xi_n)^\top$ are introduced, yielding the soft-margin primal problem
    \begin{subequations}
    \label{eq:svm_primal}
    \begin{align}
    \label{eq:svm_obj}
        \min_{\mathbf{w}, \, b, \, \boldsymbol{\xi}} \;\; & \frac{1}{2} \|\mathbf{w}\|_2^2 + C \sum_{i=1}^n \xi_i \\
    \label{eq:svm_constr}
        \textrm{s.t.} \quad & y_i(\mathbf{w}^\top \mathbf{x}_i + b) \ge 1 - \xi_i, \; i=1,\dots,n, \\
    \label{eq:svm_xi}
        & \xi_i \ge 0, \; i=1,\dots,n,
    \end{align}
    \end{subequations}
    The first term in (\ref{eq:svm_obj}) promotes a large margin, whereas the slack penalty controls violations of the unit-margin constraints. The regularization $C>0$ determines the trade-off between these two objectives. Since (\ref{eq:svm_primal}) is a convex quadratic program, its dual can be obtained through Lagrangian duality. Introducing dual variables $\boldsymbol\alpha=(\alpha_1,\ldots,\alpha_n)^\top$, the dual problem is
    \begin{subequations}
    \label{eq:svm_dual}
    \begin{align}
    \label{eq:dual_obj}
        \max_{\boldsymbol \alpha} \quad & \sum_{i=1}^n \alpha_i - \frac{1}{2} \sum_{i=1}^n \sum_{j=1}^n \alpha_i \alpha_j y_i y_j \, \mathbf{x}_i^\top \mathbf{x}_j \\
    \label{eq:dual_constr}
        \text{s.t.} \quad & \sum_{i=1}^n y_i \alpha_i = 0,\\ 
    \label{eq:dual_alpha}
        & 0 \le \alpha_i \le C, \; i=1,\dots,n.
    \end{align}
    \end{subequations}
    The equality constraint in (\ref{eq:dual_constr}) follows from stationarity with respect to the unregularized intercept $b$. Consequently, the corresponding prediction function must also retain this intercept. The dual objective depends on the training inputs only through pairwise inner products $\mathbf x_i^\top\mathbf x_j$. This is crucial, as it enables the extension of linear SVMs to nonlinear classification via the kernel trick, where the inner products are replaced by a PSD kernel function, yielding a nonlinear classifier while preserving the convex optimization structure of the SVM. Let $\varphi: \mathcal X \to \mathcal H$ be a feature map into a Hilbert space $\mathcal H$, and let $\kappa(\mathbf x_i,\mathbf x_j) = \left\langle \varphi(\mathbf x_i), \varphi(\mathbf x_j) \right\rangle_{\mathcal H}$
    be the associated positive-semidefinite kernel. Replacing the Euclidean inner products in (\ref{eq:dual_obj}) by $\kappa(\mathbf x_i,\mathbf x_j)$ yields 
    \begin{align} 
        \min_{\boldsymbol\alpha\in\mathcal A} \; J_{\mathbf K}(\boldsymbol\alpha) := \frac{1}{2} \boldsymbol\alpha^\top \mathbf Y\mathbf K\mathbf Y \boldsymbol\alpha - \mathbf 1^\top\boldsymbol\alpha, 
    \label{eq:kernel_svm_dual_min}
    \end{align}
    where $\mathbf K_{ij} = \kappa(\mathbf x_i,\mathbf x_j)$, $\mathbf Y = \operatorname{diag}(y_1,\ldots,y_n)$, and $\mathcal A := \left\{ \boldsymbol\alpha\in\mathbb R^n: 0\le\alpha_i\le C,; \mathbf y^\top\boldsymbol\alpha=0 \right\}$. The corresponding decision function is 
    \begin{align} 
        f_{\mathbf K}(\mathbf z) = \sum_{i=1}^n \alpha_i^\star y_i \kappa(\mathbf x_i,\mathbf z) + b^\star, 
    \label{eq:kernel_svm_predictor} 
    \end{align} 
    where $\boldsymbol\alpha^\star\in\arg\min_{\boldsymbol\alpha\in\mathcal A} J_{\mathbf K}(\boldsymbol\alpha)$ and $b^\star$ is the SVM bias term. This kernelized formulation provides the starting point for the quantum-kernel construction considered in the following subsection.

\subsection{Quantum Kernel Learning}
\label{subsec:QKM}
    Quantum kernel learning extends the kernel formulation of Section~\ref{subsec:CSVMs} by replacing a classical feature map with a quantum-state embedding. Let $U_\phi: \mathcal X \rightarrow \mathcal U(2^q)$ denote a data-dependent unitary acting on $q$ qubits. For an input $\mathbf x\in\mathcal X$, the corresponding quantum feature state is:
    \begin{align} 
        |\phi(\mathbf x)\rangle = U_\phi(\mathbf x) |0\rangle^{\otimes q}. 
    \label{eq:quantum_feature_state_app} 
    \end{align} 
    We consider the fidelity-based quantum kernel 
    \begin{align} 
        \kappa^Q(\mathbf x_i,\mathbf x_j) := \left| \langle \phi(\mathbf x_j) \mid \phi(\mathbf x_i) \rangle \right|^2. 
    \label{eq:fidelity_kernel} 
    \end{align} 
    For a training set ${\mathbf x_i}_{i=1}^n$, the corresponding ideal quantum Gram matrix is:
    \begin{align} 
        \mathbf K^Q := \left[ K_{ij}^Q \right]_{i,j=1}^n, \quad K_{ij}^Q := \kappa^Q(\mathbf x_i,\mathbf x_j). 
    \label{eq:Q_gram_definition} 
    \end{align}

\paragraph{Overlap estimation.} 
    We adopt an overlap-based estimator to compute each kernel entry, avoiding the ancilla qubits and controlled-SWAP operations required by SWAP-test methods \cite{Volkoff2022}.
    For a pair of inputs $(\mathbf{x}_i, \mathbf{x}_j)$, from a reference state $\ket{0}^{\otimes q}$ the overlap circuit prepares:
    \begin{align}
    \label{eq:overlap_state}
        & |\psi_{\mathrm{final}}(\mathbf{x}_i, \mathbf{x}_j) \rangle = U_\phi^\dagger(\mathbf{x}_j) U_\phi(\mathbf{x}_i)\ket{0}^{\otimes q},
    \end{align}
    with the corresponding density operator $\rho_{\mathrm{final}} (\mathbf x_i,\mathbf x_j) := |\psi_{\mathrm{final}} (\mathbf x_i,\mathbf x_j)\rangle \langle \psi_{\mathrm{final}} (\mathbf x_i,\mathbf x_j)|$. Measuring this state in the computational basis, the probability of the all-zero outcome is:
    \begin{align}
    \label{eq:overlap_probability}
        & p_0(\mathbf{x}_i,\mathbf{x}_j) = \mathrm{Tr}\!\left( \ket{0}\bra{0}^{\otimes q} \rho_{\mathrm{final}}(\mathbf{x}_i,\mathbf{x}_j) \right) = \Big| \langle 0|^{\otimes q} U_\phi^\dagger(\mathbf x_j) \, U_\phi(\mathbf x_i) |0 \rangle^{\otimes q} \Big|^2.
    \end{align}
    This Born-rule probability equals the fidelity kernel and can be estimated by repeated circuit executions and counting the frequency of the all-zero outcome.

    \begin{proposition}
    \label{prop:overlap_kernel}
        For any $\mathbf x_i, \mathbf x_j \in \mathcal X$, the probability of the all-zero measurement outcome under the overlap circuit equals the fidelity-based quantum kernel, that is:
        \begin{align}
        \label{eq:overlap_equals_kernel}
            & p_0(\mathbf x_i, \mathbf x_j) = \kappa^Q(\mathbf x_i, \mathbf x_j).
        \end{align}
    \end{proposition}
    \begin{proof}[Proof of Proposition \ref{prop:overlap_kernel}]
        Let $A := U_\phi^\dagger(\mathbf x_j)U_\phi(\mathbf x_i)$ denote the composed unitary operator, and $\Pi_0 := |0\rangle\langle 0|^{\otimes q}$ define the projector onto the all-zero state. Since $\rho_{\mathrm{final}}=A\Pi_0A^\dagger$, by the Born rule, the probability of measuring the all-zero outcome is defined by:
        \begin{align}
            & \mathbb{P} \big(0^{\otimes q} \big) = \mathrm{Tr} \big(\Pi_0 \, \rho_{\mathrm{final}} \big) = \mathrm{Tr} \big(\Pi_0 A \Pi_0 A^\dagger \big).
        \end{align}
        Additionally, let $|0_q\rangle = |0\rangle^{\otimes q}$, we have:
        \begin{align}
            \mathrm{Tr} \big(\Pi_0 A \Pi_0 A^\dagger \big) 
            & = \mathrm{Tr} \big( |0_q \rangle \langle 0_q| \, A \, |0_q \rangle \langle 0_q| \, A^\dagger \big) = \langle 0_q|A|0_q \rangle\, \mathrm{Tr} \big( |0_q \rangle \langle 0_q| \, A^\dagger \big).
        \end{align}
        Applying the identity, $\mathrm{Tr} \big(|\psi\rangle \langle\phi|, X \big) = \langle \phi| X |\psi \rangle$, with $|\psi \rangle = |\phi \rangle = |0_q \rangle$ and $X = A^\dagger$, we have:
        \begin{align}
            \mathrm{Tr} \big(\Pi_0 A \Pi_0 A^\dagger \big) 
            & = \langle 0_q|A|0_q\rangle\, \langle 0_q| A^\dagger |0_q \rangle = \langle 0_q| A |0_q \rangle \, \overline{\langle 0_q| A |0_q \rangle} = \left| \langle 0_q| A |0_q\rangle \right|^2.
        \end{align}
        Using $\Pi_0 = |0\rangle \langle 0|^{\otimes q}$, we can obtain:
        \begin{align}
            \mathbb{P} \big(0^{\otimes q} \big) 
            & = \mathrm{Tr} \big( |0\rangle\langle 0|^{\otimes q} A |0\rangle\langle 0|^{\otimes q} A^\dagger \big) = \left| \langle 0|^{\otimes q} U_\phi^\dagger (\mathbf x_j) U_\phi(\mathbf x_i) |0\rangle^{\otimes q} \right|^2. 
        \end{align}
        Since the feature map defines $|\phi(\mathbf x)\rangle = U_\phi(\mathbf x) |0\rangle^{\otimes q}$, their corresponding inner product is 
        \begin{align}
            & \langle \phi(\mathbf x_j) \mid \phi(\mathbf x_i) \rangle = \langle 0|^{\otimes q} U_\phi^\dagger(\mathbf x_j) U_\phi(\mathbf x_i) |0\rangle^{\otimes q}.
        \end{align}
        Therefore, we obtain:
        \begin{align}
            & p_0 \big(\mathbf x_i,\mathbf x_j \big) = \big|\langle \phi(\mathbf x_j) \mid \phi(\mathbf x_i)\rangle\big|^2 = \kappa^Q(\mathbf x_i, \mathbf x_j),
        \end{align}
        which proves the proposition.
    \end{proof}
    
\paragraph{Finite-shot quantum-kernel oracle.} 
    On quantum hardware, the probability in (\ref{eq:overlap_probability}) is not available exactly and must be estimated from repeated circuit executions. For a queried pair $(\mathbf x_i,\mathbf x_j)$, let 
    \begin{align} 
        X_s = \begin{cases} 1, & \text{if the $s$th measurement returns } 0^{\otimes q}, \\ 0, & \text{otherwise}, \end{cases} \quad s=1,\ldots,N_{\mathrm{s}}. 
    \end{align} 
    The finite-shot overlap estimator is 
    \begin{align} 
        \widehat\kappa^Q_{N_{\mathrm{s}}} (\mathbf x_i,\mathbf x_j) := \frac{1}{N_{\mathrm{s}}} \sum_{s=1}^{N_{\mathrm{s}}}X_s. 
    \label{eq:shot_overlap_estimator} 
    \end{align}

    \begin{proposition}[Shot-based overlap estimator]
    \label{prop:overlap_unbiased}
        For every fixed queried pair $\mathbf x_i,\mathbf x_j\in\mathcal X$, let $\widehat{\kappa}^Q_{N_{\mathrm{s}}}(\mathbf x_i, \mathbf x_j)$ denote the empirical estimator obtained from $N_{\mathrm{s}}$ independent circuit repetitions, defined as the empirical frequency of observing the all-zero outcome. Then, 
        \begin{align}
        \label{eq:overlap_unbiased}
            \mathbb E \left[ \widehat\kappa^Q_{N_{\mathrm{s}}} (\mathbf x_i,\mathbf x_j) \right] = \kappa^Q(\mathbf x_i,\mathbf x_j), \quad
            \operatorname{Var} \left[ \widehat\kappa^Q_{N_{\mathrm{s}}} (\mathbf x_i,\mathbf x_j) \right] = \frac{ \kappa^Q(\mathbf x_i,\mathbf x_j) \left[ 1- \kappa^Q(\mathbf x_i,\mathbf x_j) \right] }{ N_{\mathrm{s}} }.
        \end{align}
    \end{proposition}
    \begin{proof}[Proof of Proposition \ref{prop:overlap_unbiased}]
        For each circuit execution, by Proposition~\ref{prop:overlap_kernel}, the success probability is:
        \begin{align}
            & \mathbb{P}(X_s=1) = \kappa^Q(\mathbf x_i, \mathbf x_j).
        \end{align}
        Hence, the outcomes across repetitions $X_1, \dots, X_{N_{\mathrm{s}}}$ are i.i.d. Bernoulli random variables with parameter $p = \kappa^Q(\mathbf x_i, \mathbf x_j)$, and the empirical estimator is then expressed as $\widehat{\kappa}^Q_{N_{\mathrm{s}}}(\mathbf x_i, \mathbf x_j) = \frac{1}{N_{\mathrm{s}}} \sum_{s=1}^{N_{\mathrm{s}}} X_s$.
        Therefore, we can have: 
        \begin{align}
            & \mathbb{E} \left[ \widehat{\kappa}^Q_{N_{\mathrm{s}}}(\mathbf x_i, \mathbf x_j)\right] = \frac{1}{N_{\mathrm{s}}}\sum_{s=1}^{N_{\mathrm{s}}} \mathbb{E}[X_s] = p,
        \end{align}
        which proves (\ref{eq:overlap_unbiased}). Moreover, since
        \begin{align}
            & \mathrm{Var}(X_s) = \kappa^Q(\mathbf x_i, \mathbf x_j) \big(1 - \kappa^Q(\mathbf x_i, \mathbf x_j)\big),
        \end{align}
        independence implies that:
        \begin{align}
            \mathrm{Var} \left[\widehat{\kappa}^Q_{N_{\mathrm{s}}}(\mathbf x_i, \mathbf x_j)\right] 
            & = \frac{1}{N_{\mathrm{s}}^2} \sum_{s=1}^{N_{\mathrm{s}}} \mathrm{Var}(X_s) = \frac{\kappa^Q(\mathbf x_i, \mathbf x_j) \big[1 - \kappa^Q(\mathbf x_i, \mathbf x_j) \big]}{N_{\mathrm{s}}},
        \end{align}
        which proves the proposition.
    \end{proof}
    The estimator is unbiased with standard error scaling as $O(1/ \sqrt{N_{\mathrm{s}}})$. Achieving high accuracy may therefore require a large number of shots, especially when many kernel entries are needed. Given $n$ training samples, estimating the full kernel matrix requires $n(n+1)/2$ entries. With $N_{\mathrm{s}}$ shots per entry, the total number of circuit executions scales as $O(n^2 N_{\mathrm{s}})$. Moreover, each evaluation involves both $U_{\phi}(\mathbf{x}_i)$ and $U_{\phi}^{\dagger}(\mathbf{x}_j)$, so the runtime also depends on circuit depth and execution cost of the feature map. This quadratic scaling in $n$ is a primary computational bottleneck of QKL.

    \emph{Analytically known fidelity diagonal:} An important special property of the normalized fidelity kernel is that its diagonal is known exactly: \begin{align} 
        K_{ii}^Q = \kappa^Q(\mathbf x_i,\mathbf x_i) = \left| \langle \phi(\mathbf x_i) \mid \phi(\mathbf x_i) \rangle \right|^2 = 1. 
    \label{eq:fidelity_diagonal} 
    \end{align} 
    Equivalently, $U_\phi^\dagger(\mathbf x_i) U_\phi(\mathbf x_i) = \mathbf I$. Thus, under the finite-shot-only model considered in this paper, the diagonal entries require no circuit executions and can be initialized analytically. This property will be used explicitly by the finite-shot ICAQKL recursion.
    
\paragraph{Ideal quantum Gram matrix.}
    Accordingly, given a full training set $\{\mathbf x_i \}_{i=1}^n$, the corresponding quantum Gram matrix $\mathbf K^Q $ with $\big[ K_{ij}^Q \big]_{i,j=1}^n = \big[ \kappa^Q(\mathbf{x}_i, \mathbf{x}_j) \big]_{i,j=1}^n$ can be defined by:
    \begin{align}
    \label{eq:Q_gram_matrix}
        & \mathbf K^Q =
            \begin{bmatrix}
                \kappa^Q(\mathbf{x}_1,\mathbf{x}_1) & \kappa^Q(\mathbf{x}_1,\mathbf{x}_2) & \cdots & \kappa^Q(\mathbf{x}_1,\mathbf{x}_n)\\
                \kappa^Q(\mathbf{x}_2,\mathbf{x}_1) & \kappa^Q(\mathbf{x}_2,\mathbf{x}_2) & \cdots & \kappa^Q(\mathbf{x}_2,\mathbf{x}_n)\\
                \vdots & \vdots & \ddots & \vdots\\
                \kappa^Q(\mathbf{x}_n,\mathbf{x}_1) & \kappa^Q(\mathbf{x}_n,\mathbf{x}_2) & \cdots & \kappa^Q(\mathbf{x}_n,\mathbf{x}_n)
            \end{bmatrix}
            \in \mathbb{R}^{n\times n}.
    \end{align}
    A kernel method such as SVM accesses the quantum feature representation only through these pairwise similarities; it does not explicitly construct the exponentially large quantum feature vector. We emphasize, however, that the present work does not require or claim that the chosen quantum kernel is classically intractable. Our focus is the cost of acquiring and approximating a specified quantum kernel under finite-shot access.

    \begin{lemma}[PSD of the fidelity-based kernel]
    \label{lem:PSD_K}
        Let $\mathcal X \subseteq \mathbb{R}^d$ denote the input space. For any $\mathbf x_i, \mathbf x_j \in \mathcal X$, define the corresponding quantum kernel as:
        \begin{align}
        \label{eq:fidelity_kernel}
            & \kappa^Q(\mathbf x_i, \mathbf x_j) = \big|\langle \phi(\mathbf x_j) \mid \phi(\mathbf x_i) \rangle\big|^2,
        \end{align}
        and let $\mathbf K^Q$ be the associated Gram matrix with entries $K^Q_{ij} = \kappa^Q(\mathbf x_i,\mathbf x_j)$. Then $\mathbf K^Q$ is Hermitian PSD. That is, $K^Q_{ij} = \overline{K^Q_{ji}}$, and for every $\mathbf c \in \mathbb{C}^n$,
        \begin{align}
        \label{eq:psd_quadratic_form}
            & \mathbf c^\dagger \mathbf K^Q \mathbf c \ge 0.
        \end{align}
    \end{lemma}
    \begin{proof}
    \label{pro:lem_PSD_K}
        We first show that $\mathbf K^Q$ is Hermitian. For any $\mathbf x_i,\mathbf x_j \in \mathcal X$, we have:
        \begin{align}
        \label{eq:KQ_hermitian}
            K^Q_{ji}
            & = \big|\langle \phi(\mathbf x_i) \mid \phi(\mathbf x_j)\rangle\big|^2 = \big|\overline{\langle \phi(\mathbf x_j) \mid \phi(\mathbf x_i)\rangle}\big|^2 = \big|\langle \phi(\mathbf x_j)\mid \phi(\mathbf x_i)\rangle\big|^2 = K^Q_{ij},
        \end{align}
        implying $\mathbf K^Q$ is real symmetric, and therefore Hermitian.
    
        For each input $\mathbf x_i \in \mathcal X$, we define the lifted feature map the feature state $|\phi(\mathbf x_i) \rangle $ as follows:
        \begin{align}
        \label{eq:lifted_feature_map}
            & \Psi(\mathbf x_i) = |\phi(\mathbf x_i) \rangle \langle \phi(\mathbf x_i)|,
        \end{align}
        which is a rank-one projector on $\mathcal H$. Consider the Hilbert space of operators on $\mathcal H$ equipped with the Hilbert--Schmidt inner product, given by:
        \begin{align}
        \label{eq:HS_inner_product}
            & \langle A,B\rangle_{\mathrm{HS}}=\mathrm{Tr}(A^\dagger B).
        \end{align}
        By definition, for any $\mathbf x_i, \mathbf x_j \in \mathcal X$, we have:
        \begin{align}
        \label{eq:KQ_as_HS}
            & \langle \Psi(\mathbf x_i), \Psi(\mathbf x_j) \rangle_{\mathrm{HS}} = \mathrm{Tr}  \big(\Psi(\mathbf x_i)^\dagger \Psi(\mathbf x_j) \big).
        \end{align}
        Since $\Psi(\mathbf x_i)$ is Hermitian and rank one, we have $\Psi(\mathbf x_i)^\dagger = \Psi(\mathbf x_i)$, it follows that:
        \begin{subequations}
        \label{eq:trace_expansion}
        \begin{align}
            \langle \Psi(\mathbf x_i), \Psi(\mathbf x_j) \rangle_{\mathrm{HS}} 
            &= \mathrm{Tr}  \big(\Psi(\mathbf x_i) \Psi(\mathbf x_j) \big) = \mathrm{Tr} \left( |\phi(\mathbf x_i) \rangle\langle \phi(\mathbf x_i)| \, |\phi(\mathbf x_j) \rangle\langle \phi(\mathbf x_j)| \right) \\ 
            & = \langle \phi(\mathbf x_i) \mid \phi(\mathbf x_j)\rangle \langle \phi(\mathbf x_j) \mid \phi(\mathbf x_i)\rangle \\ 
            & = \big|\langle \phi(\mathbf x_i) \mid \phi(\mathbf x_j) \rangle\big|^2 = \big|\langle \phi(\mathbf x_j) \mid \phi(\mathbf x_i) \rangle\big|^2 \\
            & = \kappa^Q(\mathbf x_i, \mathbf x_j),
        \end{align}
        \end{subequations}
        implying that $\kappa^Q$ is the Gram kernel induced by the lifted feature map $\Psi$ in the Hilbert--Schmidt space.
        Let $\mathbf c = (c_1,\dots,c_n)^\top \in \mathbb{C}^n$ be any complex vector. Then we have:
        \begin{subequations}
        \begin{align}
        \label{eq:complex_psd}
            \mathbf c^\dagger \mathbf K^Q \mathbf c 
            & = \sum_{i=1}^n\sum_{j=1}^n \overline{c_i} \, c_j \, \kappa^Q(\mathbf x_i,\mathbf x_j) = \sum_{i=1}^n\sum_{j=1}^n \overline{c_i}\,c_j\, \langle \Psi(\mathbf x_i),\Psi(\mathbf x_j)\rangle_{\mathrm{HS}} \\ 
            & = \bigg\langle \sum_{i=1}^n c_i \Psi(\mathbf x_i), \sum_{j=1}^n c_j \Psi(\mathbf x_j) \bigg\rangle_{\mathrm{HS}} \\ 
            & = \Big\| \sum_{i=1}^n c_i \Psi(\mathbf x_i) \Big\|_{\mathrm{HS}}^2 \ge 0.
        \end{align}
        \end{subequations}
        Therefore, $\mathbf K^Q$ is \emph{Hermitian positive semidefinite}. 
    \end{proof}

    \emph{Dense finite-shot kernel acquisition.} 
    If one explicitly constructs the complete quantum kernel, symmetry reduces the number of distinct off-diagonal pairs from $n(n-1)$ to $n(n-1)/2$. Because the fidelity diagonal is analytically equal to one, the number of \emph{fresh quantum circuit queries} required for dense finite-shot construction is therefore $M_{\mathrm{full}} = \frac{n(n-1)}{2}$ rather than $n(n+1)/2$. With $N_{\mathrm{s}}$ shots per off-diagonal query, dense acquisition uses $\frac{n(n-1)}{2} N_{\mathrm{s}} = O(n^2N_{\mathrm{s}})$ quantum circuit executions. For comparison, if one counts all distinct matrix entries including the analytically known diagonal, the upper-triangular matrix contains $n(n+1)/2$ entries. We distinguish this matrix-entry count from the actual number of quantum circuit queries.

    \emph{Dense finite-shot kernel construction.} 
    To distinguish the ideal Gram matrix from its finite-shot realization, let $\widehat{\mathbf K}^{Q,(N_{\mathrm{s}})}$ denote the dense kernel matrix obtained from finite-shot measurements. Its off-diagonal entries are $\widehat K_{ij}^{Q,(N_{\mathrm{s}})} = \widehat\kappa^Q_{N_{\mathrm{s}}} (\mathbf x_i,\mathbf x_j), i\neq j$ while $\widehat K_{ii}^{Q,(N_{\mathrm{s}})} = K_{ii}^Q = 1$ is set analytically under the finite-shot-only model.

    \emph{Kernel Construction Algorithm.}
    Algorithm~\ref{alg:qkm_overlap} summarizes the construction of the quantum kernel matrix via overlap estimation. For each pair of training points, the circuit $U_{\phi}^{\dagger}(\mathbf x_j) U_{\phi}(\mathbf x_i)$ is executed, and the kernel entry is estimated from the empirical probability of the all-zero measurement outcome. By symmetry, only the upper triangular part of the matrix is computed.

    \begin{algorithm}[h]
    \caption{Dense Quantum Kernel Construction under Finite-Shot Overlap Estimation}
    \label{alg:qkm_overlap}
    \begin{algorithmic}[1]
    \REQUIRE Training set $\{\mathbf x_i\}_{i=1}^n$, feature map $U_{\phi}(\cdot)$, shot count $N_{\mathrm{s}}$
    \ENSURE Finite-shot kernel estimate $\widehat{\mathbf K}^{Q,(N_{\mathrm{s}})} \in\mathbb R^{n\times n}$ 
    \STATE Initialize $\widehat{\mathbf K}^{Q,(N_{\mathrm{s}})} \gets\mathbf I_n$ 
    \FOR{$i=1$ to $n$} 
        \FOR{$j=i+1$ to $n$}
            \STATE Prepare $|0\rangle^{\otimes q}$ 
            \STATE Apply $U_\phi(\mathbf x_i)$ followed by $U_\phi^\dagger(\mathbf x_j)$ 
            \STATE Execute the overlap circuit for $N_{\mathrm{s}}$ shots 
            \STATE Estimate $\widehat\kappa^Q_{N_{\mathrm{s}}} (\mathbf x_i,\mathbf x_j)$ as the empirical frequency of $0^{\otimes q}$ 
            \STATE Set $\widehat K_{ij}^{Q,(N_{\mathrm{s}})} \gets \widehat\kappa^Q_{N_{\mathrm{s}}} (\mathbf x_i,\mathbf x_j)$ and $\widehat K_{ji}^{Q,(N_{\mathrm{s}})} \gets \widehat K_{ij}^{Q,(N_{\mathrm{s}})}$ 
        \ENDFOR 
    \ENDFOR 
    \STATE Return $\widehat{\mathbf K}^{Q,(N_{\mathrm{s}})}$ 
    \end{algorithmic} 
    \end{algorithm}

\paragraph{Hardware and noise model.} 
    The kernel-evaluation procedure depends on the execution backend. Statevector simulation permits direct evaluation of the ideal probability in (\ref{eq:overlap_probability}) without sampling error. Shot-based simulators and quantum hardware instead return empirical measurement frequencies. The theoretical analysis in this work isolates \emph{finite-shot sampling noise}: conditional on a queried pair, the measurement outcomes are modeled as independent Bernoulli trials with mean $\kappa^Q(\mathbf x_i,\mathbf x_j)$. On physical hardware, additional coherent and incoherent effects, including gate errors, decoherence, calibration drift, and readout errors, may bias the observed probability away from the ideal kernel value. Such device-induced bias is distinct from finite-shot sampling variability and is not covered by Proposition~\ref{prop:overlap_unbiased}. Extending ICAQKL to biased or correlated hardware-noise models is therefore an important direction for future work.

\section{THEORETICAL ANALYSIS}
\label{sec:theorem}
    This section provides a detail development for the theoretical foundation of ICAQKL under finite-shot quantum-kernel access. The main difficulty is that finite-shot noise enters \emph{inside} an adaptive matrix-factorization procedure. Consequently, measurement error affects not only the numerical value of an individual queried kernel entry, but potentially also the pivot scores, the subsequent pivot sequence, the recursively constructed Cholesky factor, and ultimately the QSVM solution.

    A direct transfer of entrywise kernel-estimation error to the final low-rank surrogate is therefore insufficient. Instead, our analysis tracks how finite-shot estimation error propagates through the adaptive procedure: \emph{measurement error first perturbs the residual scores used for pivot selection, which in turn affects the recursively constructed Cholesky factor and the resulting low-rank kernel surrogate. We then propagate this kernel perturbation through the QSVM optimization problem and, finally, to the out-of-sample decision function and classification output.}

    To make this progression rigorous, we first separate the exact-entry Cholesky recursion from its finite-shot counterpart. We then establish high-probability concentration bounds for adaptively queried quantum-kernel entries and quantify how these estimation errors perturb pivot selection and propagate through the recursive factorization. Building on this analysis, we derive a path-conditioned bound on the resulting kernel approximation error and propagate that perturbation through the QSVM objective and optimizer. We finally extend the analysis to out-of-sample Cholesky evaluation, yielding explicit sufficient conditions for decision-function convergence and classification stability.

\subsection{Exact and Finite-Shot Low-Rank Approximation} \label{subsec:lowrank} 
    A central distinction in the analysis is between the \emph{exact-entry reference recursion} and the \emph{finite-shot ICAQKL recursion}. The former is the standard incomplete-Cholesky construction applied to the ideal quantum Gram matrix and therefore inherits the classical positive-semidefinite residual structure. The latter is driven by stochastic quantum-kernel estimates and is analyzed as a perturbation of an exact Cholesky construction. Maintaining this distinction is essential because finite-shot estimation preserves positive semidefiniteness of the constructed surrogate, but not necessarily of its residual with respect to the ideal kernel. 
    
\paragraph{Exact-entry reference recursion.} 
    Let $\mathbf K^Q \succeq \mathbf 0$ denote the ideal quantum Gram matrix. For a pivot sequence $\mathcal P_t = (p_1,\ldots,p_t)$, let 
    $\mathbf G_t = [ \mathbf g_1, \ldots, \mathbf g_t ] \in \mathbb R^{n\times t}$ denote the incomplete-Cholesky factor obtained when the required entries of $\mathbf K^Q$ are accessed exactly. 
    
    The associated rank-$t$ surrogate and residual matrix are defined as \begin{align} 
    \label{eq:exact_surrogate_exact_residual}
        \widetilde{\mathbf K}_t^Q := \mathbf G_t \mathbf G_t^\top, \quad 
        \mathbf R_t := \mathbf K^Q - \widetilde{\mathbf K}_t^Q,
    \end{align} 
    with residual diagonal 
    \begin{align} 
        d_{i,t} := (\mathbf R_t)_{ii}, \quad i=1,\ldots,n. \label{eq:exact_residual_diag_def} 
    \end{align} 
    At iteration $t$, exact greedy pivoting selects an index 
    \begin{align} 
        p_t \in \arg\max_{1\le i\le n} d_{i,t-1}. 
    \label{eq:exact_pivot} 
    \end{align} 
    Whenever $d_{p_t,t-1}>0$, the corresponding Cholesky column is given by 
    \begin{align} 
        G_{it} = \frac{ K_{ip_t}^Q - \sum_{\ell=1}^{t-1} G_{i\ell} G_{p_t\ell} }{ \sqrt{d_{p_t,t-1}} }, \quad i=1,\ldots,n. 
    \label{eq:exact_column_update} 
    \end{align} 
    Thus, the exact recursion is entirely determined by the residual diagonals and exact kernel entries along the selected pivot columns. 
    
\paragraph{Finite-shot ICAQKL recursion.} 
    ICAQKL does not assume direct access to the ideal entries $K_{ij}^Q$. Instead, whenever a kernel entry is required, the overlap circuit returns the finite-shot estimate $\widehat{\kappa}_{N_{\mathrm{s}}}^Q ( \mathbf x_i, \mathbf x_j )$. Let $\widehat{\mathbf G}_t^{(N_{\mathrm{s}})} = [ \widehat{\mathbf g}_1^{(N_{\mathrm{s}})}, \ldots, \widehat{\mathbf g}_t^{(N_{\mathrm{s}})} ] \in \mathbb R^{n\times t}$ denote the factor constructed from these finite-shot kernel estimates. The resulting low-rank surrogate is 
    \begin{align} 
        \widetilde{\mathbf K}_t^{Q,(N_{\mathrm{s}})} := \widehat{\mathbf G}_t^{(N_{\mathrm{s}})} \big( \widehat{\mathbf G}_t^{(N_{\mathrm{s}})} \big)^\top. 
    \label{eq:finite_surrogate} 
    \end{align} 
    For the normalized fidelity kernel considered in this work, $K_{ii}^Q = 1$. Hence, the initial diagonal information is known analytically and does not require quantum measurements. We therefore initialize $\widehat s_{i,0} = 1, i=1,\ldots,n$.
    After constructing the $t$-th finite-shot Cholesky column, ICAQKL updates 
    \begin{align} 
        \widehat s_{i,t} = \left[ \widehat s_{i,t-1} - \big( \widehat G_{it}^{(N_{\mathrm{s}})} \big)^2 \right]_+, \quad [a]_+ := \max\{a,0\}. 
    \label{eq:score_recursion} 
    \end{align} 
    The clipping operation prevents small negative values caused by finite-shot perturbations from being interpreted as admissible pivot scores. We deliberately refer to $\widehat s_{i,t}$ as a \emph{finite-shot pivot score} rather than as an exact residual diagonal. In the exact recursion, $d_{i,t} = \left( \mathbf K^Q - \mathbf G_t \mathbf G_t^\top \right)_{ii}$ is the diagonal of a PSD residual matrix. By contrast, under finite-shot access, $\mathbf K^Q - \widetilde{\mathbf K}_t^{Q,(N_{\mathrm{s}})}$ is not generally guaranteed to be PSD. Consequently, $\widehat s_{i,t}$ should be interpreted as the recursively maintained noisy analogue of the exact residual diagonal rather than identified with it.
    
    \begin{proposition}[PSD finite-shot surrogate] 
    \label{prop:ICA_psd} 
        Let $\widehat{\mathbf G}_t^{(N_{\mathrm{s}})} \in \mathbb R^{n\times t}$ denote the factor produced after $t$ iterations of the finite-shot incomplete Cholesky procedure. Then the corresponding surrogate 
        \begin{align} 
            \widetilde{\mathbf K}_t^{Q,(N_{\mathrm{s}})} := \widehat{\mathbf G}_t^{(N_{\mathrm{s}})} \big( \widehat{\mathbf G}_t^{(N_{\mathrm{s}})} \big)^\top 
        \end{align} 
        is symmetric positive semidefinite and satisfies 
        \begin{align} 
            \operatorname{rank} \big( \widetilde{\mathbf K}_t^{Q,(N_{\mathrm{s}})} \big) \le t. 
        \end{align} 
    \end{proposition}
    \begin{proof} 
        Let $\mathbf A := \widehat{\mathbf G}_t^{(N_{\mathrm{s}})}$. For any $\mathbf v\in\mathbb R^n$, it follows
        \begin{align} 
            \mathbf v^\top \mathbf A \mathbf A^\top \mathbf v = \big\| \mathbf A^\top \mathbf v \big\|_2^2 \ge 0. 
        \end{align} 
        Hence $\mathbf A \mathbf A^\top$ is symmetric PSD. By construction $\widetilde{\mathbf K}_t^{Q,(N_{\mathrm{s}})} = \mathbf A \mathbf A^\top \succeq 0$. Moreover, since $\widehat{\mathbf G}_t^{(N_{\mathrm{s}})}$ has $t$ columns, its product $\widehat{\mathbf G}_t^{(N_{\mathrm{s}})} \big( \widehat{\mathbf G}_t^{(N_{\mathrm{s}})} \big)^\top$ has rank at most $t$,
        \begin{align} 
            \operatorname{rank} \left( \widehat{\mathbf G}_t^{(N_{\mathrm{s}})} \big( \widehat{\mathbf G}_t^{(N_{\mathrm{s}})} \big)^\top \right) \le \operatorname{rank} ( \widehat{\mathbf G}_t^{(N_{\mathrm{s}})} ) \le t,
        \end{align} 
        which completes the proof.
        \end{proof} 
        
        Proposition~\ref{prop:ICA_psd} guarantees that ICAQKL always supplies the downstream QSVM with a valid PSD low-rank kernel surrogate. This property follows directly from the factorized representation and does not require the finite-shot residual to remain PSD. The distinction is important for the subsequent analysis: the classical residual identities, monotonicity properties, and full-rank exactness results are established using the exact-entry recursion, whereas the finite-shot construction is compared with this reference process through explicit perturbation bounds. This separation allows us to retain the deterministic geometry of incomplete Cholesky while accounting rigorously for stochastic quantum-kernel estimation.

\subsection{Exact Cholesky Residual Invariants} 
\label{subsec:exact_invariants}
    The exact-entry recursion provides the deterministic reference geometry against which the finite-shot ICAQKL recursion will be compared. In particular, when the Cholesky columns are constructed from exact entries of the ideal PSD kernel matrix, the residual remains PSD after every successful pivot. This property yields nonnegative and monotonically decreasing residual diagonals and, ultimately, exact recovery once the number of successful pivots reaches the rank of the kernel matrix. These invariants will later be used to define the path-conditioned exact comparator and to control both finite-shot pivot perturbations and out-of-sample approximation. 
    
    \begin{proposition}[Exact residual PSD] 
    \label{prop:exact_residual_psd} 
        Suppose that, at iteration $t-1$ of the exact-entry Cholesky recursion, $\mathbf R_{t-1}\succeq \mathbf 0$ and that the selected pivot satisfies $d_{p_t,t-1}>0$. Then the updated residual $\mathbf R_t = \mathbf R_{t-1} - \mathbf g_t\mathbf g_t^\top$ is also PSD, i.e., $\mathbf R_t\succeq\mathbf 0$.
    \end{proposition} 
    \begin{proof} 
        After permuting the pivot $p_t$ to the first coordinate, write the residual as 
        \begin{align} 
            \mathbf R_{t-1} = 
            \begin{bmatrix} 
                d_{p_t,t-1} & \mathbf r^\top \\
                \mathbf r & \mathbf B 
            \end{bmatrix}. 
        \end{align} 
        Since the pivot is strictly positive, the corresponding exact Cholesky column can be written, under the same permutation, as $\mathbf g_t = \begin{bmatrix} \sqrt{d_{p_t,t-1}} \\ \mathbf r/ \sqrt{d_{p_t,t-1}} \end{bmatrix}$.
        Therefore, we can obtain
        \begin{align} 
            \mathbf R_t = \mathbf R_{t-1} - \mathbf g_t\mathbf g_t^\top = \begin{bmatrix} 
                0 & \mathbf 0^\top \\ 
                \mathbf 0 & \mathbf B - \dfrac{ \mathbf r\mathbf r^\top }{ d_{p_t,t-1} } 
            \end{bmatrix}. 
        \end{align} 
        Since $\mathbf R_{t-1}\succeq\mathbf 0$ and $d_{p_t,t-1}>0$, the lower-right block $\mathbf B - \frac{ \mathbf r\mathbf r^\top }{ d_{p_t,t-1} }$ is the Schur complement of the pivot and is PSD. Consequently,  $\mathbf R_t \succeq \mathbf 0$. 
    \end{proof} 
    
    \begin{proposition}[Exact residual identity and monotonicity] 
    \label{prop:residual_identity} 
        For every executed iteration $t\ge 1$ of the exact-entry recursion and every $i\in\{1,\ldots,n\}$, 
        \begin{align} 
            d_{i,t} = (\mathbf R_t)_{ii} = K_{ii}^Q - \sum_{\ell=1}^{t} G_{i\ell}^2. 
        \label{eq:residual_identity} 
        \end{align} 
        Moreover, $0 \le d_{i,t} \le d_{i,t-1}$.
    \end{proposition} 
    \begin{proof} 
        From the definition $\mathbf R_t = \mathbf K^Q - \mathbf G_t\mathbf G_t^\top$ the $i$th diagonal entry satisfies 
        \begin{align} 
            (\mathbf R_t)_{ii} = K_{ii}^Q - \left( \mathbf G_t\mathbf G_t^\top \right)_{ii} = K_{ii}^Q - \sum_{\ell=1}^{t} G_{i\ell}^2. 
        \end{align} 
        Therefore, $d_{i,t} = d_{i,t-1} - G_{it}^2$. Since $G_{it}^2\ge0$, it follows immediately that $d_{i,t} \le d_{i,t-1}$. Furthermore, Proposition~\ref{prop:exact_residual_psd} gives $\mathbf R_t\succeq\mathbf0$, and every diagonal entry of a PSD matrix is nonnegative. Thus, $d_{i,t} \ge 0$. Combining the two inequalities yields $0 \le d_{i,t} \le d_{i,t-1}$,  which proves the claim. 
    \end{proof} 
    For the normalized fidelity kernel, $d_{i,0} = K_{ii}^Q = 1$ so all exact residual diagonals are initialized at one and decrease monotonically as successive Cholesky directions are incorporated.

    \begin{proposition}[Exactness at full rank] 
    \label{prop:exactness_ICA} 
        Let $\rho := \operatorname{rank} \big( \mathbf K^Q \big)$. If the exact-entry Cholesky recursion performs $\rho$ iterations with strictly positive pivots, then $\mathbf R_\rho = \mathbf 0$ and consequently, $\mathbf K^Q = \mathbf G_\rho \mathbf G_\rho^\top$. 
    \end{proposition} 
    \begin{proof} 
        Consider one successful iteration with $d_{p_t,t-1}>0$. From the Schur-complement representation in the proof of Proposition~\ref{prop:exact_residual_psd}, after permuting the pivot index to the first coordinate, the residual takes the form $\mathbf R_t = \begin{bmatrix} 0 & \mathbf 0^\top \\ \mathbf 0 & \mathbf S_t \end{bmatrix}$
        where $\mathbf S_t = \mathbf B - \frac{ \mathbf r\mathbf r^\top }{ d_{p_t,t-1} } \succeq \mathbf 0$. Since the leading pivot $d_{p_t,t-1}$ of $\mathbf R_{t-1}$ is strictly positive and therefore invertible, the block-rank identity yields 
        \begin{align} 
            \operatorname{rank} \left( \mathbf R_{t-1} \right) = 1 + \operatorname{rank} \left( \mathbf S_t \right). 
        \end{align}
        Moreover, $\operatorname{rank} \! \left( \mathbf R_t \right) \!=\! \operatorname{rank} \! \left( \mathbf S_t \right)$. Thus every successful pivot reduces residual rank by exactly one
        \begin{align} 
            \operatorname{rank} \left( \mathbf R_t \right) = \operatorname{rank} \left( \mathbf R_{t-1} \right) - 1. 
        \end{align} 
        Initially, $\mathbf R_0 = \mathbf K^Q$ so $\operatorname{rank} \left( \mathbf R_0 \right) = \operatorname{rank} \left( \mathbf K^Q \right) = \rho$. Therefore, after $\rho$ successful pivots, 
        \begin{align} 
            \operatorname{rank} \left( \mathbf R_\rho \right) = 0. 
        \end{align} 
        A matrix has rank zero if and only if it is the zero matrix, so $\mathbf R_\rho = \mathbf 0$. Finally, using 
        \begin{align} 
            \mathbf R_\rho = \mathbf K^Q - \mathbf G_\rho \mathbf G_\rho^\top, 
        \end{align} 
        we obtain $\mathbf K^Q = \mathbf G_\rho \mathbf G_\rho^\top$ which proves the proposition. 
    \end{proof}

    The exact residual diagonal $d_{i,t}$ quantifies the diagonal approximation error remaining at sample $i$ after $t$ Cholesky directions have been incorporated. Exact greedy pivoting therefore selects the sample with the largest remaining residual diagonal. Proposition~\ref{prop:residual_identity} shows that these residuals decrease monotonically along the exact-entry recursion, while Proposition~\ref{prop:exactness_ICA} establishes that they vanish once the factorization reaches the rank of the kernel matrix. 
    
    ICAQKL extends this residual-driven construction to the finite-shot setting by replacing exact quantum-kernel entries with stochastic overlap estimates. The maintained score $\widehat s_{i,t}$ therefore serves as a noisy surrogate for the exact comparator quantity $d_{i,t}$. The central analytical question is then not whether the exact residual remains PSD, which follows from Proposition~\ref{prop:exact_residual_psd}, but how finite-shot estimation errors perturb the pivot scores, alter the resulting pivot decisions, and propagate through the recursively constructed Cholesky factor. We address this question next by first establishing high-probability concentration bounds for the adaptively queried quantum-kernel entries.

\subsection{Finite-Shot Perturbation Analysis}
\label{subsec:finite_shot_perturbation} 
    We now quantify how finite-shot quantum-kernel estimation affects the adaptive Cholesky construction. Two features distinguish this setting from a standard perturbation analysis. First, the queried kernel entries are selected adaptively because each pivot depends on estimates obtained at earlier iterations. Second, the resulting entrywise errors can perturb the residual scores used for pivot selection and thereby alter the subsequent query trajectory. We therefore proceed in stages: we first establish uniform concentration over the adaptively queried entries and then translate score perturbations into guarantees on the quality and stability of the selected pivots. 
    
\paragraph{Adaptive finite-shot concentration.} 
    Let $q_m=(i_m,j_m)$ denote the $m$th fresh queried training pair, and let $\mathcal F_{m-1}$ denote the sigma-algebra generated by all queried pairs, pivot decisions, and measurement outcomes observed before the $m$th query. Since the next queried pair is selected using only previously available information, $q_m$ is $\mathcal F_{m-1}$-measurable. Conditional on this history, the $N_{\mathrm{s}}$ shots used to estimate the corresponding kernel entry are fresh independent Bernoulli measurements with success probability $\kappa^Q ( \mathbf x_{i_m}, \mathbf x_{j_m} )$.
    
    \begin{theorem}[Uniform concentration over adaptively queried entries]
    \label{thm:uniform_concentration} 
        Suppose that ICAQKL performs at most $M_r$ fresh training-kernel queries through rank $r$. For any $\delta \!\in\! (0,1)$, define $\eta_s \!:=\! \sqrt{ \frac{ \log(2M_r/\delta) }{ 2N_{\mathrm{s}} } }$. Then, with probability at least $1-\delta$,
        \begin{align} 
        \label{eq:uniform_concentration_bound}
            \max_{1 \le m \le M_r} \big| \widehat \kappa_{\!N_{\mathrm{s}}}^Q \!(\mathbf x_{i_m}, \mathbf x_{j_m}) - \kappa^Q \!(\mathbf x_{i_m}, \mathbf x_{j_m}) \big| \le \eta_s, 
        \end{align}
    \end{theorem} 
    \begin{proof} 
        For each query index $m$, define the deviation event, 
        \begin{align} 
            \mathcal E_m(\eta) := \Big\{ \big| \widehat \kappa_{\!N_{\mathrm{s}}}^Q \!(\mathbf x_{i_m}, \mathbf x_{j_m}) - \kappa^Q \!(\mathbf x_{i_m}, \mathbf x_{j_m}) \big| > \eta \Big\}. 
        \label{eq:adaptive_query_event} 
        \end{align}
        Conditional on $\mathcal F_{\!m-1}$, the queried pair $q_m$ is fixed. The corresponding estimator $\widehat \kappa_{\!N_{\mathrm{s}}}^Q \!(\mathbf x_{i_m}, \mathbf x_{j_m})$ is the empirical mean of $N_{\mathrm{s}}$ conditionally independent Bernoulli random variables with mean $\kappa^Q \!(\mathbf x_{i_m}, \mathbf x_{j_m})$. Therefore, Hoeffding's inequality yields, for every $\eta > 0$, 
        \begin{align} 
            \Pr \left( \mathcal E_m(\eta) \mid \mathcal F_{m-1} \right) \le 2 \exp \left( -2N_{\mathrm{s}}\eta^2 \right). 
        \label{eq:conditional_hoeffding} 
        \end{align}
        Taking expectations with respect to the history and using the tower property, 
        \begin{subequations}
        \label{eq:unconditional_adaptive_query_bound} 
        \begin{align}
            \Pr \left( \mathcal E_m(\eta) \right) 
            &= \mathbb E \big[ \Pr \left( \mathcal E_m(\eta) \mid \mathcal F_{m-1} \right) \big] \\
            &\le 2 \exp \left( -2N_{\mathrm{s}}\eta^2 \right).
        \end{align}
        \end{subequations} 
        Thus, although the identity of the $m$th queried pair depends on the previous finite-shot observations, its estimation error obeys the same Hoeffding bound as a fixed query.
        
        We next control all adaptive queries simultaneously. From (\ref{eq:adaptive_query_event}), the event that \emph{at least one} queried entry has estimation error at least $\eta$ can then be written as $\bigcup_{m}^{M_r} \mathcal{E}_{m}(\eta)$. Equivalently, 
        \begin{align} 
        \label{eq:max_event_union}
            \Big\{ \max_{1 \le m \le M_r} \!\big| \widehat{\kappa}^Q_{\!N_{\mathrm{s}}} \!(\mathbf x_{i_m}, \mathbf x_{j_m}) \!-\! \kappa^{\!Q}(\mathbf x_{i_m}, \mathbf x_{j_m}) \big| > \eta \Big\} = \bigcup_{m=1}^{M_r} \mathcal{E}_{m}(\eta).
        \end{align}
        Taking probabilities on both sides of (\ref{eq:max_event_union}) and applying the union bound, we obtain:
        \begin{subequations}
        \label{eq:adaptive_uniform_union_bound}
        \begin{align} 
            \Pr \! \bigg(\! \max_{1 \le m \le M_r} \!\Big| \widehat \kappa_{\!N_{\mathrm{s}}}^Q \!(\mathbf x_{i_m}, \mathbf x_{j_m}\!) \!-\! \kappa^{\!Q} \!(\mathbf x_{i_m}, \mathbf x_{j_m}\!) \Big| \!>\! \eta \!\bigg) 
            &= \Pr \bigg( \bigcup_{m=1}^{M_r} \mathcal E_m(\eta) \bigg) \\ 
            &\le \sum_{m=1}^{M_r} \Pr \left( \mathcal E_m(\eta) \right) \\ 
            &\le 2M_r \exp \left( -2N_{\mathrm{s}}\eta^2 \right).  
        \end{align}
        \end{subequations}
        To make the failure probability at most $\delta$, it is sufficient that 
        \begin{align} 
            2M_r \exp \left( -2N_{\mathrm{s}}\eta^2 \right) \le \delta. 
        \end{align} 
        Equivalently, $\eta \ge \sqrt{ \frac{ \log(2M_r/\delta) }{ 2N_{\mathrm{s}} } }$.
        Choosing $\eta=\eta_s$ therefore yields 
        \begin{align} 
            & \Pr \Big( \max_{1 \le m \le M_r} \!\Big| \widehat \kappa_{\!N_{\mathrm{s}}}^Q \!(\mathbf x_{i_m}, \mathbf x_{j_m}\!) \!-\! \kappa^{\!Q} \!(\mathbf x_{i_m}, \mathbf x_{j_m}\!) \Big| \le \eta_s \Big) \ge 1-\delta, 
        \end{align} 
        which proves the theorem.
    \end{proof}
    
    Theorem~\ref{thm:uniform_concentration} provides the first connection between the finite quantum measurement budget and the adaptive factorization. In particular, every queried training entry is simultaneously accurate, with high probability, to $\eta_s = O \Big( \sqrt{\frac{\log(M_r/\delta)}{N_{\mathrm{s}}}} \Big)$.
    This result is, however, only an entrywise oracle guarantee. It does not yet control the final low-rank surrogate because the Cholesky recursion is nonlinear: an estimation error in one queried entry can perturb the current factor, modify the pivot scores, and thereby influence later queries. 

    To quantify the aggregate finite-shot error over the entries directly queried by ICAQKL, define the sparse queried-entry estimation-error matrix 
    \begin{align}
        \left( \mathbf e_r^{Q,(N_{\mathrm{s}})} \right)_{ij} = \begin{cases} \widehat{\kappa}_{N_{\mathrm{s}}}^Q (\mathbf x_i,\mathbf x_j) - \kappa^Q(\mathbf x_i,\mathbf x_j), & (i,j)\in\mathcal Q_r, \\ 0, & \text{otherwise}, 
    \end{cases} 
    \label{eq:queried_error_matrix} 
    \end{align} 
    where $\mathcal Q_r$ denotes the set of matrix coordinates corresponding to the fresh kernel entries queried through rank $r$, with $|\mathcal Q_r| \le M_r$. Here, each fresh off-diagonal quantum query is counted once. Thus, $\mathbf e_r^{Q,(N_{\mathrm{s}})}$ is an auxiliary sparse error matrix used to summarize the directly observed oracle errors; it need not be interpreted as the symmetric perturbation of a complete Gram matrix. 
    
    \begin{lemma}[Norm bound for queried-entry estimation error] 
    \label{lem:operator_bound} 
        Under the event in Theorem~\ref{thm:uniform_concentration}, the queried-entry estimation error satisfies 
        \begin{align} 
            \big\| \mathbf e_r^{Q,(N_{\mathrm{s}})} \big\|_2 \le \big\| \mathbf e_r^{Q,(N_{\mathrm{s}})} \big\|_F \le \sqrt{M_r}\,\eta_s = \sqrt{ \frac{ M_r\log(2M_r/\delta) }{ 2N_{\mathrm{s}} } }. 
        \label{eq:operator_bound} 
        \end{align} 
        Consequently, the bound holds with probability at least $1-\delta$. 
    \end{lemma} 
    \begin{proof} 
        On the concentration event (\ref{eq:uniform_concentration_bound}), every nonzero queried entry of $\mathbf e_r^{Q,(N_{\mathrm{s}})}$ satisfies 
        \begin{align}
        \label{eq:error_non0}
            \left| \widehat{\kappa}_{N_{\mathrm{s}}}^Q (\mathbf x_i,\mathbf x_j) - \kappa^Q (\mathbf x_i,\mathbf x_j) \right| \le \eta_s. 
        \end{align} 
        The Frobenius norm of a matrix is defined by:
        \begin{align}
        \label{eq:frob_def}
            & \big\| \mathbf e_r^{Q,(N_{\mathrm{s}})} \big\|_F^2 = \sum_{i=1}^n \sum_{j=1}^n \big( \mathbf e_r^{Q, (N_{\mathrm{s}})} \big)_{ij}^2.
        \end{align}
        Since all unqueried entries are exactly zero, only the queried entries contribute to this sum. Moreover, under the concentration event, from (\ref{eq:error_non0}), each nonzero term is bounded above by $\big(\mathbf e_r^{\!Q,(N_{\mathrm{s}})}\big)_{ij}^2 \!\le\! \eta_s^2$.
        Additionally, $\mathbf e_r^{\!Q,(N_{\mathrm{s}})}$ has support contained in $\mathcal Q_r$. Thus, the matrix has at most $M_r$ nonzero entries, $|\mathcal Q_r|\le M_r$, it follows that:
        \begin{align}
        \label{eq:frob_squared_bound}
            \big\| \mathbf e_r^{Q,(N_{\mathrm{s}})} \big\|_F^2 = \sum_{i=1}^{n} \sum_{j=1}^{n} \big( \mathbf e_r^{Q,(N_{\mathrm{s}})} \big)_{ij}^2 \le M_r\eta_s^2. 
        \end{align}
        For any real matrix $A$, the spectral norm is bounded by the Frobenius norm $\|A\|_2 \le \|A\|_F$. Thus, taking the square roots of (\ref{eq:frob_squared_bound}), we can obtain:
        \begin{align} 
            \big\| \mathbf e_r^{Q,(N_{\mathrm{s}})} \big\|_2 \le \big\| \mathbf e_r^{Q,(N_{\mathrm{s}})} \big\|_F \le \sqrt{M_r} \, \eta_s. 
        \end{align} 
        This completes the proof. 
    \end{proof} 
    Lemma \ref{lem:operator_bound} bounds the aggregate finite-shot error over the kernel entries directly returned by the quantum oracle. Importantly, it is an auxiliary measurement-error bound rather than a bound on the final low-rank kernel surrogate. The queried estimates enter recursively into both the Cholesky factor and the pivot scores, so their effect on $\widetilde{\mathbf K}_r^{Q,(N_{\mathrm{s}})}$ must be propagated through the adaptive factorization itself. The subsequent analysis performs this propagation by first controlling pivot-score perturbations and then the resulting Cholesky-factor error.

\paragraph{Finite-shot pivot selection.} 
    Let $\widehat{\mathcal P}_{t-1} = ( \widehat p_1, \ldots, \widehat p_{t-1} )$ denote the pivot history selected by the finite-shot ICAQKL recursion, and let $\mathcal I_{t-1} := \{1,\ldots,n\} \setminus \{ \widehat p_1,\ldots,\widehat p_{t-1} \}$ denote the set of indices still eligible for selection at iteration $t$. To separate the effect of the selected pivot path from the numerical error induced by finite-shot kernel estimation, consider an exact-entry Cholesky comparator that follows the same previously selected pivot history $\widehat{\mathcal P}_{t-1}$. Let $d_{i,t-1}^{\widehat{\mathcal P}_{t-1}}$ denote the corresponding exact residual diagonal for $i\in\mathcal I_{t-1}$. We define the pivot-score discrepancy by 
    \begin{align} 
        \zeta_{t-1} := \big\| \widehat{\mathbf s}_{t-1} - \mathbf d_{t-1}^{\widehat{\mathcal P}_{t-1}} \big\|_{\infty,\mathcal I_{t-1}}, 
    \label{eq:score_error} 
    \end{align} 
    where $|\mathbf v|_{\infty,\mathcal I_{t-1}} := \max_{i\in\mathcal I_{t-1}} |v_i|$. Thus, $\zeta_{t-1}$ measures the largest deviation between the finite-shot pivot scores and the exact residual diagonals associated with the same pivot history. 
    
    \begin{lemma}[Approximate-greedy finite-shot pivot] 
    \label{lem:approx_greedy}
        Let the finite-shot pivot be selected according to $\widehat p_t \in \arg\max_{i\in\mathcal I_{t-1}} \widehat s_{i,t-1}$. Then 
        \begin{align} 
            d_{\widehat p_t,t-1}^{\widehat{\mathcal P}_{t-1}} \ge \max_{i\in\mathcal I_{t-1}} d_{i,t-1}^{\widehat{\mathcal P}_{t-1}} - 2\zeta_{t-1}. 
        \label{eq:approx_greedy_bound} 
        \end{align} 
        \end{lemma} 
        \begin{proof} 
            Let $p_t^\star \in \arg\max_{i\in\mathcal I_{t-1}} d_{i,t-1}^{\widehat{\mathcal P}_{t-1}}$ be an exact greedy pivot conditional on the same previous pivot history. By the definition of $\zeta_{t-1}$, 
            \begin{align} 
                d_{\widehat p_t,t-1}^{\widehat{\mathcal P}_{t-1}} \ge \widehat s_{\widehat p_t,t-1} - \zeta_{t-1}. 
            \end{align} 
            Since $\widehat p_t$ maximizes the finite-shot score over $\mathcal I_{t-1}$, 
            \begin{align} 
                \widehat s_{\widehat p_t,t-1} \ge \widehat s_{p_t^\star,t-1}. 
            \end{align} 
            Applying the score-discrepancy bound once more yields 
            \begin{align} 
                \widehat s_{p_t^\star,t-1} \ge d_{p_t^\star,t-1}^{\widehat{\mathcal P}_{t-1}} - \zeta_{t-1}. 
            \end{align} 
            Combining the three inequalities yields 
            \begin{align} 
                d_{\widehat p_t,t-1}^{\widehat{\mathcal P}_{t-1}} \ge d_{p_t^\star,t-1}^{\widehat{\mathcal P}_{t-1}} - 2\zeta_{t-1} = \max_{i\in\mathcal I_{t-1}} d_{i,t-1}^{\widehat{\mathcal P}_{t-1}} - 2\zeta_{t-1}, 
            \end{align} 
            which completes the proof. 
        \end{proof} 
        Lemma~\ref{lem:approx_greedy} does not require finite-shot ICAQKL to reproduce the exact greedy pivot. Instead, it establishes that the selected pivot is approximately greedy with respect to the exact residual geometry associated with the same previously selected path. In particular, the exact residual value of the finite-shot-selected pivot is at most $2\zeta_{t-1}$ below the best admissible exact residual. This approximate-greedy characterization is preferable to requiring exact pivot agreement when several candidate samples have nearly identical residual diagonals. In such cases, finite-shot noise may alter the ordering of the candidates without materially degrading the quality of the selected pivot. A stronger result is available when the best exact pivot is separated from all competing candidates by a sufficiently large residual gap.

    \begin{corollary}[Pivot-gap stability] 
    \label{cor:pivot_gap} 
        Suppose the exact same-history greedy pivot $p_t^\star$ is unique, and define its residual gap over the remaining admissible indices as $\gamma_t \!:=\! d_{p_t^\star,t-1}^{\widehat{\mathcal P}_{t-1}} - \max_{j \in \mathcal I_{t-1} \!\setminus \{p_t^\star\}} d_{j,t-1}^{\widehat{\mathcal P}_{t-1}}$.
        If $\gamma_t \!>\! 2\zeta_{t-1}$ then the finite-shot recursion selects the same pivot as the exact same-history greedy rule $\widehat p_t \!=\! p_t^\star$.
    \end{corollary} 
    \begin{proof} 
        Suppose, for contradiction, that $\widehat p_t \!\neq\! p_t^\star$. Since $\widehat p_t \!\in\! \mathcal I_{t-1} \!\setminus\! \{ p_t^\star \}$, pivot gap's definition implies 
        \begin{align} 
            d_{\widehat p_t,t-1}^{\widehat{\mathcal P}_{t-1}} \le d_{p_t^\star,t-1}^{\widehat{\mathcal P}_{t-1}} - \gamma_t. 
        \label{eq:pivot_gap_upper} 
        \end{align} 
        On the other hand, Lemma~\ref{lem:approx_greedy} yields 
        \begin{align} 
            d_{\widehat p_t,t-1}^{\widehat{\mathcal P}_{t-1}} \ge \max_{i\in\mathcal I_{t-1}} d_{i,t-1}^{\widehat{\mathcal P}_{t-1}} - 2\zeta_{t-1} = d_{p_t^\star,t-1}^{\widehat{\mathcal P}_{t-1}} - 2\zeta_{t-1}.
        \label{eq:pivot_gap_lower} 
        \end{align} 
        Combining (\ref{eq:pivot_gap_upper}) and (\ref{eq:pivot_gap_lower}) yields 
        \begin{align} 
            \gamma_t \le 2\zeta_{t-1}, 
        \end{align}
        which contradicts the stated condition $\gamma_t>2\zeta_{t-1}$. Therefore, $\widehat p_t = p_t^\star$.
    \end{proof}
    Corollary~\ref{cor:pivot_gap} identifies a sufficient condition for exact pivot recovery. When the separation between the largest and second-largest exact residual diagonals exceeds twice the finite-shot score error, measurement noise cannot change their ordering. Thus, pivot-path deviations can occur only when the exact residual gap becomes comparable to the uncertainty in the finite-shot scores.

\emph{From pivot stability to factor stability:}
    The preceding concentration and pivot-selection results characterize how finite-shot uncertainty can influence the adaptive query path, but they do not yet quantify the numerical error in the Cholesky factor itself. Even when the finite-shot recursion selects the same pivot as its exact counterpart, the corresponding Cholesky column is still constructed from noisy kernel estimates and therefore differs from the exact column. To isolate this effect, we compare the finite-shot factor with an exact-entry factor constrained to follow the same selected pivot sequence. This path-conditioned comparison separates the structural quality of the chosen pivot path from the recursive numerical perturbation induced by finite-shot kernel estimation, enabling a direct stability analysis of the Cholesky factor.

\paragraph{Path-conditioned factor perturbation.} 
    A direct comparison between the finite-shot factor and the ordinary exact greedy Cholesky factor would confound two distinct effects. First, finite-shot estimation may alter the pivot sequence itself. Second, even when the pivot sequence is fixed, the queried kernel values used to construct each Cholesky column remain noisy. To separate these effects, we introduce an exact-entry comparator that follows the same pivot sequence selected by the finite-shot recursion. Specifically, for the finite-shot pivot sequence through iteration $t$, $\widehat{\mathcal P}_t = ( \widehat p_1, \ldots, \widehat p_t )$, let $\mathbf G_t^{\widehat{\mathcal P}_t}$ be the exact-entry Cholesky factor obtained by following this same sequence while evaluating every required kernel entry exactly. This path-conditioned comparator isolates the numerical perturbation induced by finite-shot estimation from the structural approximation quality of the selected pivot path. 
    
    \begin{assumption}[Stable executed path] 
    \label{ass:stable_path} 
        For every executed iteration $t\le r$, there exist constants $\tau \!>\! 0, B \!<\! \infty, \kappa \!<\! \infty$ such that the exact and finite-shot pivot quantities satisfy $d_{\widehat p_t, t-1}^{\widehat{\mathcal P}_{t-1}} \!\ge\! \tau$, $\widehat s_{\widehat p_t,t-1}\! \ge\! \tau$ and the exact and finite-shot factor rows satisfy 
        \begin{align} 
            \max_i \big\| \mathbf G_{i,1:t}^{\widehat{\mathcal P}_t} \big\|_2 \le B, \quad 
            \max_i \left\| \widehat{\mathbf G}_{i,1:t}^{(N_{\mathrm{s}})} \right\|_2 \le B.
        \label{eq:row_factor_bound} 
        \end{align}
        Moreover, $K_{ii}^Q \le \kappa, i=1,\ldots,n$. For the normalized fidelity kernel considered in this work, $\kappa=1$. 
    \end{assumption} 
    The constant $\tau$ is an analysis condition on the executed pivot path rather than a modification of the ICAQKL algorithm. It prevents the reciprocal square-root normalization in the Cholesky update from becoming arbitrarily ill-conditioned. Define the rowwise factor perturbation 
    \begin{align}
        E_t := \max_i \big\| \widehat{\mathbf G}_{i,1:t}^{(N_{\mathrm{s}})} - \mathbf G_{i,1:t}^{\widehat{\mathcal P}_t} \big\|_2, 
    \label{eq:E_t} 
    \end{align} 
    the maximum error in the newly constructed column 
    \begin{align} 
        a_t := \max_i \big| \widehat G_{it}^{(N_{\mathrm{s}})} - G_{it}^{\widehat{\mathcal P}_t} \big|, 
    \label{eq:a_t} 
    \end{align} 
    and the pivot-score discrepancy 
    \begin{align} 
        \zeta_t := \big\| \widehat{\mathbf s}_t - \mathbf d_t^{\widehat{\mathcal P}_t} \big\|_{\infty, \mathcal I_t}, 
    \label{eq:zeta_t} 
    \end{align} 
    where $\mathcal I_t$ denotes the indices still admissible after iteration $t$. Because the fidelity-kernel diagonal is known exactly and both recursions are initialized from $K_{ii}^Q = \widehat s_{i,0} = 1$, we have $E_0 = \zeta_0 = 0$.
    
    \begin{theorem}[Recursive finite-shot factor perturbation] 
    \label{thm:factor_perturbation} 
        Under Assumption~\ref{ass:stable_path} and the concentration event \ref{eq:uniform_concentration_bound}, the finite-shot perturbations satisfy, for every $t\le r$, 
        \begin{align}
        \label{eq:a_recursion_E_recursion_zeta_recursion}
            a_t \le \frac{ \eta_s + 2B E_{t-1} }{ \sqrt{\tau} } + \frac{ \kappa\zeta_{t-1} }{ 2\tau^{3/2} }, \quad
            E_t \le \sqrt{ E_{t-1}^2 + a_t^2 }, \quad
            \zeta_t \le \zeta_{t-1} + 2B a_t. 
        \end{align} 
        Consequently, for fixed $r$, $\tau$, $B$, and $\kappa$, there exists a finite constant $C_{r,\tau,B,\kappa}>0$ such that \begin{align} E_r 
            \le C_{r,\tau,B,\kappa}\eta_s. 
        \label{eq:E_r_rate} 
        \end{align}
    \end{theorem} 
    \begin{proof} 
        Fix an iteration $t$ and the common pivot $\widehat p_t$. For each row $i$, define the exact and finite-shot Cholesky numerators by 
        \begin{align}
        \label{eq:exact_numerator_noisy_numerator}
            u_{i,t} = K_{i\widehat p_t}^Q - \sum_{\ell=1}^{t-1} G_{i\ell}^{\widehat{\mathcal P}_{t-1}} G_{\widehat p_t\ell}^{\widehat{\mathcal P}_{t-1}}, \quad
            \widehat u_{i,t} = \widehat K_{i\widehat p_t}^Q - \sum_{\ell=1}^{t-1} \widehat G_{i\ell}^{(N_{\mathrm{s}})} \widehat G_{\widehat p_t\ell}^{(N_{\mathrm{s}})}. 
        \end{align} 
        On the concentration event, $\big| \widehat K_{i\widehat p_t}^Q - K_{i\widehat p_t}^Q \big| \le \eta_s$ for every queried entry. For the recursive inner-product term, by the triangle inequality and Cauchy--Schwarz inequality, 
        \begin{subequations}
        \begin{align}
            \big| \big( \widehat{\mathbf G}_{i,1:t-1}^{(N_{\mathrm{s}})} \big)^{\!\top} \!\widehat{\mathbf G}_{\widehat p_t,1:t-1}^{(N_{\mathrm{s}})} \!-\! \big( \mathbf G_{i,1:t-1}^{\widehat{\mathcal P}_{t-1}} \big)^{\!\top} \! \mathbf G_{\widehat p_t,1:t-1}^{\widehat{\mathcal P}_{t-1}} \big| 
            &\le \big\| \widehat{\mathbf G}_{i,1:t-1}^{(N_{\mathrm{s}})} - \mathbf G_{i,1:t-1}^{\widehat{\mathcal P}_{t-1}} \big\|_2 \big\| \widehat{\mathbf G}_{\widehat p_t,1:t-1}^{(N_{\mathrm{s}})} \big\|_2 \\ 
            &\;+ \big\| \mathbf G_{i,1:t-1}^{\widehat{\mathcal P}_{t-1}} \big\|_2 \big\| \widehat{\mathbf G}_{\widehat p_t,1:t-1}^{(N_{\mathrm{s}})} - \mathbf G_{\widehat p_t,1:t-1}^{\widehat{\mathcal P}_{t-1}} \big\|_2 \nonumber\\
            &\le 2B E_{t-1}.
            \end{align}
        \end{subequations} 
        Therefore, we can obtain
        \begin{align} 
            \left| \widehat u_{i,t} - u_{i,t} \right| \le \eta_s + 2B E_{t-1}. 
        \label{eq:numerator_error} 
        \end{align} 
        Next, the map $h(x) = x^{-1/2}$ has derivative $|h'(x)| = \frac{1}{2x^{3/2}} \le \frac{1}{2\tau^{3/2}}$ on $[\tau,\infty)$. Therefore, 
        \begin{align} 
            \bigg| \frac{ 1 }{ \sqrt{ \widehat s_{\widehat p_t,t-1} } } - \frac{ 1 }{ \sqrt{ d_{\widehat p_t,t-1}^{\widehat{\mathcal P}_{t-1}} } } \bigg| 
            \le \frac{\big| \widehat s_{\widehat p_t,t-1} - d_{\widehat p_t,t-1}^{\widehat{\mathcal P}_{t-1}} \big| }{2\tau^{3/2} } 
            \le \frac{ \zeta_{t-1} }{ 2\tau^{3/2} }. 
        \label{eq:denom_error} 
        \end{align} 
        The exact numerator is an entry of the exact residual, i.e., $u_{i,t} = \big( \mathbf R_{t-1}^{\widehat{\mathcal P}_{t-1}} \big)_{i,\widehat p_t}$. Since this residual is PSD, Cauchy--Schwarz for PSD matrices yields 
        \begin{align} 
            |u_{i,t}| \le \sqrt{ d_{i,t-1}^{\widehat{\mathcal P}_{t-1}} d_{\widehat p_t,t-1}^{\widehat{\mathcal P}_{t-1}} }.
        \end{align}
        By residual monotonicity and $K_{ii}^Q\le\kappa$, we have
        \begin{align} 
            d_{i,t-1}^{\widehat{\mathcal P}_{t-1}} \le K_{ii}^Q \le \kappa, 
        \end{align} 
        and similarly for the pivot diagonal. Hence, $|u_{i,t}| \le \kappa$. By using 
        \begin{align} 
            \widehat G_{it}^{(N_{\mathrm{s}})} = \frac{ \widehat u_{i,t} }{ \sqrt{ \widehat s_{\widehat p_t,t-1} } }, \quad 
            G_{it}^{\widehat{\mathcal P}_t} = \frac{ u_{i,t} }{ \sqrt{ d_{\widehat p_t,t-1}^{\widehat{\mathcal P}_{t-1}} } }, 
        \end{align} 
        we obtain 
        \begin{align} 
            \big| \widehat G_{it}^{(N_{\mathrm{s}})} - G_{it}^{\widehat{\mathcal P}_t} \big| 
            \!\le\! \frac{ |\widehat u_{i,t}-u_{i,t}| }{ \sqrt{ \widehat s_{\widehat p_t,t-1} } } + |u_{i,t}| \bigg| \frac{ 1 }{ \sqrt{ \widehat s_{\widehat p_t,t-1} } } - \frac{ 1 }{ \sqrt{ d_{\widehat p_t,t-1}^{\widehat{\mathcal P}_{t-1}} } } \bigg| 
            \!\le\! \frac{ \eta_s \!+\! 2B E_{t-1} }{ \sqrt{\tau} } + \frac{ \kappa\zeta_{t-1} }{ 2\tau^{3/2} }.
        \end{align} 
        Taking the maximum over $i$ proves for $a_t$ in (\ref{eq:a_recursion_E_recursion_zeta_recursion}). For each row, 
        \begin{align} 
            \big\| \widehat{\mathbf G}_{i,1:t}^{(N_{\mathrm{s}})} - \mathbf G_{i,1:t}^{\widehat{\mathcal P}_t} \big\|_2^2 
            = \big\| \widehat{\mathbf G}_{i,1:t-1}^{(N_{\mathrm{s}})} - \mathbf G_{i,1:t-1}^{\widehat{\mathcal P}_{t-1}} \big\|_2^2 + \big\| \widehat G_{it}^{(N_{\mathrm{s}})} - G_{it}^{\widehat{\mathcal P}_t} \big\|^2. 
        \end{align} 
        Taking the maximum over $i$ yields 
        \begin{align} 
            E_t \le \sqrt{ E_{t-1}^2 + a_t^2 }, 
        \end{align} 
        which proves for $E_t$ in (\ref{eq:a_recursion_E_recursion_zeta_recursion}). For the score recursion, let $P_+(x):=[x]_+$. Since projection onto $[0,\infty)$ is nonexpansive and the exact residual diagonal is nonnegative,
        \begin{align} 
            d_{i,t}^{\widehat{\mathcal P}_t} = P_+ \Big( d_{i,t-1}^{\widehat{\mathcal P}_{t-1}} - \big( G_{it}^{\widehat{\mathcal P}_t} \big)^2 \Big). 
        \end{align} 
        Therefore, we can obtain
        \begin{align} 
            \big| \widehat s_{i,t} - d_{i,t}^{\widehat{\mathcal P}_t} \big| 
            \le \big| \widehat s_{i,t-1} - d_{i,t-1}^{\widehat{\mathcal P}_{t-1}} \big| + \left| \big( \widehat G_{it}^{(N_{\mathrm{s}})} \big)^2 - \big( G_{it}^{\widehat{\mathcal P}_t} \big)^2 \right|. 
        \end{align} 
        Using $|a^2-b^2| = |a-b||a+b|$ and the row bounds in Assumption~\ref{ass:stable_path}, 
        \begin{align} 
            \big| \big( \widehat G_{it}^{(N_{\mathrm{s}})} \big)^2 - \big( G_{it}^{\widehat{\mathcal P}_t} \big)^2 \big| \le 2B a_t. 
        \end{align} 
        Taking the maximum over admissible indices yields $\zeta_t \le \zeta_{t-1} + 2B a_t$, which proves $\zeta_t$ in (\ref{eq:a_recursion_E_recursion_zeta_recursion}). 
            
        It remains to establish the final proportionality to $\eta_s$. Define deterministic constants recursively by $C_E^{(0)} = C_\zeta^{(0)} = 0$ and, for $t\ge1$, 
        \begin{align} 
            C_a^{(t)} := \frac{1 \!+\! 2B C_E^{(t\!-\!1)} }{ \sqrt{\tau} } \!+\! \frac{ \kappa C_\zeta^{(t\!-\!1)} }{ 2\tau^{3/2} }, 
            C_E^{(t)} \!:=\! \sqrt{\! \big(\! C_E^{(t-1)} \!\big)^2 \!+\! \big(\! C_a^{(t)} \!\big)^2 },
            C_\zeta^{(t)} \!:=\! C_\zeta^{(t\!-\!1)} \!+\! 2B C_a^{(t)}. 
        \end{align} 
        An induction using (\ref{eq:a_recursion_E_recursion_zeta_recursion}) shows that 
        \begin{align} 
            a_t \le C_a^{(t)} \eta_s, \quad 
            E_t \le C_E^{(t)} \eta_s, \quad 
            \zeta_t \le C_\zeta^{(t)} \eta_s. 
        \end{align}
        For fixed $r, \tau, B, \kappa$, all of these constants are finite. Taking $C_{r,\tau,B,\kappa} := C_E^{(r)}$ establishes (\ref{eq:E_r_rate}). 
    \end{proof} 
    Theorem~\ref{thm:factor_perturbation} identifies the recursive mechanism through which finite-shot quantum-kernel error propagates through ICAQKL. The oracle error $\eta_s$ directly perturbs the numerator of each Cholesky update, while the accumulated factor error $E_{t-1}$ feeds back into subsequent inner products and the score discrepancy $\zeta_{t-1}$ perturbs the pivot normalization. Thus, finite-shot error is propagated recursively rather than added only once at the matrix level. The lower pivot bound $\tau$ quantifies the conditioning of this recursion. As a selected residual approaches zero, the normalization by its square root becomes increasingly sensitive to perturbations. Assumption~\ref{ass:stable_path} therefore characterizes the regime in which the executed finite-shot factorization remains stable. To convert the rowwise perturbation into a matrix-level factor error, define 
    \begin{align} 
        \Delta_r := \big\| \widehat{\mathbf G}_r^{(N_{\mathrm{s}})} - \mathbf G_r^{\widehat{\mathcal P}_r} \big\|_2. 
    \label{eq:Delta_r} 
    \end{align} 
    Using Frobenius norm property, $|\mathbf A|_2\le|\mathbf A|_F$, 
    \begin{align} 
        \Delta_r \le \big| \widehat{\mathbf G}_r^{(N_{\mathrm{s}})} - \mathbf G_r^{\widehat{\mathcal P}_r} \big|_F \le \sqrt n E_r. 
    \label{eq:Delta_E} 
    \end{align} 
    Consequently, for fixed stable rank and fixed $\tau,B,\kappa$, \begin{align} 
        \Delta_r = O(\eta_s) = O \Big( \sqrt{ \frac{ \log(M_r/\delta) }{ N_{\mathrm{s}} } } \Big). 
    \label{eq:Delta_rate} 
    \end{align} 
    This result establishes the required bridge from finite-shot quantum-kernel estimation error to perturbation of the recursively constructed Cholesky factor. The remaining step is to combine this numerical factor error with the structural approximation error of the selected pivot path to obtain a bound on the final low-rank kernel surrogate.

\paragraph{Total finite-shot kernel approximation error.} 
    The preceding analysis allows us to separate two conceptually distinct sources of error in the finite-shot ICAQKL surrogate. The first is the \emph{structural rank/path error}. Even if every kernel entry required along the selected pivot sequence were evaluated exactly, a rank-$r$ Cholesky factor generally does not recover the complete kernel unless the selected rank is sufficiently large. The second is the \emph{finite-shot factorization error}. Conditional on the same pivot sequence, the factor constructed from finite-shot kernel estimates differs from the corresponding exact-entry factor. The following result separates these two effects. 
    
    \begin{theorem}[Path-conditioned total kernel error] \label{thm:total_kernel_error} 
        Define the exact-path rank-$r$ approximation error by $\varepsilon_r \big( \widehat{\mathcal P}_r \big) \!:=\! \big\| \mathbf K^Q \!-\! \mathbf G_r^{\!\widehat{\mathcal P}_r} \big( \mathbf G_r^{\!\widehat{\mathcal P}_r} \big)^{\!\top} \!\big\|_2$,
        and let $\beta_r \!:=\! \big\| \mathbf G_r^{\!\widehat{\mathcal P}_r} \big\|_2$. Then the finite-shot kernel-surrogate error $\mathcal E_r^{(N_{\mathrm{s}})} \!:=\! \big\| \mathbf K^Q \!-\! \widetilde{\mathbf K}_r^{\!Q,(N_{\mathrm{s}})} \big\|_2$ satisfies 
        \begin{align} 
            \mathcal E_r^{(N_{\mathrm{s}})} \le \varepsilon_r \big( \widehat{\mathcal P}_r \big) + \left( 2\beta_r + \Delta_r \right) \Delta_r. 
        \label{eq:correct_total_kernel_error} 
        \end{align} 
    \end{theorem} 
    \begin{proof} 
        Recall that 
        \begin{align} 
            \widetilde{\mathbf K}_r^{Q,(N_{\mathrm{s}})} = \widehat{\mathbf G}_r^{(N_{\mathrm{s}})} \big( \widehat{\mathbf G}_r^{(N_{\mathrm{s}})} \big)^\top. 
        \end{align} 
        Add and subtract the exact-entry surrogate following the same selected pivot path: 
        \begin{subequations}
        \begin{align} 
            \big\| \mathbf K^Q - \widetilde{\mathbf K}_r^{Q,(N_{\mathrm{s}})} \big\|_2 
            & \le \big\| \mathbf K^Q - \mathbf G_r^{\widehat{\mathcal P}_r} \big( \mathbf G_r^{\widehat{\mathcal P}_r} \big)^\top \big\|_2 + \big| \mathbf G_r^{\widehat{\mathcal P}_r} \big( \mathbf G_r^{\widehat{\mathcal P}_r} \big)^\top - \widehat{\mathbf G}_r^{(N_{\mathrm{s}})} \left( \widehat{\mathbf G}_r^{(N_{\mathrm{s}})} \right)^\top \big\|_2 \\
            &= \varepsilon_r \big( \widehat{\mathcal P}_r \big) + \big\| \mathbf G_r^{\widehat{\mathcal P}_r} \big( \mathbf G_r^{\widehat{\mathcal P}_r} \big)^\top - \widehat{\mathbf G}_r^{(N_{\mathrm{s}})} \big( \widehat{\mathbf G}_r^{(N_{\mathrm{s}})} \big)^\top \big\|_2.
        \label{eq:kernel_error_triangle} 
        \end{align} 
        \end{subequations}
        Define the factor perturbation matrix $\mathbf D_r := \widehat{\mathbf G}_r^{(N_{\mathrm{s}})} - \mathbf G_r^{\widehat{\mathcal P}_r}$ so that, by definition, $\|\mathbf D_r\|_2 = \Delta_r$. Since 
        \begin{align} 
            \widehat{\mathbf G}_r^{(N_{\mathrm{s}})} = \mathbf G_r^{\widehat{\mathcal P}_r} + \mathbf D_r, 
        \end{align} 
        it follows that
        \begin{align} 
            \widehat{\mathbf G}_r^{(N_{\mathrm{s}})} \big( \widehat{\mathbf G}_r^{(N_{\mathrm{s}})} \big)^\top - \mathbf G_r^{\widehat{\mathcal P}_r} \big( \mathbf G_r^{\widehat{\mathcal P}_r} \big)^\top 
            = \mathbf D_r \big( \mathbf G_r^{\widehat{\mathcal P}_r} \big)^\top + \mathbf G_r^{\widehat{\mathcal P}_r} \mathbf D_r^\top + \mathbf D_r\mathbf D_r^\top. 
        \label{eq:factor_product_expansion} 
        \end{align} 
        Therefore, by the triangle inequality and submultiplicativity of the spectral norm, 
        \begin{subequations}
        \begin{align} 
            \big\| \mathbf G_r^{\widehat{\mathcal P}_r} \big( \mathbf G_r^{\widehat{\mathcal P}_r} \big)^\top - \widehat{\mathbf G}_r^{(N_{\mathrm{s}})} \big( \widehat{\mathbf G}_r^{(N_{\mathrm{s}})} \big)^\top \big\|_2 
            &\le \big\| \mathbf D_r \big( \mathbf G_r^{\widehat{\mathcal P}_r} \big)^\top \big\|_2 + \big\| \mathbf G_r^{\widehat{\mathcal P}_r} \mathbf D_r^\top \big\|_2 + \big\| \mathbf D_r\mathbf D_r^\top \big\|_2 \\
            &\le \|\mathbf D_r\|_2 \big\| \mathbf G_r^{\widehat{\mathcal P}_r} \big\|_2 + \big\| \mathbf G_r^{\widehat{\mathcal P}_r} \big\|_2 \|\mathbf D_r\|_2 + \|\mathbf D_r\|_2^2 \\
            &= 2\beta_r\Delta_r + \Delta_r^2 = \left( 2\beta_r+\Delta_r \right) \Delta_r.
        \label{eq:factor_product_bound} 
        \end{align}
        \end{subequations}
        Substituting (\ref{eq:factor_product_bound}) into (\ref{eq:kernel_error_triangle}) proves (\ref{eq:correct_total_kernel_error}). 
    \end{proof} 
    
    \begin{remark}
    \label{rem:exact_path} 
        Suppose the pivot-gap condition in Corollary~\ref{cor:pivot_gap} holds at every executed iteration $t=1,\ldots,r$. Then, inductively, the finite-shot pivot sequence coincides with the exact greedy sequence, $\widehat{\mathcal P}_r = \mathcal P_r^\star$. Consequently, $\varepsilon_r \big( \widehat{\mathcal P}_r \big) \!=\! \big\| \mathbf K^Q \!-\! \mathbf G_r^{\!\mathcal P_r^\star} \big( \mathbf G_r^{\!\mathcal P_r^\star} \big)^{\!\top} \big\|_2$,
        which is precisely the rank-$r$ approximation error generated by the exact greedy Cholesky recursion.
    \end{remark} 
    
    Theorem~\ref{thm:total_kernel_error} is deterministic once the selected path and the two factors are fixed. Combining it with Theorem~\ref{thm:factor_perturbation}, however, yields the finite-shot consequence $\Delta_r = O(\eta_s)$ for fixed stable $r, \tau, B, \kappa$. Therefore, $\mathcal E_r^{(N_{\mathrm{s}})} \!\le\! \varepsilon_r \big( \widehat{\mathcal P}_r \big) \!+\! O(\eta_s)$. 
    Thus, at a fixed stable rank, increasing the shot budget suppresses the numerical finite-shot contribution, while the remaining error is determined by the approximation quality of the selected rank-$r$ pivot path. 
    
    Additionally, Theorem~\ref{thm:total_kernel_error} provides the interface between quantum-kernel acquisition and the subsequent QSVM analysis. The overall kernel error separates into the structural approximation error associated with the selected rank-$r$ path and the numerical factorization error induced by finite-shot measurements, $\mathcal E_r^{(N_{\mathrm{s}})} \!\le\! \varepsilon_r \big( \widehat{\mathcal P}_r \big) \!+\! \big( 2\beta_r+\Delta_r \big) \Delta_r$. The downstream learning analysis, therefore, need not track individual measurement outcomes, pivot-score errors, or Cholesky-column perturbations separately. Once $\mathcal E_r^{(N_{\mathrm{s}})}$ is controlled, this matrix-level error can be propagated through the QSVM objective and ultimately to the learned classifier.

\subsection{QSVM Objective and Optimizer Stability} 
\label{subsec:qsvm_stability} 
    Having controlled the finite-shot kernel-surrogate error, we now propagate this perturbation through the QSVM training problem. The key observation is that the kernel matrix enters the dual objective only through a quadratic form. Consequently, spectral-norm control of the kernel perturbation yields a uniform perturbation bound over the entire feasible set, which in turn allows us to compare the exact and approximate optimizers. Let 
    \begin{align} 
        \mathcal A := \big\{\boldsymbol\alpha\in\mathbb R^n : 0 \le \alpha_i \le C, \mathbf y^\top \boldsymbol \alpha = 0 \big\}
    \label{eq:qsvm_feasible_set} 
    \end{align} 
    denote the feasible set of the soft-margin SVM dual, and define $\mathbf Y \!:=\! \operatorname{diag}(y_1, \ldots, y_n)$.
    Since $y_i \in \{-1,+1\}$, we have 
    \begin{align} 
        \mathbf Y^\top\mathbf Y = \mathbf I. 
    \label{eq:Y_orthogonal} 
    \end{align} 
    For a kernel matrix $\mathbf K$, write the SVM dual in minimization form as 
    \begin{align} 
        J_{\mathbf K}(\boldsymbol\alpha) := \frac12 \boldsymbol\alpha^\top \mathbf Y \mathbf K \mathbf Y \boldsymbol\alpha - \mathbf 1^\top \boldsymbol\alpha. 
    \label{eq:dual_objective_general} 
    \end{align} 
    
    \begin{lemma}[Uniform perturbation of the QSVM dual] 
    \label{lem:uniform_perturb} 
        For every $\boldsymbol\alpha\in\mathcal A$, 
        \begin{align} 
            \big| J_{\mathbf K^Q}(\boldsymbol\alpha) - J_{\widetilde{\mathbf K}_r^{Q,(N_{\mathrm{s}})}} (\boldsymbol\alpha) \big| 
            \le \frac12 |\boldsymbol\alpha|_2^2 \mathcal E_r^{(N_{\mathrm{s}})}. 
        \label{eq:uniform_dual_perturb} 
        \end{align} 
        Consequently, 
        \begin{align} 
            \sup_{\boldsymbol\alpha\in\mathcal A} \big| J_{\mathbf K^Q}(\boldsymbol\alpha) - J_{\widetilde{\mathbf K}_r^{Q,(N_{\mathrm{s}})}} (\boldsymbol\alpha) \big| 
            \le \frac{nC^2}{2} \mathcal E_r^{(N_{\mathrm{s}})}. 
        \label{eq:uniform_dual_perturb_box} 
        \end{align} 
    \end{lemma} 
    \begin{proof} 
        Subtracting the two objectives gives 
        \begin{align} 
            J_{\mathbf K^Q}(\boldsymbol\alpha) - J_{\widetilde{\mathbf K}_r^{Q,(N_{\mathrm{s}})}} (\boldsymbol\alpha) 
            = \frac12 \boldsymbol\alpha^\top \mathbf Y \left( \mathbf K^Q - \widetilde{\mathbf K}_r^{Q,(N_{\mathrm{s}})} \right) \mathbf Y \boldsymbol\alpha. 
        \label{eq:objective_difference} 
        \end{align} 
        Let $\mathbf v \!:=\! \mathbf Y \boldsymbol \alpha$. Since $\mathbf Y^\top \mathbf Y \!=\! \mathbf I$, then $\|\mathbf v\|_2 = \|\boldsymbol\alpha\|_2$. Using the standard quadratic-form bound 
        \begin{align} 
            \big|\mathbf v^\top \mathbf A \mathbf v \big| \le \|\mathbf A\|_2 \|\mathbf v\|_2^2 
        \end{align} 
        for symmetric $\mathbf A$, we obtain 
        \begin{align} 
            \big| J_{\mathbf K^Q}(\boldsymbol\alpha) - J_{\widetilde{\mathbf K}_r^{Q,(N_{\mathrm{s}})}} (\boldsymbol\alpha) \big| 
            \le \frac12 \|\boldsymbol\alpha\|_2^2 \big\| \mathbf K^Q - \widetilde{\mathbf K}_r^{Q,(N_{\mathrm{s}})} \big\|_2 
            = \frac12 \|\boldsymbol \alpha\|_2^2 \mathcal E_r^{(N_{\mathrm{s}})}, 
        \end{align} 
        which proves (\ref{eq:uniform_dual_perturb}). Moreover, the box constraints imply 
        \begin{align} 
            \|\boldsymbol \alpha\|_2^2 = \sum_{i=1}^n\alpha_i^2 \le nC^2. 
        \end{align} 
        Taking the supremum over $\mathcal A$ therefore yields (\ref{eq:uniform_dual_perturb_box}). 
    \end{proof} 
    
    Lemma~\ref{lem:uniform_perturb} is uniform over the feasible set rather than being evaluated at a particular optimizer. This distinction is important because the exact and finite-shot QSVM problems generally have different minimizers. Uniform control allows the two optimization problems to be compared without assuming that their solutions coincide. 
    
    Define the exact optimal set by 
    \begin{align} 
        \mathcal A^{Q,\star} := \arg\min_{\boldsymbol\alpha\in\mathcal A} J_{\mathbf K^Q}(\boldsymbol\alpha), 
    \label{eq:exact_optimal_set} 
    \end{align} 
    and let 
    \begin{align} 
        \widetilde{\boldsymbol\alpha}_r^{(N_{\mathrm{s}})} \in \arg\min_{\boldsymbol\alpha\in\mathcal A} J_{\widetilde{\mathbf K}_r^{Q,(N_{\mathrm{s}})}} (\boldsymbol\alpha) 
    \label{eq:approx_optimizer} 
    \end{align} 
    be any optimizer of the finite-shot low-rank QSVM. 
    
    \begin{theorem}[Exact-objective suboptimality] \label{thm:suboptimality} 
        For every $\boldsymbol\alpha^\star\in\mathcal A^{Q,\star}$, \begin{align} 
            0 \le J_{\mathbf K^Q} \big( \widetilde{\boldsymbol\alpha}_r^{(N_{\mathrm{s}})} \big) - J_{\mathbf K^Q} \left( \boldsymbol\alpha^\star \right) 
            \le nC^2 \mathcal E_r^{(N_{\mathrm{s}})}. 
        \label{eq:suboptimality} 
        \end{align} 
    \end{theorem} 
    \begin{proof} 
        The lower bound follows immediately from the optimality of $\boldsymbol \alpha^\star$ for the exact objective. For the upper bound, insert the approximate objective: 
        \begin{align} 
            J_{\mathbf K^Q} \big( \widetilde{\boldsymbol\alpha}_r^{(N_{\mathrm{s}})} \big) 
            &- J_{\mathbf K^Q} \left( \boldsymbol\alpha^\star \right) 
            = \Big[ J_{\mathbf K^Q} \big( \widetilde{\boldsymbol\alpha}_r^{(N_{\mathrm{s}})} \big) - J_{\widetilde{\mathbf K}_r^{Q,(N_{\mathrm{s}})}} \big( \widetilde{\boldsymbol\alpha}_r^{(N_{\mathrm{s}})} \big) \Big] \\
            & + \Big[ J_{\widetilde{\mathbf K}_r^{Q,(N_{\mathrm{s}})}} \big( \widetilde{\boldsymbol\alpha}_r^{(N_{\mathrm{s}})} \big) - J_{\widetilde{\mathbf K}_r^{Q,(N_{\mathrm{s}})}} \left( \boldsymbol\alpha^\star \right) \Big] + \Big[ J_{\widetilde{\mathbf K}_r^{Q,(N_{\mathrm{s}})}} \left( \boldsymbol\alpha^\star \right) - J_{\mathbf K^Q} \left( \boldsymbol\alpha^\star \right) \Big]. \nonumber
        \label{eq:suboptimality_decomposition} 
        \end{align} 
        The middle term is nonpositive by the optimality of $\widetilde{\boldsymbol \alpha}_r^{(N_{\mathrm{s}})}$ for the approximate problem. Lemma~\ref{lem:uniform_perturb} bounds the magnitude of each of the remaining terms by $\frac{nC^2}{2} \mathcal E_r^{(N_{\mathrm{s}})}$. Therefore, 
        \begin{align}
            J_{\mathbf K^Q} \big( \widetilde{\boldsymbol\alpha}_r^{(N_{\mathrm{s}})} \big) - J_{\mathbf K^Q} \left( \boldsymbol\alpha^\star \right) \le nC^2 \mathcal E_r^{(N_{\mathrm{s}})}, 
        \end{align} 
        which proves (\ref{eq:suboptimality}). 
    \end{proof} 
    
    Theorem~\ref{thm:suboptimality} shows that optimization with the finite-shot low-rank kernel produces a solution that remains approximately optimal for the exact QSVM objective whenever the kernel-surrogate error is small. This is an objective-value guarantee; controlling the optimizer itself requires an additional regularity condition. 
    
    \begin{assumption}[Quadratic growth] 
    \label{ass:quadratic_growth} 
        There exists $\mu>0$ such that, for every $\boldsymbol\alpha\in\mathcal A$, 
        \begin{align} J
            _{\mathbf K^Q}(\boldsymbol\alpha) - J_{\mathbf K^Q}^{\star} \ge \frac{\mu}{2} \operatorname{dist} \left( \boldsymbol\alpha, \mathcal A^{Q,\star} \right)^2, 
        \label{eq:quadratic_growth} 
        \end{align} 
        where $J_{\mathbf K^Q}^{\star} := \min_{\boldsymbol\alpha\in\mathcal A} J_{\mathbf K^Q}(\boldsymbol\alpha)$.
    \end{assumption} 
    
    \begin{theorem}[Distance to the exact optimal set] 
    \label{thm:distance_to_optset} 
        Under Assumption~\ref{ass:quadratic_growth}, 
        \begin{align}
            \operatorname{dist} \left( \widetilde{\boldsymbol \alpha}_r^{(N_{\mathrm{s}})}, \mathcal A^{Q,\star} \right) \le \sqrt{ \frac{2nC^2}{\mu} \mathcal E_r^{(N_{\mathrm{s}})} }. 
        \label{eq:distance_optset} 
        \end{align} 
    \end{theorem} 
    \begin{proof} 
        Since every $\boldsymbol\alpha^\star\in\mathcal A^{Q,\star}$ attains $J_{\mathbf K^Q}^{\star}$, Theorem~\ref{thm:suboptimality} yields
        \begin{align}
            J_{\mathbf K^Q} \big( \widetilde{\boldsymbol\alpha}_r^{(N_{\mathrm{s}})} \big) - J_{\mathbf K^Q}^{\star} \quad\le nC^2 \mathcal E_r^{(N_{\mathrm{s}})}. 
        \label{eq:exact_objective_gap} 
        \end{align} 
        On the other hand, applying Assumption~\ref{ass:quadratic_growth} at $\widetilde{\boldsymbol\alpha}_r^{(N_{\mathrm{s}})}$ yields 
        \begin{align}
            \frac{\mu}{2} \operatorname{dist} \big( \widetilde{\boldsymbol\alpha}_r^{(N_{\mathrm{s}})}, \mathcal A^{Q,\star} \big)^2 \le J_{\mathbf K^Q} \big( \widetilde{\boldsymbol\alpha}_r^{(N_{\mathrm{s}})} \big) - J_{\mathbf K^Q}^{\star}. 
        \end{align} 
        Combining the two inequalities, we obtain
        \begin{align} 
            \frac{\mu}{2} \operatorname{dist} \left( \widetilde{\boldsymbol\alpha}_r^{(N_{\mathrm{s}})}, \mathcal A^{Q,\star} \right)^2 \le nC^2 \mathcal E_r^{(N_{\mathrm{s}})}. 
        \end{align} 
        Taking square roots completes the proof. 
    \end{proof} 

    The exact QSVM dual optimizer need not be unique. Theorem~\ref{thm:distance_to_optset} therefore controls the distance from the finite-shot optimizer to the complete exact optimal set $\mathcal A^{Q,\star}$, rather than to an arbitrarily selected exact optimizer. This formulation avoids an unnecessary uniqueness assumption. Combining the preceding results yields the training-side perturbation chain 
    \begin{align} 
        \mathcal E_r^{(N_{\mathrm{s}})} \longrightarrow \text{QSVM objective stability} \longrightarrow \text{optimizer-set stability}. 
    \end{align} 
    Particularly, if $\mathcal E_r^{(N_{\mathrm{s}})} \!\to\! 0$ then $J_{\!\mathbf K^{\!Q}} \!\big(\! \widetilde{\boldsymbol \alpha}_r^{(N_{\mathrm{s}})} \!\big) \!-\! J_{\!\mathbf K^{\!Q}}^{\star} \!\to\! 0$
    , under quadratic growth $\operatorname{dist} \!\big(\! \widetilde{\boldsymbol\alpha}_r^{(N_{\mathrm{s}})}, \mathcal A^{Q,\star} \!\big) \!\to\! 0$.
    Prediction requires one additional ingredient, however: control of the kernel similarities between the training samples and a previously unseen test point. We address this out-of-sample extension next.

\subsection{Out-of-Sample and Prediction Stability} 
\label{subsec:oos_prediction} 
    The preceding analysis controls the error of the finite-shot ICAQKL surrogate on the training Gram matrix and establishes stability of the corresponding QSVM optimizer. Prediction on a previously unseen point requires an additional step. In particular, the learned low-rank representation must be extended from the training samples to a new test point using fresh quantum-kernel evaluations. We therefore distinguish two sources of out-of-sample error: the structural error of the exact rank-$r$ Cholesky extension along the selected training path and the additional finite-shot error introduced by noisy training factors and fresh test-time measurements. Throughout this subsection, let $\mathbf z \in \mathcal X$ be a fixed test point and define the exact train--test quantum-kernel vector
    \begin{align} 
        \mathbf k_{\mathbf z}^Q := \left( \kappa^Q(\mathbf x_1,\mathbf z), \ldots, \kappa^Q(\mathbf x_n,\mathbf z) \right)^\top. 
    \label{eq:exact_test_vector} 
    \end{align} 
    Let $\widehat{\mathcal P}_r = ( \widehat p_1,\ldots,\widehat p_r )$ denote the pivot path selected during finite-shot ICAQKL training. 
    
\paragraph{Exact out-of-sample Cholesky extension.} 
    Consider first an exact-entry extension that follows the same selected training path $\widehat{\mathcal P}_r$. Let $\mathbf g_r(\mathbf z) = \left( g_1(\mathbf z),\ldots,g_r(\mathbf z) \right)^\top$ denote the resulting exact Cholesky feature for $\mathbf z$. Its components are generated recursively according to 
    \begin{align} 
        g_\ell(\mathbf z) = \frac{ \kappa^Q ( \mathbf x_{\widehat p_\ell}, \mathbf z ) - \sum_{m=1}^{\ell-1} g_m(\mathbf z) G_{\widehat p_\ell m}^{\widehat{\mathcal P}_r} }{ G_{\widehat p_\ell\ell}^{\widehat{\mathcal P}_r} }, \quad \ell=1,\ldots,r. 
    \label{eq:oos_exact_recursion} 
    \end{align} 
    The corresponding exact-entry rank-$r$ train--test approximation is 
    \begin{align} 
        \widetilde{\mathbf k}_{r,\mathbf z}^{Q,\widehat{\mathcal P}_r} := \mathbf G_r^{\widehat{\mathcal P}_r} \mathbf g_r(\mathbf z). 
    \label{eq:oos_exact_vector} 
    \end{align} 
    The following result relates the accuracy of this extension directly to the training residual. 
    
    \begin{lemma}[Exact out-of-sample residual bound] 
    \label{lem:exact_train_test} 
    Suppose $\kappa^Q(\mathbf z,\mathbf z) \le \kappa$. Then 
    \begin{align} 
        \big\| \mathbf k_{\mathbf z}^Q - \widetilde{\mathbf k}_{r,\mathbf z}^{Q,\widehat{\mathcal P}_r} \big\|_2^2 
        \le \kappa \operatorname{tr} \big( \mathbf R_r^{\widehat{\mathcal P}_r} \big) 
        \le n\kappa \big\| \mathbf R_r^{\widehat{\mathcal P}_r} \big\|_2. 
    \label{eq:exact_train_test_bound} 
    \end{align} 
    Moreover, 
    \begin{align} 
        \big\| \mathbf k_{\mathbf z}^Q - \widetilde{\mathbf k}_{r,\mathbf z}^{Q,\widehat{\mathcal P}_r} \big\|_2 \le \sqrt{n\kappa \max_i d_{i,r}^{\widehat{\mathcal P}_r} }. 
    \label{eq:exact_train_test_diag}
    \end{align} 
    \end{lemma} 
    \begin{proof} 
        Consider the augmented Gram matrix containing the $n$ training samples and the test point $\mathbf z$: 
        \begin{align} 
        \mathbf K_{\mathrm{aug}}^Q(\mathbf z) = 
        \begin{bmatrix} 
            \mathbf K^Q & \mathbf k_{\mathbf z}^Q \\ (\mathbf k_{\mathbf z}^Q)^\top & \kappa^Q(\mathbf z,\mathbf z) 
        \end{bmatrix}. 
        \label{eq:augmented_kernel} 
        \end{align} 
        Since $\kappa^Q$ is a positive-semidefinite kernel, $\mathbf K_{\mathrm{aug}}^Q(\mathbf z) \succeq \mathbf 0$. Apply the same $r$ exact Cholesky pivots $\widehat{\mathcal P}_r$, restricted to the training indices. The corresponding augmented residual takes the form 
        \begin{align} 
            \begin{bmatrix} 
                \mathbf R_r^{\widehat{\mathcal P}_r} & \mathbf e_{r,\mathbf z} \\ \mathbf e_{r,\mathbf z}^\top & \rho_{r,\mathbf z} 
            \end{bmatrix} 
            \succeq \mathbf 0, 
        \label{eq:augmented_residual} 
        \end{align} 
        where $\mathbf e_{r,\mathbf z} \!:=\! \mathbf k_{\mathbf z}^Q \!-\! \widetilde{\mathbf k}_{r,\mathbf z}^{Q,\widehat{\mathcal P}_r}$. For every training index $i$, positive semidefiniteness of the corresponding $2\times2$ principal minor implies $|e_{r, \mathbf z,i}|^2 \le \big( \mathbf R_r^{\widehat{\mathcal P}_r} \big)_{ii} \rho_{r, \mathbf z}$. The test-point residual diagonal satisfies
        \begin{align} 
            0 \le \rho_{r,\mathbf z} \le \kappa^Q(\mathbf z,\mathbf z) \le \kappa. 
        \end{align} 
        Therefore, we obtain
        \begin{align} 
            \big\|\mathbf e_{r,\mathbf z}\big\|_2^2 = \sum_{i=1}^n |e_{r,\mathbf z,i}|^2 \le \kappa \sum_{i=1}^n \big( \mathbf R_r^{\widehat{\mathcal P}_r} \big)_{ii} = \kappa \operatorname{tr} \big( \mathbf R_r^{\widehat{\mathcal P}_r} \big). 
        \end{align} 
        Since the exact residual is PSD, 
        \begin{align} 
            \operatorname{tr} \big( \mathbf R_r^{\widehat{\mathcal P}_r} \big) \le n \big\| \mathbf R_r^{\widehat{\mathcal P}_r} \big\|_2, 
        \end{align} 
        which proves (\ref{eq:exact_train_test_bound}). Similarly,
        \begin{align} 
            \operatorname{tr} \big( \mathbf R_r^{\widehat{\mathcal P}_r} \big) = \sum_{i=1}^n d_{i,r}^{\widehat{\mathcal P}_r} \le n \max_i d_{i,r}^{\widehat{\mathcal P}_r}. 
        \end{align} 
        Substitution and taking square roots yield (\ref{eq:exact_train_test_diag}). 
    \end{proof} 
    Lemma~\ref{lem:exact_train_test} provides a direct link between the training Cholesky residual and the accuracy of the exact out-of-sample extension. In particular, if the training recursion terminates when $\max_i \widehat s_{i,r} \le \varepsilon$ and the score discrepancy satisfies $\big\| \widehat{\mathbf s}_r - \mathbf d_r^{\widehat{\mathcal P}_r} \big\|_\infty \le \zeta_r$ then $\max_i d_{i,r}^{\widehat{\mathcal P}_r} \le \varepsilon+\zeta_r$. Hence, 
    \begin{align} 
        \big\| \mathbf k_{\mathbf z}^Q - \widetilde{\mathbf k}_{r,\mathbf z}^{Q,\widehat{\mathcal P}_r} \big\|_2 \le \sqrt{ n\kappa ( \varepsilon+\zeta_r ) }. 
    \label{eq:oos_stopping} 
    \end{align} 
    Thus, the same residual scores used to terminate the training factorization also provide a sufficient criterion for controlling the exact-entry out-of-sample approximation error.
    
\paragraph{Finite-shot pivot--test measurements.} 
    Training and prediction incur different quantum-query costs. Constructing the rank-$r$ training factor requires at most 
    \begin{align} 
        M_r = nr-\frac{r(r+1)}{2} = O(nr) 
    \end{align} 
    fresh off-diagonal training-kernel evaluations under symmetry, caching, and analytic fidelity diagonals. Once the factor has been constructed, however, evaluating a single test point requires only the $r$ pivot--test similarities $\widehat\kappa_{N_{\mathrm{s}}}^Q \left( \mathbf x_{\widehat p_\ell}, \mathbf z \right), \ell=1,\ldots,r$. Conditioned on the completed training phase, these are fresh quantum measurements. Define $\eta_z := \sqrt{ \frac{ \log(2r/\delta_z) }{ 2N_{\mathrm{s}} } }$.
    Applying Hoeffding's inequality and a union bound over the $r$ pivot--test queries gives, with conditional probability at least $1-\delta_z$, 
    \begin{align} 
        \Big( \sum_{\ell=1}^r \left| \widehat\kappa_{N_{\mathrm{s}}}^Q \left( \mathbf x_{\widehat p_\ell}, \mathbf z \right) - \kappa^Q \left( \mathbf x_{\widehat p_\ell}, \mathbf z \right) \right|^2 \Big)^{1/2} \le \sqrt r \eta_z = \sqrt{ \frac{ r\log(2r/\delta_z) }{ 2N_{\mathrm{s}} } }.
    \label{eq:test_query_concentration} 
    \end{align} 
    The appearance of $r$, rather than $M_r$, reflects the different roles of training and prediction: $M_r$ counts the distinct quantum-kernel evaluations needed to construct the training factor, whereas only $r$ fresh pivot--test evaluations are required for each new test point. 
    
\paragraph{Finite-shot out-of-sample perturbation.} 
    Finite-shot out-of-sample feature is computed recursively 
    \begin{align} 
        \widetilde g_\ell^{(N_{\mathrm{s}})}(\mathbf z) = \frac{ \widehat\kappa_{N_{\mathrm{s}}}^Q ( \mathbf x_{\widehat p_\ell}, \mathbf z ) - \sum_{m=1}^{\ell-1} \widetilde g_m^{(N_{\mathrm{s}})}(\mathbf z) \widehat G_{\widehat p_\ell m}^{(N_{\mathrm{s}})} }{ \widehat G_{\widehat p_\ell\ell}^{(N_{\mathrm{s}})} }, \quad \ell=1,\ldots,r. 
    \label{eq:oos_recursion} 
    \end{align} 
    Let $\widehat{\mathbf g}_r(\mathbf z) := \widetilde{\mathbf g}^{(N_{\mathrm{s}})}(\mathbf z)$ and define the finite-shot approximate train--test vector by $\widetilde{\mathbf k}_{r,\mathbf z}^{Q,(N_{\mathrm{s}})} \!:=\! \widehat{\mathbf G}_r^{(N_{\mathrm{s}})} \widehat{\mathbf g}_r(\mathbf z)$.
    Let $\mathbf S_{\widehat{\mathcal P}_r}$ denote the row-selection operator extracting the selected pivots in their order of selection, and define $\mathbf T_r := \mathbf S_{\widehat{\mathcal P}_r} \mathbf G_r^{\widehat{\mathcal P}_r}$, and $\widehat{\mathbf T}_r := \mathbf S_{\widehat{\mathcal P}_r} \widehat{\mathbf G}_r^{(N_{\mathrm{s}})}$.
    Along a successful Cholesky path, $\mathbf T_r$ is lower triangular with positive diagonal entries. Define $\nu_r := \sigma_{\min}(\mathbf T_r)$
    and recall $\beta_r := \big\| \mathbf G_r^{\widehat{\mathcal P}_r} \big\|_2$.
    
    \begin{theorem}[Finite-shot out-of-sample perturbation] 
    \label{thm:oos_finite} 
        Suppose $\nu_r>0, \Delta_r < \nu_r$.
        Conditioned on the completed training phase, on the event (\ref{eq:test_query_concentration}), 
        \begin{align} 
            \big\| \widetilde{\mathbf k}_{r,\mathbf z}^{Q,\widehat{\mathcal P}_r} - \widetilde{\mathbf k}_{r,\mathbf z}^{Q,(N_{\mathrm{s}})} \big\|_2  
            \le \sqrt{\kappa}\Delta_r + (\beta_r+\Delta_r) \frac{ \sqrt r\eta_z + \sqrt{\kappa}\Delta_r }{ \nu_r-\Delta_r }. 
        \label{eq:oos_finite_bound} 
        \end{align} 
        Consequently, 
        \begin{align} 
            \big\| \mathbf k_{\mathbf z}^Q - \widetilde{\mathbf k}_{r,\mathbf z}^{Q,(N_{\mathrm{s}})} \big\|_2 
            \le \sqrt{ \kappa \operatorname{tr} \big( \mathbf R_r^{\widehat{\mathcal P}_r} \big) } + \sqrt{\kappa}\Delta_r + (\beta_r+\Delta_r) \frac{ \sqrt r\eta_z + \sqrt{\kappa}\Delta_r }{ \nu_r-\Delta_r }. 
        \label{eq:oos_total_bound}
        \end{align} 
    \end{theorem} 
    \begin{proof} 
        Define the exact and finite-shot vectors of pivot--test kernel values by 
        \begin{subequations}
        \begin{align} 
            \mathbf v_r(\mathbf z) &:= \left( \kappa^Q ( \mathbf x_{\widehat p_1}, \mathbf z ), \ldots, \kappa^Q ( \mathbf x_{\widehat p_r}, \mathbf z ) \right)^\top, \\ \widehat{\mathbf v}_r(\mathbf z) &:= \left( \widehat\kappa_{N_{\mathrm{s}}}^Q ( \mathbf x_{\widehat p_1}, \mathbf z ), \ldots, \widehat\kappa_{N_{\mathrm{s}}}^Q ( \mathbf x_{\widehat p_r}, \mathbf z ) \right)^\top. 
        \end{align} 
        \end{subequations}
        The exact and finite-shot recursive extensions can be written as lower-triangular systems: 
        \begin{subequations}
        \begin{align} 
            \mathbf T_r \mathbf g_r(\mathbf z) &= \mathbf v_r(\mathbf z), \label{eq:exact_triangular}\\
            \widehat{\mathbf T}_r \widehat{\mathbf g}_r(\mathbf z) &= \widehat{\mathbf v}_r(\mathbf z). \label{eq:noisy_triangular} 
        \end{align} 
        \end{subequations}
        Since row selection is a contraction, 
        \begin{align} 
            \big\| \widehat{\mathbf T}_r - \mathbf T_r \big\|_2 \le \Delta_r. \label{eq:pivot_block_error} 
        \end{align} 
        Weyl's singular-value perturbation inequality therefore gives $\sigma_{\min} \big( \widehat{\mathbf T}_r \big) \!\ge\! \nu_r-\Delta_r \!>\! 0$. Hence,
        \begin{align} 
            \big\| \widehat{\mathbf T}_r^{-1} \big\|_2 \le \frac{1}{ \nu_r-\Delta_r }. 
        \label{eq:T_inverse_bound} 
        \end{align} 
        From (\ref{eq:test_query_concentration}), we have
        \begin{align} 
            \big\| \widehat{\mathbf v}_r(\mathbf z) - \mathbf v_r(\mathbf z) \big\|_2 \le \sqrt r\eta_z. 
        \label{eq:v_test_bound} 
        \end{align} 
        Moreover, the exact augmented Cholesky residual implies $\rho_{r,\mathbf z} = \kappa^Q(\mathbf z,\mathbf z) - \|\mathbf g_r(\mathbf z)\|_2^2 \ge 0$, it follows that
        $\|\mathbf g_r(\mathbf z)\|_2^2 \le \kappa^Q(\mathbf z,\mathbf z) \le \kappa$ so 
        \begin{align} 
            \big\|\mathbf g_r(\mathbf z)\big\|_2 \le \sqrt{\kappa}. 
        \label{eq:g_exact_norm} 
        \end{align}
        Subtracting (\ref{eq:exact_triangular}) from (\ref{eq:noisy_triangular}) yields
        \begin{align} 
            \widehat{\mathbf g}_r(\mathbf z) - \mathbf g_r(\mathbf z) 
            = \widehat{\mathbf T}_r^{-1} \left[ \widehat{\mathbf v}_r(\mathbf z) - \mathbf v_r(\mathbf z) - \big( \widehat{\mathbf T}_r - \mathbf T_r \big) \mathbf g_r(\mathbf z) \right]. 
        \end{align} 
        By using (\ref{eq:T_inverse_bound}), (\ref{eq:v_test_bound}), (\ref{eq:pivot_block_error}), and (\ref{eq:g_exact_norm}), we obtain 
        \begin{align} 
            \big\| \widehat{\mathbf g}_r(\mathbf z) - \mathbf g_r(\mathbf z) \big\|_2 \le \frac{ \sqrt r \eta_z + \sqrt{\kappa}\Delta_r }{ \nu_r-\Delta_r }. 
        \label{eq:g_oos_error} 
        \end{align} 
        Finally, we have
            \begin{align} 
                \big\| \widehat{\mathbf G}_r^{(N_{\mathrm{s}})} \widehat{\mathbf g}_r(\mathbf z) \!-\! \mathbf G_r^{\widehat{\mathcal P}_r} \mathbf g_r(\mathbf z) \big\|_2 
                \!\le\! \big\| \widehat{\mathbf G}_r^{(N_{\mathrm{s}})} \!-\! \mathbf G_r^{\widehat{\mathcal P}_r} \big\|_2 \|\mathbf g_r(\mathbf z)\|_2 \!+\! \big\| \widehat{\mathbf G}_r^{(N_{\mathrm{s}})} \big\|_2 \big\| \widehat{\mathbf g}_r(\mathbf z) \!-\! \mathbf g_r(\mathbf z) \big\|_2. 
            \end{align} 
        Since 
        \begin{align} 
            \big\| \widehat{\mathbf G}_r^{(N_{\mathrm{s}})} \big\|_2 
            \le \big\| \mathbf G_r^{\widehat{\mathcal P}_r} \big\|_2 + \Delta_r = \beta_r+\Delta_r, 
        \end{align} 
        substituting (\ref{eq:g_exact_norm}) and (\ref{eq:g_oos_error}) proves (\ref{eq:oos_finite_bound}). The total bound (\ref{eq:oos_total_bound}) then follows by the triangle inequality together with Lemma~\ref{lem:exact_train_test}. 
    \end{proof} 
    
    The conditioning factor $\nu_r$ is essential. Although positive pivot diagonals ensure that the triangular system is nonsingular, they do not by themselves provide a uniform lower bound on its smallest singular value. If $\nu_r$ is small, perturbations in either the learned factor or the fresh test-time kernel estimates can be strongly amplified. We therefore retain the conditioning requirement explicitly rather than attempting to infer it from the pivot floor alone. 
    
\paragraph{Bias-aware decision-function stability.} 
    The equality $\mathbf y^\top \boldsymbol \alpha = 0$ in the QSVM dual corresponds to an unregularized intercept. Consequently, prediction stability requires explicit control of the bias term in addition to the dual variables and train--test kernel vector. Choose $\boldsymbol\alpha^\circ \in \mathcal A^{Q,\star}$ such that 
    \begin{align} 
        \big\| \widetilde{\boldsymbol\alpha}_r^{(N_{\mathrm{s}})} - \boldsymbol\alpha^\circ \big\|_2 = \operatorname{dist} \big( \widetilde{\boldsymbol\alpha}_r^{(N_{\mathrm{s}})}, \mathcal A^{Q,\star} \big). 
    \label{eq:nearest_exact_optimizer} 
    \end{align} 
    The corresponding exact predictor is 
    \begin{align} 
        f_{\mathbf K^Q}(\mathbf z) = \sum_{i=1}^n \alpha_i^\circ y_i \kappa^Q(\mathbf x_i,\mathbf z) + b^\circ, 
    \label{eq:exact_predictor} 
    \end{align} 
    whereas the finite-shot ICAQKL predictor is 
    \begin{align} 
        f_r^{Q,(N_{\mathrm{s}})}(\mathbf z) = \sum_{i=1}^n \widetilde\alpha_{r,i}^{(N_{\mathrm{s}})} y_i \widetilde\kappa_r^{Q,(N_{\mathrm{s}})} (\mathbf x_i,\mathbf z) + \widetilde b_r^{(N_{\mathrm{s}})}. 
    \label{eq:approx_predictor} 
    \end{align} 
    We first record a simple norm bound for the exact train--test vector. 
    
    \begin{lemma}[Exact train--test norm] 
    \label{lem:kernel_vector_bound} 
        Suppose $\kappa^Q(\mathbf x,\mathbf x) \!\le\! \kappa$ for every $\mathbf x \!\in\! \mathcal X$. Then $\|\mathbf k_{\mathbf z}^Q\|_2 \!\le\! \kappa \sqrt n$.
    \end{lemma} 
    \begin{proof} 
        Kernel Cauchy--Schwarz yields 
        \begin{align} 
            \left| \kappa^Q(\mathbf x_i,\mathbf z) \right| &\le \sqrt{ \kappa^Q(\mathbf x_i,\mathbf x_i) \kappa^Q(\mathbf z,\mathbf z) } \le \kappa. 
        \end{align} 
        Therefore, $\|\mathbf k_{\mathbf z}^Q\|_2^2 \le n\kappa^2$ which proves the claim. 
    \end{proof} 
    
    \begin{theorem}[Decision-function stability] 
    \label{thm:predictor_stability} 
        The exact and finite-shot predictors satisfy 
        \begin{align} 
            \big| f_{\!\mathbf K^Q}(\mathbf z) \!-\! f_r^{Q,(N_{\mathrm{s}})}(\mathbf z) \big| \!\le\! \kappa\sqrt n \operatorname{dist} \! \big( \widetilde{\boldsymbol\alpha}_r^{(N_{\mathrm{s}})}\!, \mathcal A^{Q,\star} \big) \!+\! C\sqrt n \big\| \mathbf k_{\mathbf z}^Q - \widetilde{\mathbf k}_{r,\mathbf z}^{Q,(N_{\mathrm{s}})} \big\|_2 \!+\! \big| b^\circ \!-\! \widetilde b_r^{(N_{\mathrm{s}})} \big|. 
        \label{eq:predictor_stability} 
        \end{align} 
    \end{theorem} 
    \begin{proof} 
        Using vector notation, we write 
        \begin{subequations}
        \begin{align} 
            & f_{\mathbf K^Q}(\mathbf z) = (\boldsymbol\alpha^\circ)^\top \mathbf Y \mathbf k_{\mathbf z}^Q + b^\circ, \\ 
            & f_r^{Q,(N_{\mathrm{s}})}(\mathbf z) = \big( \widetilde{\boldsymbol\alpha}_r^{(N_{\mathrm{s}})} \big)^\top \mathbf Y \widetilde{\mathbf k}_{r,\mathbf z}^{Q,(N_{\mathrm{s}})} + \widetilde b_r^{(N_{\mathrm{s}})}. 
        \end{align} 
        \end{subequations}
        Adding and subtracting $\big( \widetilde{\boldsymbol \alpha}_r^{(N_{\mathrm{s}})} \big)^\top \mathbf Y \mathbf k_{\mathbf z}^Q$,
        we then obtain
        \begin{align} 
            f_{\mathbf K^Q}(\mathbf z) \!-\! f_r^{Q,(N_{\mathrm{s}})}(\mathbf z) \!=\! \big( \boldsymbol \alpha^\circ \!-\! \widetilde{\boldsymbol\alpha}_r^{(N_{\mathrm{s}})} \big)^\top \mathbf Y \mathbf k_{\mathbf z}^Q \!+\! \big( \widetilde{\boldsymbol\alpha}_r^{(N_{\mathrm{s}})} \big)^\top \mathbf Y \big( \mathbf k_{\mathbf z}^Q - \widetilde{\mathbf k}_{r,\mathbf z}^{Q,(N_{\mathrm{s}})} \big) + b^\circ \!-\! \widetilde b_r^{(N_{\mathrm{s}})}. 
        \end{align} 
        Since $\mathbf Y^\top \mathbf Y = \mathbf I$, Cauchy--Schwarz yields
        \begin{subequations}
        \begin{align} 
            \big| \big( \boldsymbol \alpha^\circ - \widetilde{\boldsymbol \alpha}_r^{(N_{\mathrm{s}})} \big)^\top \mathbf Y \mathbf k_{\mathbf z}^Q \big| 
            &\le \big\| \boldsymbol \alpha^\circ - \widetilde{\boldsymbol \alpha}_r^{(N_{\mathrm{s}})} \big\|_2 \big\|\mathbf k_{\mathbf z}^Q\big\|_2 \\
            &\le \kappa \sqrt n \operatorname{dist} \big( \widetilde{\boldsymbol \alpha}_r^{(N_{\mathrm{s}})}, \mathcal A^{Q, \star} \big). 
        \end{align} 
        \end{subequations}
        Moreover, the box constraints imply $\big\| \widetilde{\boldsymbol \alpha}_r^{(N_{\mathrm{s}})} \big\|_2 \le C\sqrt n$. Applying Cauchy--Schwarz to the second term and then the triangle inequality completes the proof. 
    \end{proof} 
    
    Theorem~\ref{thm:predictor_stability} separates prediction perturbation into three distinct channels: perturbation of the learned dual variables, approximation of the train--test kernel vector, and perturbation of the SVM intercept. The first two are controlled by the preceding kernel and optimization analysis. The remaining intercept term can also be controlled under a standard support-vector stability condition. 
    
    \begin{corollary}[Bias perturbation]
    \label{cor:intercept} 
        Suppose there exists an index $j$ satisfying $0 \!<\! \alpha_j^\circ \!<\! C$ and $0 \!<\! \widetilde\alpha_{r,j}^{(N_{\mathrm{s}})} \!<\! C$. Then the following bound holds
        \begin{align} 
            \big| b^\circ - \widetilde b_r^{(N_{\mathrm{s}})} \big| \le \kappa\sqrt n \operatorname{dist} \big( \widetilde{\boldsymbol \alpha}_r^{(N_{\mathrm{s}})}, \mathcal A^{Q,\star} \big) + C\sqrt n \big\| \mathbf k_{\mathbf x_j}^Q - \widetilde{\mathbf k}_{r,\mathbf x_j}^{Q,(N_{\mathrm{s}})} \big\|_2. 
        \label{eq:intercept_perturbation} 
        \end{align} 
    \end{corollary} 
    \begin{proof} 
        Since $j$ is a non-bound support vector for both solutions, the KKT conditions imply 
        \begin{subequations}
        \begin{align} 
            & b^\circ = y_j - \sum_{i=1}^n \alpha_i^\circ y_i \kappa^Q(\mathbf x_i,\mathbf x_j), \\
            & \widetilde b_r^{(N_{\mathrm{s}})} = y_j - \sum_{i=1}^n \widetilde\alpha_{r,i}^{(N_{\mathrm{s}})} y_i \widetilde\kappa_r^{Q,(N_{\mathrm{s}})} (\mathbf x_i,\mathbf x_j). 
        \end{align}
        \end{subequations}
        Adding and subtracting $\sum_{i=1}^n \widetilde\alpha_{r,i}^{(N_{\mathrm{s}})} y_i \kappa^Q(\mathbf x_i,\mathbf x_j)$ and then applying the same Cauchy--Schwarz argument used in Theorem~\ref{thm:predictor_stability} yields (\ref{eq:intercept_perturbation}).
    \end{proof} 
    
\paragraph{Margin preservation and classification stability.} 
    Exact recovery of the decision value is stronger than what is required for classification. A prediction remains unchanged whenever the perturbation is smaller than the exact classification margin. 
    
    \begin{corollary}[Margin preservation] 
    \label{cor:margin} 
        Fix $\mathbf z\in\mathcal X$ with label $y\in{-1,+1}$. Suppose the exact classifier satisfies $y f_{\mathbf K^Q}(\mathbf z) \!\ge\! \gamma \!>\! 0$. If 
        \begin{align} 
            \kappa\sqrt n \operatorname{dist} \big( \widetilde{\boldsymbol\alpha}_r^{(N_{\mathrm{s}})}, \mathcal A^{Q,\star} \big) + C\sqrt n \big\| \mathbf k_{\mathbf z}^Q - \widetilde{\mathbf k}_{r,\mathbf z}^{Q,(N_{\mathrm{s}})} \big\|_2 + \big| b^\circ - \widetilde b_r^{(N_{\mathrm{s}})} \big| < \gamma, 
        \label{eq:margin_condition} 
        \end{align} 
        then $\operatorname{sign} \big( f_r^{Q,(N_{\mathrm{s}})}(\mathbf z) \big) = \operatorname{sign} \big( f_{\mathbf K^Q}(\mathbf z) \big) = y$.
    \end{corollary} 
    \begin{proof}
        By Theorem~\ref{thm:predictor_stability} and condition (\ref{eq:margin_condition}) imply 
        \begin{align} 
            \big| f_{\mathbf K^Q}(\mathbf z) - f_r^{Q,(N_{\mathrm{s}})}(\mathbf z) \big| < \gamma. 
        \end{align} 
        Therefore, we can obtain
        \begin{align} 
            y f_r^{Q,(N_{\mathrm{s}})}(\mathbf z) \ge y f_{\mathbf K^Q}(\mathbf z) - \big| f_{\mathbf K^Q}(\mathbf z) - f_r^{Q,(N_{\mathrm{s}})}(\mathbf z) \big| > 0. 
        \end{align} 
        Consequently, the two classifiers assign the same label.
    \end{proof} 
    
    Corollary~\ref{cor:margin} explains why sizeable kernel-reconstruction errors need not translate directly into comparable losses in classification accuracy. Exact reproduction of every kernel value is unnecessary; only perturbations sufficiently large to cross the decision boundary can change the predicted label. 
    
\paragraph{Joint shot--rank convergence.} 
    We now combine the training-side and out-of-sample perturbation results into an end-to-end convergence statement. We first state the result for a fixed finite training set of size $n$. If the training sample size is also allowed to increase, the explicit $n$-dependent factors appearing in the preceding bounds must additionally be controlled. Let $\left\{ \left( r_\ell, N_{\mathrm{s},\ell} \right) \right\}_{\ell \ge 1}$ denote a sequence of rank and shot-budget configurations with $r_\ell \le \rho := \operatorname{rank}(\mathbf K^Q)$.
    
    \begin{corollary}[Joint shot--rank convergence]
    \label{cor:joint_convergence}
    
        Suppose, along this sequence, 
        \begin{align}
        \label{eq:conv_structural_conv_factor_conv_testshots}
            \varepsilon_{r_\ell} \big( \widehat{\mathcal P}_{r_\ell} \big) \to 0, \quad
            \Delta_{r_\ell} \to 0, \quad
            \frac{ r_\ell \log \left( 2r_\ell/\delta_{z,\ell} \right) }{ N_{\mathrm{s},\ell} } \to 0. 
        \end{align} 
        Assume furthermore that the out-of-sample pivot blocks and exact path-conditioned factors remain uniformly controlled: 
        \begin{align}
        \label{eq:uniform_oos_condition_uniform_factor_condition}
            \inf_\ell \nu_{r_\ell} \ge \nu_0 > 0, \quad \sup_\ell \beta_{r_\ell} \le \beta_0 < \infty. 
        \end{align} 
        Assume additionally either $\big| b^\circ - \widetilde b_{r_\ell}^{(N_{\mathrm{s},\ell})} \big| \to 0$
        or that the shared non-bound support-vector condition of Corollary~\ref{cor:intercept} holds and the corresponding train--test kernel-vector error vanishes. Then, for every fixed $\mathbf z\in\mathcal X$, 
        \begin{align} 
            f_{r_\ell}^{Q,(N_{\mathrm{s},\ell})} (\mathbf z) \longrightarrow f_{\mathbf K^Q}(\mathbf z) 
        \label{eq:decision_convergence} 
        \end{align} 
        on the joint training and test-time concentration events. If, in addition, 
        \begin{align} 
            y f_{\mathbf K^Q}(\mathbf z) \ge \gamma > 0, 
        \end{align} 
        then for all sufficiently large $\ell$, $\operatorname{sign} \big( f_{r_\ell}^{Q,(N_{\mathrm{s},\ell})} (\mathbf z) \big) = y$. If the training concentration event has probability at least $1-\delta_\ell$ and the conditional test-time event has probability at least $1-\delta_{z,\ell}$, then the corresponding joint event has probability at least $1-\delta_\ell-\delta_{z,\ell}$.
        In particular, if $\delta_\ell+\delta_{z,\ell} \to 0$ then (\ref{eq:decision_convergence}) holds in probability.
    \end{corollary} 
    \begin{proof} 
        By Theorem~\ref{thm:total_kernel_error}, we have
        \begin{align} 
            \mathcal E_{r_\ell}^{(N_{\mathrm{s},\ell})} \le \varepsilon_{r_\ell} \left( \widehat{\mathcal P}_{r_\ell} \right) + \left( 2\beta_{r_\ell} + \Delta_{r_\ell} \right) \Delta_{r_\ell}. 
        \end{align} 
        Conditions (\ref{eq:conv_structural_conv_factor_conv_testshots}) therefore imply $\mathcal E_{r_\ell}^{(N_{\mathrm{s},\ell})} \to 0$.
        Under Assumption~\ref{ass:quadratic_growth}, Theorem~\ref{thm:distance_to_optset} then yields 
        \begin{align} 
            \operatorname{dist} \!\big( \widetilde{\boldsymbol\alpha}_{r_\ell}^{(N_{\mathrm{s},\ell})}, \mathcal A^{Q,\star} \big) \to 0. 
        \label{eq:optimizer_converges} 
        \end{align} 
        For the exact out-of-sample term, Lemma~\ref{lem:exact_train_test} and $\varepsilon_{r_\ell} \!\big( \widehat{\mathcal P}_{\!r_\ell} \big) \!=\! \big\| \mathbf R_{r_\ell}^{\widehat{\mathcal P}_{r_\ell}} \big\|_2$ yields, because $n$ is fixed, 
        \begin{align} 
            \big\| \mathbf k_{\mathbf z}^Q - \widetilde{\mathbf k}_{r_\ell,\mathbf z}^{Q,\widehat{\mathcal P}_{r_\ell}} \big\|_2 \le \sqrt{ n\kappa \varepsilon_{r_\ell} \left( \widehat{\mathcal P}_{r_\ell} \right) } \to 0. 
        \label{eq:exact_oos_converges} 
        \end{align} 
        Condition (\ref{eq:conv_structural_conv_factor_conv_testshots}) implies 
        \begin{align} 
            \sqrt{r_\ell} \eta_{z,\ell} = \sqrt{ \frac{ r_\ell \log \left( 2r_\ell/\delta_{z,\ell} \right) }{ 2N_{\mathrm{s},\ell} } } \to 0. 
        \end{align} 
        Together with $\Delta_{r_\ell}\to 0,  \nu_{r_\ell} \ge \nu_0 > 0, \beta_{r_\ell} \le \beta_0$ Theorem~\ref{thm:oos_finite} therefore yields 
        \begin{align} 
            \big\| \widetilde{\mathbf k}_{r_\ell, \mathbf z}^{Q,\widehat{\mathcal P}_{r_\ell}} - \widetilde{\mathbf k}_{r_\ell,\mathbf z}^{Q,(N_{\mathrm{s},\ell})} \big\|_2 \to 0. 
        \end{align} 
        Combining this with (\ref{eq:exact_oos_converges}) we can obtain
        \begin{align} 
            \big\| \mathbf k_{\mathbf z}^Q - \widetilde{\mathbf k}_{r_\ell,\mathbf z}^{Q,(N_{\mathrm{s},\ell})} \big\|_2 \to 0. 
        \label{eq:total_oos_converges} 
        \end{align} 
        The optimizer term (\ref{eq:optimizer_converges}), the kernel-vector term (\ref{eq:total_oos_converges}), and the assumed intercept convergence together imply through Theorem~\ref{thm:predictor_stability} that
        \begin{align} 
            \big| f_{\mathbf K^Q}(\mathbf z) - f_{r_\ell}^{Q,(N_{\mathrm{s},\ell})}(\mathbf z) \big| \longrightarrow 0. 
        \end{align} 
        This proves (\ref{eq:decision_convergence}). If the exact decision value has margin $\gamma>0$, the classification statement follows from Corollary~\ref{cor:margin}. Finally, the training concentration event fails with probability at most $\delta_\ell$, while the test-time event fails conditionally with probability at most $\delta_{z,\ell}$. A union bound gives a joint success probability of at least $1-\delta_\ell-\delta_{z,\ell}$.
    \end{proof} 
    
\paragraph{Relation to the training shot requirement.} 
    Under the stable-path conditions of Theorem~\ref{thm:factor_perturbation}, $\Delta_r \!\le\! \sqrt n C_{r, \tau, B, \kappa} \eta_s$. For a fixed finite training set and a fixed stable rank, therefore, $N_{\mathrm{s}} \!\to\! \infty$ implies $\eta_s \!\to\! 0$ and hence $\Delta_r \!\to\! 0$. If the approximation rank or training-set size is also allowed to vary, it is not sufficient to control only $\eta_s$. The rank-dependent perturbation constant $C_{r,\tau,B,\kappa}$ and the explicit $n$-dependent factors in the preceding bounds must remain controlled as well. A sufficient training-measurement requirement is therefore 
    \begin{align} 
        C_{r, \tau, B, \kappa} \sqrt{ \frac{ n\log(2M_r/\delta) }{ N_{\mathrm{s}} } } \longrightarrow 0. 
    \label{eq:growing_training_requirement} 
    \end{align} 
    For the fixed-data regime emphasized here, $r \!\le\! \rho \!\le\! n$ and $M_r$ remain finite. Consequently, increasing the shot budget removes the finite-shot contribution at every fixed executed rank. 
    
\paragraph{Full-rank special case.} 
    Let $\rho \!=\! \operatorname{rank}(\mathbf K^Q)$. If an exact comparator path executes $\rho$ strictly positive pivots, then Proposition~\ref{prop:exactness_ICA} gives $\mathbf R_\rho \!=\! \mathbf 0$ and hence $\varepsilon_\rho \big( \widehat{\mathcal P}_\rho \big) \!=\! 0$. Lemma~\ref{lem:exact_train_test} then implies $\mathbf k_{\mathbf z}^Q \!=\! \widetilde{\mathbf k}_{\rho,\mathbf z}^{Q,\widehat{\mathcal P}_\rho}$ for every fixed test point satisfying the kernel-diagonal bound. Thus, at full exact rank, the structural training and out-of-sample errors vanish completely. Under the stability and conditioning assumptions, increasing $N_{\mathrm{s}}$ then removes the remaining finite-shot factor and pivot--test measurement errors. 
    
    The preceding results complete the error-propagation analysis of ICAQKL. Finite-shot quantum measurements first perturb the adaptively queried training-kernel entries. These perturbations affect the residual scores and recursively constructed Cholesky factor, which determine the spectral error of the low-rank training kernel. The resulting kernel error controls both the QSVM objective and the distance of the finite-shot optimizer to the exact optimal set. Prediction introduces two additional ingredients: fresh pivot--test quantum measurements and the conditioning of the triangular Cholesky extension. Once these terms and the SVM intercept are controlled, the finite-shot decision function converges to its exact-kernel counterpart, and any test point separated from the decision boundary by a nonzero margin eventually retains the same classification. This decomposition exposes the two principal resources controlled by ICAQKL. The approximation rank determines \emph{which} quantum-kernel similarities are acquired, whereas the shot budget determines \emph{how accurately} those selected similarities are measured. The theory therefore connects adaptive kernel acquisition and quantum measurement precision directly to the stability of the final QSVM classifier.

\subsection{Reproducibility Details}
\paragraph{Datasets and binary tasks.} 
    We evaluate the proposed ICD-based quantum kernel SVM on tabular, handwritten-digit, fashion-image, and biomedical-image classification benchmarks. The standard benchmarks include the Wisconsin Diagnostic Breast Cancer (WDBC) dataset~\cite{WDBC1995}, MNIST/Digits, and FashionMNIST~\cite{xiao2017fashionmnist}. Biomedical experiments use MedMNIST v2~\cite{yang2023medmnistv2}, including BreastMNIST, PneumoniaMNIST, PathMNIST, and OrganAMNIST. To keep a consistent binary-classification setting, multiclass datasets are restricted to specified class pairs in the main benchmark. For example, PathMNIST and OrganAMNIST are evaluated on classes \(0\) and \(1\). The selected labels are then remapped to \(\{0,1\}\). For datasets without predefined validation partitions, we use train/validation/test splits that preserve class proportions. MedMNIST experiments retain the official train/validation/test partitions before any optional computational subsampling. When dataset caps are needed because of the cost of quantum-kernel evaluation, samples are selected without replacement using the experiment seed, and the same partition is used by all methods compared in that run.

\paragraph{PCA-Based Quantum Input Construction}
    The original datasets generally have substantially more input dimensions than the number of qubits available to the quantum feature map. For example, image datasets contain hundreds or thousands of pixel features after flattening, whereas our quantum circuits use only a small number of qubits. We therefore employ principal component analysis (PCA) as a deterministic, unsupervised dimensionality-reduction step that maps each input to a $q$-dimensional representation, where $q$ is the number of qubits used by the quantum kernel circuit.

    \emph{Training-only preprocessing.} 
    To prevent information leakage, all preprocessing transformations are estimated exclusively from the training partition. Let 
    \begin{align} 
        X_{\mathrm{tr}} = [x_1^\top,\ldots,x_n^\top]^\top \in\mathbb{R}^{n\times d} 
    \end{align} 
    denote the training data, where $d$ is the original input dimension. We first standardize every feature using the training-set mean and standard deviation, 
    \begin{align} 
        \widehat{x}_{ij} = \frac{x_{ij}-\mu_j}{\sigma_j}, 
    \end{align} 
    where 
    \begin{align} 
        \mu_j = \frac{1}{n}\sum_{i=1}^{n}x_{ij}, \quad \sigma_j^2 = \frac{1}{n}\sum_{i=1}^{n}(x_{ij}-\mu_j)^2. 
    \end{align} 
    The same $(\mu_j,\sigma_j)$ estimated from the training partition are subsequently applied to validation and test examples. 
    
    \emph{Principal-component projection.} 
    PCA is then fitted to the standardized training matrix. Let $\widehat X_{\mathrm{tr}} \in\mathbb{R}^{n\times d}$ denote the standardized data and consider the empirical covariance matrix 
    \begin{align} 
        \Sigma = \frac{1}{n-1} \widehat X_{\mathrm{tr}}^\top \widehat X_{\mathrm{tr}}. 
    \end{align} 
    Let $\lambda_1\ge\lambda_2\ge\cdots\ge\lambda_d\ge 0$ be the eigenvalues of $\Sigma$ with corresponding orthonormal eigenvectors $v_1,\ldots,v_d$. PCA retains the first $q$ eigenvectors, 
    \begin{align} 
        V_q=[v_1,\ldots,v_q]\in\mathbb{R}^{d\times q}, 
    \end{align} 
    and maps each standardized input $\widehat x$ to
    \begin{align} 
        z = V_q^\top \widehat x \in\mathbb{R}^{q}. 
    \end{align} 
    Thus, the PCA dimension is chosen to equal the number of qubits: $d_{\mathrm{PCA}}=q$. Unless otherwise stated, we use $q=6$. Experiments employing a different circuit width use the corresponding PCA dimension; for example, a four-qubit configuration uses four principal components.

    \emph{Explained variance.}
    The fraction of training-set variance retained by the $q$-dimensional representation is 
    \begin{align} 
        \mathrm{EVR}(q) = \frac{\sum_{j=1}^{q}\lambda_j} {\sum_{j=1}^{d}\lambda_j}. 
    \end{align} 
    We regard this quantity as a preprocessing diagnostic rather than an optimization objective: the number of retained components is determined by the quantum circuit width, not by selecting a variance threshold. Reporting $\mathrm{EVR}(q)$ for each dataset provides a direct measure of how much information is compressed before the quantum kernel is evaluated. 
    
    \emph{Angle rescaling.} 
    The PCA coordinates generally take values on an unbounded real scale, whereas the quantum feature map encodes them as rotation angles. We therefore fit a component-wise affine rescaling on the PCA-transformed training set. For the $j$th retained principal component, let 
    \begin{align} 
        a_j = \min_{i\in\mathrm{train}} z_{ij}, \quad b_j = \max_{i\in\mathrm{train}} z_{ij}. 
    \end{align} 
    The corresponding quantum input coordinate is 
    \begin{align} 
        \theta_j = \pi \frac{z_j-a_j}{b_j-a_j}, 
    \end{align}
    so that training examples are mapped into $\theta\in[0,\pi]^q$. The same training-derived values $(a_j,b_j)$ are used to transform validation and test data. The complete preprocessing map is therefore 
    \begin{align} 
        x \longmapsto \widehat x \longmapsto V_q^\top\widehat x \longmapsto \theta. 
    \end{align} 
    
    \emph{Connection to the quantum kernel.} 
    The resulting angle vector 
    \begin{align} 
        \theta(x) = [\theta_1(x),\ldots,\theta_q(x)]^\top 
    \end{align} 
    is supplied to the $q$-qubit quantum feature map. In our implementation, the PCA coordinates are encoded through single-qubit $Y$ rotations followed by entangling layers. The fidelity kernel is then 
    \begin{align} 
        k(x,x') = \left| \langle \phi(\theta(x)) \mid \phi(\theta(x')) \rangle \right|^2. 
    \end{align} 
    Consequently, PCA is not used as a classifier and does not access class labels. Its role is solely to construct a low-dimensional, training-derived representation compatible with the available number of qubits.

\paragraph{Pivoted incomplete Cholesky training.} 
    Given the exact training kernel \(K\in\mathbb{R}^{n\times n}\), ICD constructs 
    \begin{align} 
        K \approx \widetilde K_r = G_rG_r^\top, \quad G_r\in\mathbb{R}^{n\times r}, 
    \end{align} 
    using greedy residual-diagonal pivoting. At iteration \(t\), the pivot is selected as 
    \begin{align} 
        p_t=\arg\max_i (R_t)_{ii}, \quad R_t=K-G_tG_t^\top, 
    \end{align} 
    and the procedure terminates either after the prescribed maximum rank \(r\) or when the maximum residual diagonal falls below the numerical tolerance. Unless otherwise stated, we use \(r=30\) and tolerance \(10^{-8}\). Rank-ablation experiments evaluate $r\in\{5,10,15,20,25,30,40\}$. The resulting rows of \(G_r\) serve as explicit low-dimensional kernel features, on which a linear SVM is trained. Unless stated otherwise, the SVM regularization parameter is \(C=1\). 
    
\paragraph{Exact-kernel comparison protocol.} 
    Because construction of the exact quantum kernel requires \(O(n^2)\) fidelity evaluations, Full-QSVM is additionally evaluated on a common capped training subset when a full dense quantum kernel is computationally prohibitive. We refer to this setting as \emph{Full-QSVM (subset)}. ICD-QSVM and Nystr\"om-QSVM are also rerun on the \emph{same samples} when evaluating approximation error against the exact kernel. Consequently, quantities such as 
    \begin{align} 
        \frac{\|K-\widetilde K_r\|_F}{\|K\|_F} 
    \end{align} 
    are computed on identical matrices and identical training samples. In contrast, the ``ICD-full'' result denotes ICD-QSVM trained on the complete available training partition and is intended to measure the practical scalable performance of the proposed method rather than to provide a sample-size-matched comparison with the exact dense kernel. 
    
\paragraph{Kernel-evaluation accounting.} 
    We distinguish kernel-evaluation cost from classical SVM optimization cost. For the exact Full-QSVM on \(n\) training examples, direct dense construction requires \(n^2\) kernel evaluations in our implementation. For rank-\(r\) pivoted Cholesky, kernel entries are evaluated lazily and cached. Kernel symmetry is exploited, and diagonal entries are not queried because fidelity kernels satisfy \(k(x_i,x_i)=1\). Therefore, the number of \emph{new} training-kernel evaluations is 
    \begin{align} 
        N_{\mathrm{ICD}} = nr-\frac{r(r+1)}{2}. 
    \end{align} 
    For example, with \(n=603\) and \(r=30\), 
    \begin{align} 
        N_{\mathrm{ICD}} = 603\times 30-\frac{30\times31}{2} = 17{,}625. 
    \end{align} 
    For a capped subset with \(n=120\) and \(r=30\), the corresponding count is \(3{,}135\), compared with \(14{,}400\) entries for direct dense construction. We report wall-clock fit time separately because simulator runtime also depends on the quantum-circuit backend and software overhead. 
    
\paragraph{Finite-shot experiments.} 
    To assess sensitivity to measurement noise, we additionally replace analytic fidelity probabilities with finite-shot estimates. For a shot budget \(S\), every required kernel value is estimated from \(S\) circuit measurements. We study several shot budgets, e.g., $S\in\{64,128,256,512\}$, together with the analytic-kernel reference. Because finite-shot estimation introduces stochastic perturbations into the kernel matrix, these experiments report both predictive accuracy and approximation quality as functions of \(S\). Analytic and finite-shot results are kept distinct in all tables and figures. 
    
\paragraph{Reproducibility and randomness control.} 
    Randomness arises from dataset splitting or computational subsampling, capped-subset selection, and, where applicable, finite-shot measurement. Every experiment is indexed by an explicit random seed. For multi-seed experiments, the complete data-splitting and sampling procedure is repeated independently for each seed rather than changing only the downstream classifier initialization. We report the mean and standard deviation across seeds,
    \begin{align} 
        \bar m = \frac{1}{S_{\mathrm{seed}}} \sum_{s=1}^{S_{\mathrm{seed}}}m_s, \quad \sigma_m = \sqrt{ \frac{1}{S_{\mathrm{seed}}-1} \sum_{s=1}^{S_{\mathrm{seed}}} (m_s-\bar m)^2 }, 
    \end{align} 
    for accuracy, balanced accuracy, approximation error, and runtime where appropriate. All competing methods within a seed use the same training, validation, and test samples. The PCA transformation is refit independently on the corresponding training partition for each seed, thereby reproducing the complete learning pipeline rather than reusing preprocessing statistics across repeated trials.
    
    \begin{table}[h]
    \centering
    \caption{Default experimental and reproducibility settings. Dataset-specific deviations are stated with the corresponding experiment.}
    \label{tab:reproducibility}
    \small
    \begin{tabular}{ll}
    \toprule
        \textbf{Setting} & \textbf{Value} \\
    \midrule
        Quantum kernel & State fidelity \\
        Default number of qubits & \(q=6\) \\
        Quantum feature-map depth & \(2\) layers \\
        Input preprocessing & Standardization + PCA + rescaling \\
        PCA fitting data & Training partition only \\
        Quantum angle range & \([0,\pi]\) \\
        Default ICD rank & \(r=30\) \\
        Rank ablation & \(\{5,10,15,20,25,30,40\}\) \\
        ICD tolerance & \(10^{-8}\) \\
        SVM regularization & \(C=1\) unless otherwise stated \\
        Finite-shot budgets & \(\{64,128,256,512\}\) \\
        Analytic reference & \(S=\infty\) / analytic probability \\
        ICD diagonal convention & \(K_{ii}=1\) \\
        ICD train-query count & \(nr-r(r+1)/2\) \\
        Exact subset size & \(120\) or dataset-specific value \\
        Reported predictive metrics & Accuracy, balanced accuracy, F1 \\
        Approximation metric & Relative Frobenius error \\
        Repeated-run reporting & Mean \(\pm\) standard deviation \\
    \bottomrule
    \end{tabular}
    \end{table}

\end{document}